\documentclass[preprintnumbers,prd,floatfix,nofootinbib,superscriptaddress]{revtex4}

\usepackage[a4paper,
			left=20mm,
			right=20mm,
			top=20mm,
			bottom=20mm]
			{geometry} 

\usepackage[tbtags]{amsmath}  
\usepackage{mathrsfs}
\usepackage{tabularx}
\usepackage{amssymb}          
\usepackage{bm}               
\usepackage{graphicx}         
\usepackage[export]{adjustbox}
\usepackage{hhline,multirow}  
\usepackage{makecell}  
\usepackage{dcolumn}          
\usepackage{slashed}
\usepackage{datetime}
\usepackage{color}
\usepackage[dvipsnames]{xcolor}
\usepackage{slashed}
\usepackage{subfloat}

\usepackage{amscd}            
\usepackage{epsfig}
\usepackage{dcolumn}          
\usepackage[dvipsnames]{xcolor}
\usepackage[normalem]{ulem}
\usepackage{mathtools}
\usepackage{esint}
\usepackage{comment}
\usepackage[utf8]{inputenc}

\RequirePackage{color}
\RequirePackage[colorlinks=true
,urlcolor=black
,anchorcolor=black
,citecolor=black
,filecolor=black
,linkcolor=black
,menucolor=black
,pagecolor=black
,linktocpage=black
,pdfproducer=medialab
,pdfa=true
]{hyperref}

\newcommand{\nocontentsline}[3]{}
\newcommand{\tocless}[2]{\bgroup\let\addcontentsline=\nocontentsline#1{#2}\egroup}

\newcommand{\hermes}{\rm HERMES}
\newcommand{\compass}{\rm COMPASS}

\newcommand{\bb}{b}

\newcommand{\F}{\hat{f}_1}

\newcommand{\kperp}{\boldsymbol{k}_\perp}
\newcommand{\modkperp}{|\boldsymbol{k}_\perp|}
\newcommand{\bT}{\boldsymbol{b}_T}
\newcommand{\modbT}{|\boldsymbol{b}_T|}

\newcommand{\Pperp}{\boldsymbol{P}_\perp}
\newcommand{\modPperp}{|\boldsymbol{P}_\perp|}
\newcommand{\qT}{\bm{q}_{T}}
\newcommand{\T}{\perp}
\newcommand{\PhT}{\bm{P}_{hT}}
\newcommand{\modqT}{|\bm{q}_{T}|}

\allowdisplaybreaks[2]

\begin{document}

\title{Flavor dependence of unpolarized quark \\ Transverse Momentum Distributions from a global fit \\ \vspace{0.2cm}
\normalsize{\textmd{The \textbf{MAP} (Multi-dimensional Analyses of Partonic distributions) Collaboration}}}


\preprint{JLAB-THY-24-4066}

\author{Alessandro Bacchetta}
\thanks{E-mail: alessandro.bacchetta@unipv.it -- \href{https://orcid.org/0000-0002-8824-8355}{ORCID: 0000-0002-8824-8355}}
\affiliation{Dipartimento di Fisica, Universit\`a di Pavia, via Bassi 6, I-27100 Pavia, Italy}
\affiliation{INFN - Sezione di Pavia, via Bassi 6, I-27100 Pavia, Italy}

\author{Valerio Bertone}
\thanks{E-mail: valerio.bertone@cea.fr -- \href{https://orcid.org/0000-0003-0148-0272}{ORCID: 0000-0003-0148-0272}}
\affiliation{IRFU, CEA, Universit\'e Paris-Saclay, F-91191 Gif-sur-Yvette, France}

\author{Chiara Bissolotti}
\thanks{E-mail: cbissolotti@anl.gov -- \href{https://orcid.org/0000-0003-3061-0144}{ORCID: 0000-0003-3061-0144}}
\affiliation{Argonne National Laboratory, Lemont, IL, USA}

\author{Giuseppe Bozzi}
\thanks{E-mail: giuseppe.bozzi@unica.it -- \href{https://orcid.org/0000-0002-2908-6077}{ORCID: 0000-0002-2908-6077}}
\affiliation{Dipartimento di Fisica, Universit\`a di Cagliari, Cittadella Universitaria, I-09042, Monserrato (CA), Italy}
\affiliation{INFN - Sezione di Cagliari, Cittadella Universitaria, I-09042, Monserrato (CA), Italy}

\author{Matteo Cerutti}
\thanks{E-mail: mcerutti@jlab.org -- \href{https://orcid.org/0000-0001-7238-5657}{ORCID: 0000-0001-7238-5657}}
\affiliation{Hampton University, Hampton, Virginia 23668, USA}
\affiliation{Jefferson Lab, Newport News, Virginia 23606, USA}

\author{Filippo Delcarro}
\thanks{E-mail: filippo.delcarro@cern.ch -- \href{https://orcid.org/0000-0001-7636-5493}{ORCID: 0000-0001-7636-5493}}
\affiliation{Dipartimento di Fisica, Universit\`a di Pavia, via Bassi 6, I-27100 Pavia, Italy}
\affiliation{INFN - Sezione di Pavia, via Bassi 6, I-27100 Pavia, Italy}

\author{Marco Radici}
\thanks{E-mail: marco.radici@pv.infn.it -- \href{https://orcid.org/0000-0002-4542-9797}{ORCID: 0000-0002-4542-9797}}
\affiliation{INFN - Sezione di Pavia, via Bassi 6, I-27100 Pavia, Italy}

\author{Lorenzo Rossi}
\thanks{E-mail: lorenzo.rossi@pv.infn.it -- \href{https://orcid.org/0000-0002-8326-3118}{ORCID: 0000-0002-8326-3118}}
\affiliation{Dipartimento di Fisica, Universit\`a di Pavia, via Bassi 6, I-27100 Pavia, Italy}
\affiliation{INFN - Sezione di Pavia, via Bassi 6, I-27100 Pavia, Italy}

\author{Andrea Signori}
\thanks{E-mail: andrea.signori@unito.it -- \href{https://orcid.org/0000-0001-6640-9659}{ORCID: 0000-0001-6640-9659}}
\affiliation{Department of Physics, University of Turin, via Pietro Giuria 1, I-10125 Torino, Italy}
\affiliation{INFN, Section of Turin, via Pietro Giuria 1, I-10125 Torino, Italy}

\begin{abstract}
We present an extraction of the unpolarized transverse-momentum-dependent
parton distribution and fragmentation functions that takes into account
possible differences between quark flavors and final-state hadrons.
The extraction is based on experimental measurements from Drell-Yan processes
and semi-inclusive deep-inelastic scattering, whose combination is essential
to distinguish flavor differences. The analysis is carried out at N$^3$LL
accuracy. The extracted flavor-dependent distributions give a very good
description of the data ($\chi^2/N_{\rm dat} = 1.08$). The resulting
error bands take fully into account also the uncertainties in the
determination of the corresponding collinear distributions.
\end{abstract}

\maketitle
\newpage
\tableofcontents

\section{Introduction}
\label{s:intro}

The transverse-momentum distributions (TMDs) provide insights into the three-dimensional structure of hadrons in momentum space, and are fundamental in understanding the world at the subatomic level. Thanks to the wealth of experimental measurements and the development of a robust theoretical framework, the study of TMDs has witnessed remarkable progress in recent years, and accurate phenomenological extractions for unpolarized quark TMDs in the proton are available~\cite{Bacchetta:2017gcc,Scimemi:2017etj,Bertone:2019nxa,Scimemi:2019cmh,Bacchetta:2019sam,Bury:2022czx,Bacchetta:2022awv,Moos:2023yfa}. TMDs were also studied in a different framework, the so called parton-branching approach~\cite{BermudezMartinez:2018fsv,BermudezMartinez:2019anj,BermudezMartinez:2020tys}. The outcomes of these studies are partly available in the public {\tt TMDlib} library~\cite{Hautmann:2014kza,Abdulov:2021ivr} (for a review, see also Ref.~\cite{Angeles-Martinez:2015sea}).
Despite this advancement, there is still a lack of knowledge regarding the transverse momentum distribution of different quark flavors, and we are unable to clearly answer the question: do certain quark flavors carry more transverse momentum than others?

The question is legitimate because global extractions of collinear parton distribution functions (PDFs) clearly show that the distribution of longitudinal fractional momentum of partons strongly depends on their flavor (see Ref.~\cite{Ethier:2020way} for a recent review); similarly, for collinear fragmentation functions (FFs)~\cite{Albino:2008aa,Metz:2016swz}. Moreover, there is no theoretical principle that prevents the transverse-momentum distribution of partons from having a similar behavior.

In this article, we aim to shed light on the variations in TMDs across different quark flavors. To achieve this goal, we compare theoretical predictions with experimental data from two distinct processes: Drell-Yan (DY) lepton-pair production and semi-inclusive deep-inelastic scattering (SIDIS). In relation to our goal, the two processes are highly complementary. On the one side, DY interactions do not involve hadrons in the final state and do not depend on TMD fragmentation functions (TMD FFs), but they offer valuable insight into TMD distribution functions (TMD PDFs) of quark-antiquark pairs. On the other side, SIDIS processes imply detecting final-state hadrons, and through TMD FFs they are particularly sensitive to flavor differences.
The combination of these two processes is essential for our global analysis that incorporates for the first time all the necessary ingredients to reach a full N$^3$LL accuracy in the theoretical description of both DY and SIDIS processes. We remark also that TMDs depend on collinear PDFs and FFs: in this analysis, we take fully into account the uncertainties on these quantities by using all members of Monte Carlo PDF and FF sets. This procedure was already applied to DY in Ref.~\cite{Bury:2022czx} and is applied here for the first time to SIDIS. We obtain more realistic estimates of the uncertainties on the extracted TMDs.

In the literature, the problem of flavor-dependent TMDs has been addressed through models, lattice QCD calculations, and data-driven extractions. Some model calculations (see Ref.~\cite{Burkardt:2015qoa} for a review) predict different TMDs for different  quarks~\cite{Bacchetta:2008af,Bacchetta:2010si,Wakamatsu:2009fn,Efremov:2010mt,Bourrely:2010ng,Matevosyan:2011vj,Schweitzer:2012hh}, although others do not~\cite{Pasquini:2008ax,Lorce:2011dv,Avakian:2010br}. The only pioneering work in lattice QCD on the subject indicates that down quarks carry higher transverse momentum than up quarks~\cite{Musch:2010ka}.

Earlier phenomenological extractions of flavor-dependent TMDs have been attempted in Refs.~\cite{Signori:2013mda,Bury:2022czx,Moos:2023yfa}. Ref.~\cite{Signori:2013mda} considered only a limited amount of data from SIDIS in a parton-model framework and concluded that there was room for a flavor dependence of TMDs, especially for the TMD FFs, but it was not possible to constrain it well, given the mentioned limitations. Refs.~\cite{Bury:2022czx,Moos:2023yfa} considered only data from DY, which has a reduced sensitivity to flavor differences.

By unraveling flavor-specific differences in transverse-momentum distributions, improving the theoretical accuracy of both DY and SIDIS cross sections to a full N$^3$LL level, and taking fully into account the uncertainties on collinear distributions, we take a significant step towards a more complete and precise understanding of the fundamental building blocks of matter. Our study not only contributes to the understanding of the internal structure of hadrons but also has broader implications for the interpretation of high-energy physics phenomena, such as the determination of the $W$ mass in hadronic collisions~\cite{Bacchetta:2018lna,Bozzi:2019vnl,Rottoli:2023xdc}. It also paves the way for a deeper understanding of SIDIS experimental results at the future Electron-Ion Collider (EIC)~\cite{AbdulKhalek:2021gbh,AbdulKhalek:2022erw,Burkert:2022hjz,Abir:2023fpo}.

\section{Formalism}
\label{s:formalism}

\subsection{Drell--Yan}
\label{ss:DY_Z}
\begin{figure}
\centering
\includegraphics[width=0.6\textwidth]{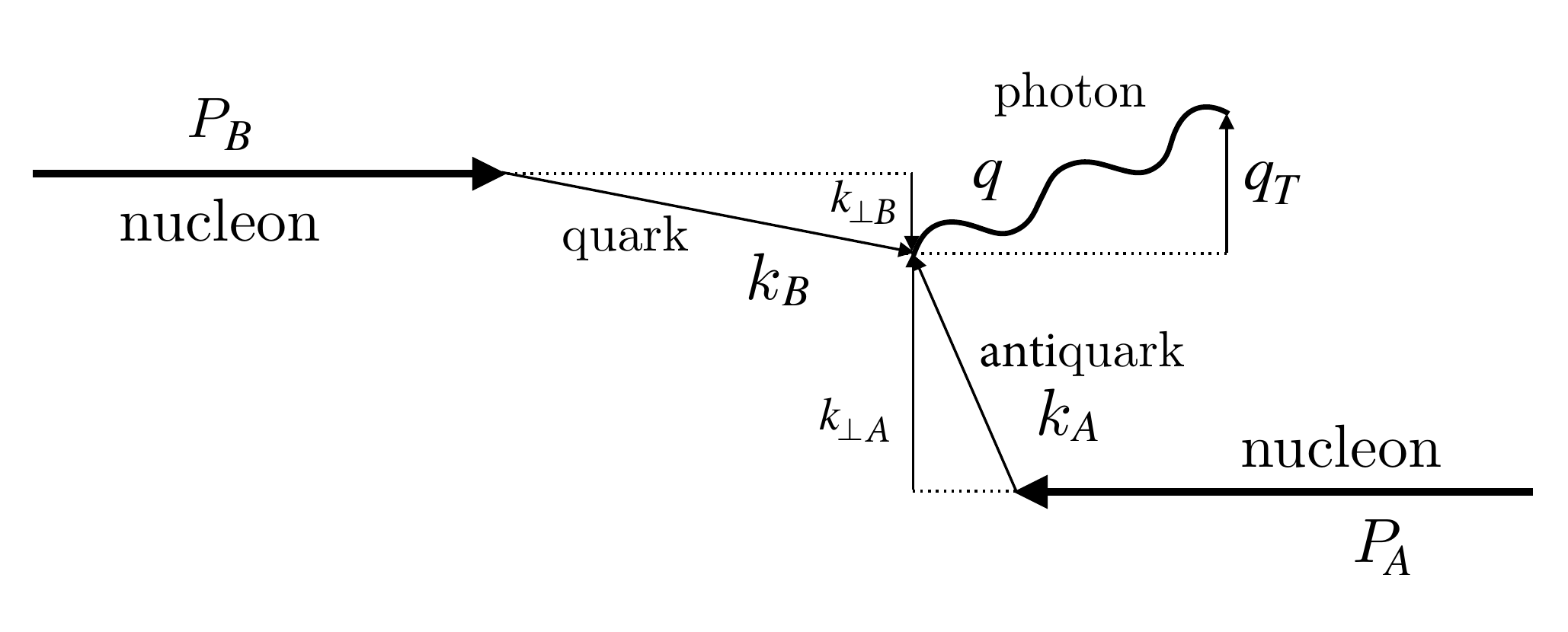}
\caption{Diagram describing the relevant momenta involved in a DY event. In the collision of two nucleons with momenta $P_A$, $P_B$, a quark and an antiquark, with intrinsic (unmeasured) transverse momenta ${\kperp}_A$ and  ${\kperp}_B$, annihilate and produce a virtual vector boson with (measured) transverse momentum $\qT = {\kperp}_A + {\kperp}_B$ with respect to the collision axis.}
\label{f:trans_momenta_DY}
\end{figure}

The inclusive Drell--Yan (DY) process
\begin{equation}
\label{e:DY_Z}
h_A(P_A) + h_B(P_B)\ \longrightarrow\ \gamma^*/Z(q) + X \longrightarrow \ell^+(l)+ \ell^-(l^\prime) + X \; ,
\end{equation}
is the production of a lepton pair with four-momenta $l,l^{\prime}$ from the collision of two hadrons with four-momenta $P_A,P_B$ via an intermediate neutral vector boson $\gamma^*/Z$ with four-momentum $q$ and large invariant mass $Q=\sqrt{q^2}$. The center-of-mass energy squared of the collision is $s=(P_A+P_B)^2$ and the conservation of momentum implies $q=l+l^{\prime}$. The transverse momentum $|\qT|=\sqrt{q_{x}^{2}+q_{y}^{2}}$ of the intermediate boson with respect to the collision axis can be expressed in terms of the intrinsic transverse momenta of the incoming quarks $\qT={\kperp}_{A}+{\kperp}_{B}$, while its rapidity is given by $y=\ln\sqrt{\frac{q_{0}+q_{z}}{q_{0}-q_{z}}}$. The relevant kinematic quantities are schematically depicted in Fig.~\ref{f:trans_momenta_DY}.

We are interested in the inclusive cross section differential with respect to the transverse momentum of the vector boson in the region of small $|\qT|$ ($|\qT| \ll Q$), which can be written as
\begin{equation}
\begin{split}
\label{e:DYZ_xsec}
\frac{d\sigma^{\text{DY}}}{d |\qT|\, dy\, dQ} & = \frac{16 \pi^2 \alpha^2 |\qT|}{9 Q^3}\, {\cal P}\, x_A\, x_B \, {\cal H}^{\text{DY}}(Q,\mu)\, \sum_a c_a(Q^2)
\\
& \times \int d^2 {\kperp}_A\, d^2 {\kperp}_B\, f_1^a(x_A,{\kperp^2}_A;\mu,\zeta_A)\, f_1^{\bar{a}}(x_B,{\kperp^2}_B;\mu,\zeta_B)\, \delta^{(2)}({\kperp}_A + {\kperp}_B - \qT) \; .
\end{split}
\end{equation}
In the first line of Eq.~(\ref{e:DYZ_xsec}), $\alpha$ is the electromagnetic coupling, ${\cal P}$ is a phase-space-reduction factor accounting for possible lepton cuts,\footnote{See Appendix C of Ref.~\cite{Bacchetta:2019sam} for details.} $x_{A}=Qe^{y}/\sqrt{s}$ and $x_{B}=Qe^{-y}/\sqrt{s}$ are the longitudinal momentum fractions carried by the incoming partons, ${\cal H}^{\text{DY}}$ is a perturbative hard factor encoding the virtual part of the scattering and depending on $Q$ and on a renormalization scale $\mu$. The sum runs over all active quark flavors and $c_{a}$ are the  electroweak charges given by
\begin{equation}
\label{e:EW_charges}
c_a(Q^2) = e_a^2 - 2 e_a V_a V_\ell \, \chi_1(Q^2) + (V_\ell^2 + A_\ell^2)\, (V_a^2 + A_a^2)\, \chi_2(Q^2)\; ,
\end{equation}
with
\begin{align}
\label{e:EW_chi_functions}
\chi_1(Q^2) &= \frac{1}{4 \sin^2\theta_W \cos^2\theta_W } \frac{Q^2 ( Q^2 -  M_Z^2 )}{ (Q^2 - M_Z^2)^2 + M_Z^2 \Gamma_Z^2} \; ,\\
\chi_2(Q^2) &= \frac{1}{16 \sin^4\theta_W\cos^4\theta_W} \frac{Q^4}{ (Q^2 - M_Z^2)^2 + M_Z^2 \Gamma_Z^2} \; ,
\end{align}
where $e_a$, $V_a$, and $A_a$ are the electric, vector, and axial charges of the flavor $a$, $V_\ell$ and $A_\ell$ are the vector and axial charges of the lepton $\ell$, $\sin\theta_W$ is the weak mixing angle, $M_Z$ and $\Gamma_Z$ are mass and width of the $Z$ boson.
The second line of Eq.~(\ref{e:DYZ_xsec}) contains the convolution of the unpolarized TMDs $f_1^{a}$ and $f_1^{\bar{a}}$, each one depending on the longitudinal and transverse momenta of the incoming quark/antiquark, and on the renormalization ($\mu$) and rapidity ($\zeta$) scales. The arbitrary choice made for the latter has to satisfy the kinematic constraint $\zeta_A \zeta_B = Q^4$: we will set $\mu^2=\zeta_A= \zeta_B=Q^2$. Finally, the delta function in the second line of Eq.~(\ref{e:DYZ_xsec}) guarantees the conservation of transverse momentum.

The evolution of the TMD PDFs will be addressed in Sec.~\ref{ss:TMDs}. As usual, we work in the conjugate position space ($\bT$ space) by defining the Fourier transform of the TMD PDFs:
\begin{equation}
\label{eq:FTdef}
  \begin{split}
\F^a \big( x, \modbT; \mu, \zeta \big) &= \int d^2 \bm{k}_\T \, e^{i
      \bm{\bb}_T \cdot \bm{k}_\T  } \, f_1^a \big( x, \bm{k}_\T^2; \mu,
    \zeta \big) \,
    \\
      &= 2 \pi \int_0^{\infty} d \modkperp \,\modkperp  J_0(\modbT \modkperp) \, f_1^a \big( x, \kperp^2; \mu,
    \zeta \big), \,
 \end{split}
\end{equation}
where $J_{0}$ is the Bessel function of the first kind. This allows to rewrite the convolution in the second line of Eq.~(\ref{e:DYZ_xsec}) as
\begin{equation}
\frac{1}{2\pi}\, \int_0^{+\infty} d|\bT| |\bT| J_0\big( |\bT| |\qT| \big) \hat{f}_1^a(x_A,\bT^2;\mu,\zeta_A)\, \hat{f}_1^{\bar{a}}(x_B,\bT^2;\mu,\zeta_B).
\end{equation}

\subsection{Semi-inclusive deep-inelastic scattering}
\label{ss:SIDIS}
In the SIDIS process, a lepton with momentum $l$ scatters off a hadron target
$N$ with mass $M$ and four-momentum $P$, and the final state contains the
scattered lepton with momentum $l^\prime$ and the hadron $h$ with mass $M_h$
and four-momentum $P_h$, \textit{i.e.},
\begin{equation}
\label{e:SIDIS}
\ell(l) +  N(P) \rightarrow\ \ell(l^\prime) + h(P_h) + X\, .
\end{equation}
The (space-like) four-momentum transfer $q = l-l^\prime$, with $Q^2 \equiv -q^2 > 0$, is carried by a virtual photon and we consider the standard SIDIS kinematic invariants~\cite{Bacchetta:2006tn,Boglione:2019nwk}:
\begin{align}
\label{e:kin_invariants}
x &= \frac{Q^2}{2\,P\cdot q} \, , &
y &= \frac{P \cdot q}{P \cdot l} \, , &
z &= \frac{P \cdot P_h}{P\cdot q} \, ,
\end{align}
with $s = (P+l)^2$ the invariant mass squared of the process.

\begin{figure}
\centering
\includegraphics[width=0.6\textwidth]{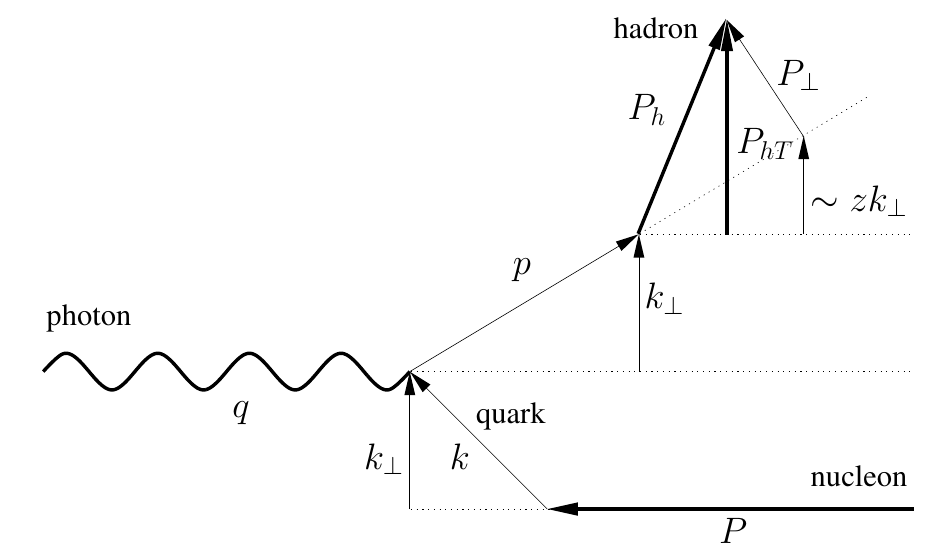}
\caption{Diagram describing the relevant momenta involved in a SIDIS event in the Breit (nucleon-photon) frame. A virtual photon with momentum $q$ (defining the reference axis) strikes a parton with momentum $k$ and (unmeasured) transverse momentum $\kperp$ inside a nucleon with momentum $P$. The struck parton with momentum $p=k+q$ fragments into a hadron with momentum $P_h$, which acquires a further (unmeasured) transverse momentum $\Pperp$ with respect to the fragmenting quark axis. The total (measured) transverse momentum of the final hadron is $\PhT$. In the large $Q^2$ limit, $\PhT \approx z \kperp + \Pperp$.}
\label{f:trans_momenta_SIDIS}
\end{figure}

As for the transverse momenta, we consider the transverse component ($|\bm{P}_{hT}|$) of
the final hadron momentum with respect to $P$ and $q$ or, equivalently, the
transverse component ($|\qT|$) of the virtual photon momentum with respect to $P$ and $P_h$
(see, for instance, Refs.~\cite{Mulders:1995dh,Bacchetta:2004jz,Boer:2011fh}).
The two momenta are related by~\cite{Bacchetta:2008xw,Bacchetta:2019qkv}
\begin{equation}
\label{e:qT_vs_PhT_general}
q_T^\mu = -\frac{P^\mu_{hT}}{z} - 2 x \frac{\modqT^2}{Q^2} P^\mu \, \approx -\frac{P_{hT}^{\mu}}{z} \, ,
\end{equation}
where the last approximation is valid assuming that the invariant mass of the photon is large compared to its transverse momentum ($|\qT| \ll Q$) and the hadron masses involved in the process can be neglected. The relevant kinematic quantities are schematically depicted in Fig.~\ref{f:trans_momenta_SIDIS}.

We are interested in the hadron multiplicity, \textit{i.e.}, the differential number of hadrons of a given species $h$ produced per corresponding inclusive DIS event:
\begin{equation}
\label{e:mult_def}
M(x, z, |\PhT|=z |\qT|, Q)\ =\
\frac{1}{z}\, \frac{d\sigma^{\text{SIDIS}}}{dx\, dz\, d |\qT|\, dQ} \bigg/ \frac{d\sigma^{\text{DIS}}}{dx\, dQ} \, .
\end{equation}

The differential cross section at small transverse momenta, neglecting target mass corrections, reads~\cite{Bacchetta:2006tn,Bacchetta:2017gcc}
\begin{equation}
\label{e:SIDIS_xsec}
\begin{split}
\frac{d\sigma^{\text{SIDIS}}}{dx\, dz\, d |\qT|\, dQ} & = \frac{8\pi^2\, \alpha^2\, z^2\, |\qT|}{2\, x\, Q^3}\, \left[1+\left(1-\frac{Q^2}{xs}\right)^2\right] \, x\,{\cal H}^{\text{SIDIS}}(Q,\mu) \sum_a e_a^2 \\
&\times \int d^2 \kperp\, \int \frac{d^2 \Pperp}{z^2}\,f_1^a(x,\kperp^2;\mu,\zeta_A)\, D_1^{a \to h}(z,\Pperp^2;\mu,\zeta_B)\,\delta^{(2)}(\kperp + \Pperp/z + \qT).
\end{split}
\end{equation}
In the first line of Eq.~(\ref{e:SIDIS_xsec}), the sum runs over all active quark flavors. The hard factor ${\cal H}^{\text{SIDIS}}$ is perturbatively computable and depends on $Q$ and a renormalization scale $\mu$.
The second line contains the convolution of the unpolarized TMD PDF $f_1^a$ as function of the rapidity scale $\zeta_{A}$ and of the transverse momentum $|\kperp|$ of the struck quark with respect to the nucleon axis, and the TMD FF $D_1^{a \to h}$ as function of the rapidity scale $\zeta_{B}$ and of the transverse momentum $|\Pperp|$ of the produced hadron $h$ with respect to the fragmenting quark axis.

Also in this case, it is convenient to work in the conjugate position ($\bT$) space by defining the Fourier transform of the TMD FF:
\begin{align}
 \begin{split}
\label{eq:FTdefFF}
\hat{D}_1^{a \to h}\big( z, \bT^2; \mu, \zeta \big) &=
  \int \frac{d^2 \Pperp}{z^2} \, e^{-i
      \bT  \cdot  \Pperp/z } \, D_1^a \big( z, \Pperp^2; \mu,
    \zeta \big) \,
    \\
      &= 2 \pi \int_0^{\infty} \frac{d \modPperp}{z^2} \,\modPperp  J_0(\modbT \modPperp/z) \, D_1^a \big( z, \Pperp^2; \mu,
    \zeta \big) \, .
 \end{split}
\end{align}
The convolution in the second line of Eq.~(\ref{e:SIDIS_xsec}) can be rewritten as
\begin{align}
\label{e:FUUT_def}
\frac{1}{2\pi}\, \int_0^{+\infty} d|\bT| |\bT| J_0\big( |\bT| |\qT| \big) \hat{f}_1^a(x,\bT^2;\mu,\zeta_A)\, \hat{D}_1^{a \to h}(z,\bT^2;\mu,\zeta_B) \, .
\end{align}

In the TMD extraction of Ref.~\cite{Bacchetta:2022awv}, it was noted that a good description of low transverse-momentum SIDIS data can be achieved in a theoretical formalism where the TMD factorization formula contains the resummation of transverse-momentum logarithms up to the next-to-leading logarithmic (NLL) accuracy. However, it was also remarked that the quality of the description deteriorates when increasing the accuracy beyond NLL, because the predictions undershoot the data by approximately a $q_T$-independent factor.

In Ref.~\cite{Bacchetta:2022awv}, the problem was fixed by incorporating into the definition of the SIDIS multiplicity in Eq.~\eqref{e:mult_def} the normalization factor
\begin{equation}
\label{e:norm_SIDIS_def}
\omega(x,z,Q) = \frac{d\sigma^{\rm nomix}}{dx\, dz\, dQ}  \bigg/  \int d^2 \qT\, W  \, ,
\end{equation}
where the symbol $W$, commonly known as ``W-term'', denotes the differential cross section in Eq.~\eqref{e:SIDIS_xsec}. In other words, the normalization factor $\omega$ is meant to compensate for all contributions in the collinear SIDIS cross section (numerator of Eq.~\eqref{e:norm_SIDIS_def}) that are not included by simply integrating upon transverse momentum the corresponding differential SIDIS cross section (denominator of Eq.~\eqref{e:norm_SIDIS_def}). The collinear SIDIS cross section includes only the terms that do not mix initial- and final-state contributions, hence the ``nomix" label (see Ref.~\cite{Bacchetta:2022awv} for a more complete explanation). Since in our theoretical framework we reach N$^3$LL accuracy (see Sec.~\ref{ss:TMDs}), we consistently include in the numerator terms up to second order in the strong coupling constant $\alpha_s$, {\it i.e.} including ${\cal O}(\alpha_s^2)$ corrections as computed in Ref.~\cite{Abele:2021nyo}. Alternative approaches to the normalization problem are available in the literature~\cite{Vladimirov:2023aot}.

In conclusion, in our analysis we adopt the following expression for the fully differential SIDIS cross-section:
\begin{equation}
\label{e:sidis_xsec_expr_w_norm}
\frac{d\sigma_\omega^{\text{SIDIS}}}{dx\, dz\, d |\qT|\, dQ} = \omega(x,z,Q)\, \frac{d\sigma^{\text{SIDIS}}}{dx\, dz\, d |\qT|\, dQ} \, .
\end{equation}

\subsection{TMD evolution}
\label{ss:TMDs}
The dependence of TMD PDFs and TMD FFs on the renormalization scale $\mu$ and
the rapidity scale $\zeta$ arises from the removal of ultraviolet and rapidity
divergences~\cite{Collins:2011zzd,Echevarria:2011epo,Grewal:2020hoc}. Each
dependence is controlled by an evolution equation.\footnote{In this subsection,
we briefly describe the evolution of TMD PDFs (an analogous description applies
to TMD FFs): a more detailed treatment can be found in
Sec.~2 of Ref.~\cite{Bacchetta:2019sam} (see also Refs.~\cite{Rogers:2015sqa,Scimemi:2018xaf}).}
The complete set of equations (omitting the $x$ and $\bT$ dependencies for simplicity) is given by
\begin{align}
\frac{\partial \hat{f}_{1}}{\partial\ln\mu} &= \gamma(\mu,\zeta)
&
\frac{\partial \hat{f}_{1}}{\partial\ln\sqrt{\zeta}} &= K(\mu)
&
\frac{\partial K}{\partial\ln\mu} &=  \frac{\partial \gamma}{\partial\ln\sqrt{\zeta}} = -\gamma_{K}(\alpha_{s}(\mu))
\label{e:evoleq}
\end{align}
where $\gamma$ and $K$ are the anomalous dimensions of the
renormalisation-group and of the Collins-Soper evolution
equations, respectively, and $\gamma_{K}$ is the so-called cusp
anomalous dimension.

Given a set of initial conditions at the scales ($\mu_i,\zeta_i$), the solution to these differential equations allows us to determine the TMD at any final pair of scales ($\mu_f,\zeta_f$). In addition, in the region of small $|\bT|$ the TMD $\hat{f}_{1}$ can be matched onto its corresponding collinear PDF $f_1$ through a convolution with suitable perturbative matching coefficients $C$. The resulting expression for the TMD PDF at the final scales ($\mu_f,\zeta_f$) is
\begin{equation}
\label{e:evolved_TMDs}
\hat{f}_{1}(x,\,\bT;\,\mu_f,\zeta_f) =
\big[C\otimes f_1\big] (x,\,\bT;\,\mu_i,\zeta_i) \,
\exp\bigg\{ K(\mu_{i})\ln\frac{\sqrt{\zeta_{f}}}{\sqrt{\zeta_{i}}}+\int_{\mu_i}^{\mu_f} \frac{d\mu}{\mu}\, \bigg[\gamma_{F}(\alpha_{s}(\mu))-\gamma_{K}(\alpha_{s}(\mu))\ln\frac{\sqrt{\zeta_{f}}}{\mu}\bigg]\bigg\} \; ,
\end{equation}
where $\gamma_F(\alpha_s(\mu)) = \gamma(\mu,\mu^2)$. A convenient choice for the scales $\mu_i$ and $\zeta_i$ is $\mu_i=\sqrt{\zeta_i} \equiv \mu_b = 2e^{-\gamma_E}/|\bT|$, with $\gamma_E$ the Euler constant, since it avoids the insurgence of large logarithms in the rapidity evolution kernel $K$ and the matching coefficients $C$.

A given accuracy in the resummation of large logarithms of $|\bT|$ implies that each ingredient in Eq.~\eqref{e:evolved_TMDs} must be
computed to the perturbative accuracies summarized in Tab.~\ref{t:logcountings}.
After a careful benchmark of the perturbative expressions in our code against other well-known
codes~\cite{Bizon:2017rah,Camarda:2019zyx}, we introduced some small modifications
in some of the ingredients at the N$^3$LL level compared to what we used in the MAPTMD22 extraction~\cite{Bacchetta:2022awv}.
We stress that the present extraction incorporates for the first time all the necessary
ingredients in the TMD PDFs and TMD FFs to reach a full N$^{3}$LL accuracy.
\begin{table}[h!]
\begin{center}
\begin{tabular}{|c|c|c|c|c|c|}
 \hline
   &  \multicolumn{5}{c|}{${\cal O}(\alpha_s^m)$ perturbative order} \\
 \hline
 Accuracy N$^n$LL  & $H$ and $C$    &  $K$   and  $\gamma_F$  &  $\gamma_K$  & PDF and $\alpha_s$ evolution & FF evolution   \\
\hline
\hline
NLL          & 0     & 1         &    2   & LO & LO\\
 \hline
N$^2$LL       & 1     & 2         &   3    & NLO &  NLO\\
 \hline
N$^3$LL & 2      & 3    &  4   &  NNLO & NNLO \\
 \hline
\end{tabular}
\caption{Logarithmic accuracies of the TMD evolution vs. ${\cal O}(\alpha_s^m)$ corrections in TMD ingredients.}
\label{t:logcountings}
\end{center}
\end{table}

The introduction of $\mu_{b}$ as the initial scale of the TMD evolution
implies a prescription to avoid hitting
the QCD Landau pole in the large-$|\bT|$ region ($|\bT| \gtrsim 1 / \Lambda_{QCD}$)
and to smoothly match the TMD formula onto the fixed-order calculation at large transverse momentum ($|\qT|\sim Q$)~\cite{Bozzi:2003jy,Bozzi:2005wk,Bizon:2018foh} in the small-$|\bT|$ region ($|\bT|\to 0$).
Here, we adopt the same choice of Refs.~\cite{Bacchetta:2017gcc,Bacchetta:2022awv} and we replace $\mu_{b}$ with $\mu_{b_*} = 2e^{-\gamma_E}/b_*$, where
\begin{equation}
\label{e:bTstar}
b_*(|\bT|,b_{\text{min}},b_{\text{max}}) = b_{\text{max}}\, \bigg( \frac{1 - e^{ -|\bT|^4 / b_{\text{max}}^4 }}{1 - e^{ -|\bT|^4 / b_{\text{min}}^4 }} \bigg)^{1/4} \, ,
\end{equation}
with
\begin{align}
b_{\text{max}} &= 2 e^{-\gamma_E}  \text{  GeV}^{-1} \approx 1.123 \text{  GeV}^{-1}\, ,
&
b_{\text{min}} &= 2 e^{-\gamma_E}/\mu_f \ .
\label{e:bminmax}
\end{align}
This choice guarantees that the new variable $b_{*}$ rapidly saturates to $b_{\text{max}}$ ($b_{\text{min}}$) at large (small) values of $|\bT|$ (see Refs.~\cite{Bacchetta:2017gcc,Bacchetta:2022awv} for more details).
At the same time, the upper limit $b_{\text{max}}$ introduces power corrections scaling like ${ \cal O}( (\Lambda_{\text{QCD}}/|\qT|)^k)$~\cite{Catani:1996yz}, with $k>0$, that in the region $|\qT| \simeq \Lambda_{\text{QCD}}$ need to be accounted for
by introducing nonperturbative corrections to the Collins--Soper kernel $K$ and to the TMD formula of Eq.~\eqref{e:evolved_TMDs}. Following Refs.~\cite{Bacchetta:2017gcc,Bacchetta:2022awv}, we split the Collins--Soper kernel $K$ into a perturbative part $K(b_*, \mu_{b_*})$ and a nonperturbative part $g_K (|\bT|)$ that must vanish in the limit $|\bT| \to 0$. The final expression for the evolved TMD PDF is 
\begin{eqnarray}
\hat{f}_{1}(x,\,\bT;\,\mu_f,\zeta_f) &=
&\big[C\otimes f_1\big] (x,\,\bT;\,\mu_{b_*},\mu_{b_*}^2)  \nonumber \\
& &\hspace{-2cm} \times \exp\bigg\{ K(b_*, \mu_{b_*}) \ln\frac{\sqrt{\zeta_{f}}}{\sqrt{\mu_{b_*}^2}}+\int_{\mu_{b_*}}^{\mu_f} \frac{d\mu}{\mu}\, \bigg[\gamma_{F}(\alpha_{s}(\mu))-\gamma_{K}(\alpha_{s}(\mu))\ln\frac{\sqrt{\zeta_{f}}}{\mu}\bigg]\bigg\} \, f_{1NP} (x, \bT; \, \zeta_f, Q_0) \; ,
\label{e:evolved_TMDs_b*}
\end{eqnarray}
where $f_{1NP}$ is a correction term that contains the nonperturbative part of the Collins--Soper kernel $g_K$, as well as other parameters (see Sec.~\ref{s:results}). The function $f_{1NP}$ must satisfy the boundary condition $f_{1NP} \to 1$ for $|\bT| \to 0$, and it depends on an arbitrary scale $Q_0$ at which this correction is parametrized.

\section{Analysis framework}
\label{s:analysis}

\subsection{Data}
\label{s:data}

\begin{figure}
\centering
\includegraphics[width=0.6\textwidth]{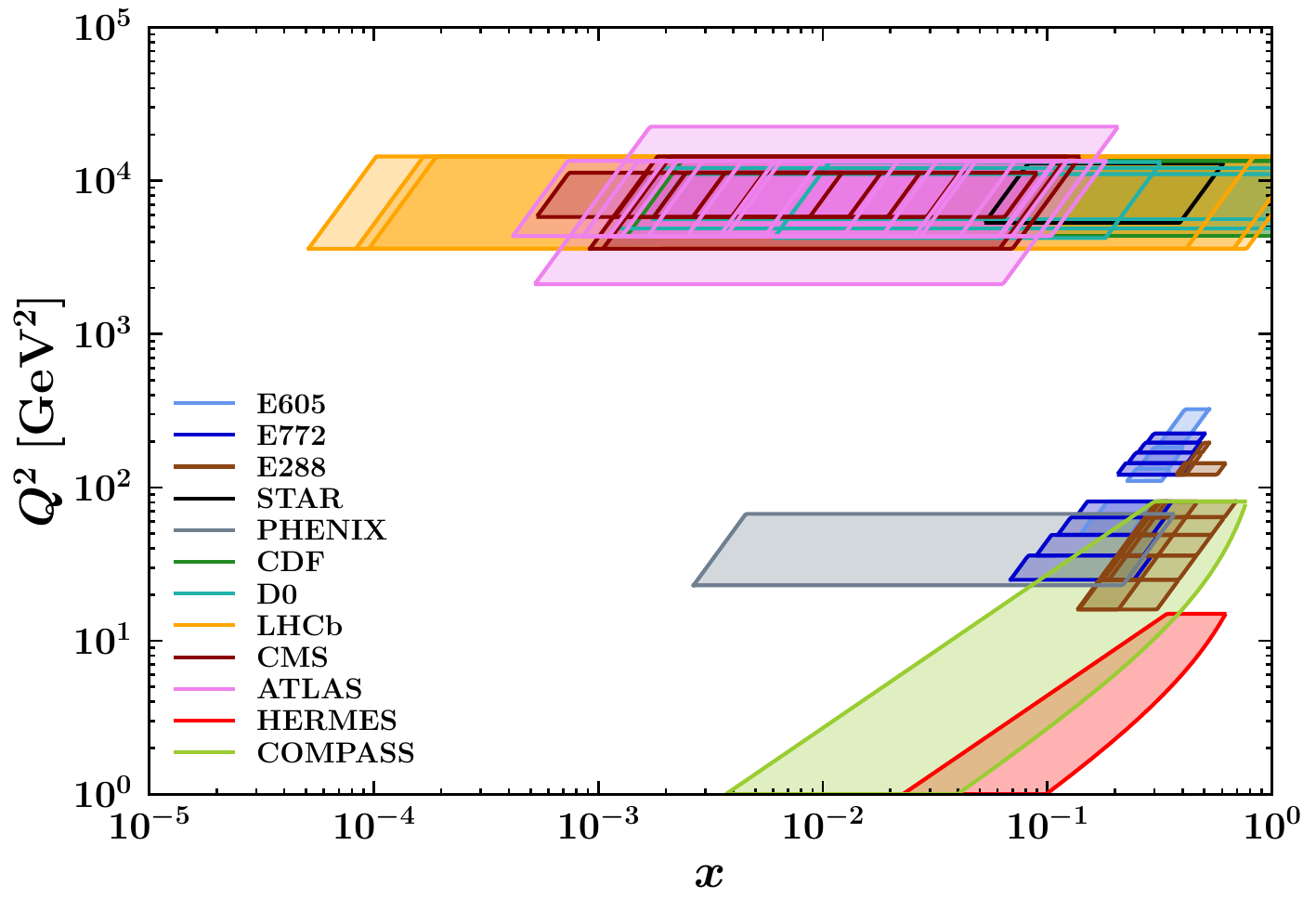}
\caption{Coverage in the $(x, Q^{2})$ plane of the full experimental data set included in this global fit.}
\label{f:KinematicCoveragexQ2}
\end{figure}

The set of experimental data used in the present analysis is identical to our previous MAPTMD22 global fit~\cite{Bacchetta:2022awv}. The total number of data points is 2031, of which 484 are from DY and 1547 from SIDIS measurements. In Tabs.~\ref{t:dataDY}-\ref{t:dataSIDIS}, we collect the relevant information on each data set. We emphasize that by combining data sets coming from a large number of different experimental collaborations, we are able to cover a wide range in the $(x, Q^2)$ plane, as shown in Fig.~\ref{f:KinematicCoveragexQ2}.

In order to focus on the region of phase space relevant for the TMD formalism, it is necessary to impose appropriate kinematic cuts on the data set. For DY data, we consider vector-boson transverse momenta that satisfy $|\qT |<0.2 \, Q$ to match the conditions for TMD factorization, and we further exclude the bins in $Q$ that contain the $\Upsilon$ resonance.
For SIDIS data, identifying the kinematic region where TMD factorization holds is
more involved. First of all, we impose that $Q > 1.4$ GeV in order to
match the conditions for collinear QCD factorization.
Moreover, we require that $0.2 < z < 0.7$ in order to include only data points in the SIDIS current fragmentation region and
avoid contamination from exclusive processes. Finally, we
adopt a kinematic cut in the detected hadron transverse momentum, $|\PhT| < \min[\,\min[ \, 0.2 \, Q,\,0.5 \, z \, Q\,]+0.3 \text{ GeV}, z\,  Q\, ] $.
In this way, we can safely assume that $|\qT| \ll Q$ without
excluding too many bins, consistently with our previous study~\cite{Bacchetta:2022awv}.

We refer to Ref.~\cite{Bacchetta:2022awv} and references therein for more extensive details on the kinematic cuts and the treatment of systematic and statistical uncertainties.

\begin{table}[t]
\footnotesize
\begin{center}
\renewcommand{\tabcolsep}{0.4pc}
\renewcommand{\arraystretch}{1.2}
\begin{tabular}{|c|c|c|c|c|c|c|c|}
  \hline
  Experiment & $N_{\rm dat}$ & Observable  &  $\sqrt{s}$ [GeV]& $Q$ [GeV] &  $y$ or $x_F$ & Lepton cuts & Ref. \\
  \hline
  \hline
  E605 & 50 & $E d^3\sigma/d^3 \bm{q}$ & 38.8 & 7 - 18  & $x_F=0.1$ & - & \cite{Moreno:1990sf} \\
  \hline
  E772 & 53 & $E d^3\sigma/d^3 \bm{q}$ & 38.8 & 5 - 15  & $0.1 < x_F < 0.3$ & - & \cite{E772:1994cpf} \\
  \hline
  E288 200 GeV & 30 & $E d^3\sigma/d^3 \bm{q}$ &  19.4  & 4 - 9  & $y=0.40$ & - & \cite{Ito:1980ev} \\
  \hline
  E288 300 GeV & 39 & $E d^3\sigma/d^3 \bm{q}$ &  23.8  & 4 - 12  & $y=0.21$ & - & \cite{Ito:1980ev} \\
  \hline
  E288 400 GeV & 61 & $E d^3\sigma/d^3 \bm{q}$ &  27.4  & 5 - 14  & $y=0.03$ & - & \cite{Ito:1980ev} \\
  \hline
  STAR 510 & 7 & $d\sigma/d |\qT|$ & 510  & 73 - 114  & $|y|<1$ & \makecell{$p_{T\ell} > 25$~GeV\\ $|\eta_\ell|<1$} & \cite{STAR2023} \\
  \hline
  PHENIX200 & 2 & $d\sigma/d |\qT|$ & 200 & 4.8 - 8.2  & $1.2 < y < 2.2$ & - & \cite{PHENIX:2018dwt} \\
  \hline
  CDF Run I & 25 & $d\sigma/d |\qT|$ & 1800 & 66 - 116  & Inclusive & - & \cite{Affolder:1999jh} \\
  \hline
  CDF Run II & 26 & $d\sigma/d |\qT|$ & 1960 & 66 - 116  &  Inclusive & - & \cite{Aaltonen:2012fi} \\
  \hline
  D0 Run I & 12 & $d\sigma/d |\qT|$ & 1800 & 75 - 105  &  Inclusive & - & \cite{Abbott:1999wk} \\
  \hline
  D0 Run II & 5 & $(1/\sigma)d\sigma/d |\qT|$ & 1960 & 70 - 110  & Inclusive & - & \cite{Abazov:2007ac} \\
  \hline
  D0 Run II $(\mu)$ & 3 & $(1/\sigma)d\sigma/d |\qT|$ & 1960 & 65 - 115  & $|y|<1.7$ & \makecell{$p_{T\ell} > 15$~GeV\\$|\eta_\ell|<1.7$} & \cite{Abazov:2010kn} \\
  \hline
  LHCb 7 TeV & 7 & $d\sigma/d |\qT|$ & 7000 & 60 - 120  & $2<y<4.5$ & \makecell{$p_{T\ell} > 20$ GeV\\$2<\eta_\ell<4.5$} & \cite{Aaij:2015gna} \\
  \hline
  LHCb 8 TeV & 7 & $d\sigma/d |\qT|$ & 8000 & 60 - 120  & $2<y<4.5$ & \makecell{$p_{T\ell} > 20$ GeV\\$2<\eta_\ell<4.5$} & \cite{Aaij:2015zlq} \\
  \hline
  LHCb 13 TeV & 7 &  $d\sigma/d |\qT|$ & 13000 & 60 - 120  & $2<y<4.5$ & \makecell{$p_{T\ell} > 20$ GeV\\$2<\eta_\ell<4.5$} & \cite{Aaij:2016mgv} \\
  \hline
  CMS 7 TeV & 4 & $(1/\sigma)d\sigma/d |\qT|$  & 7000 & 60 - 120  & $|y|<2.1$ & \makecell{$p_{T\ell} > 20$ GeV\\$|\eta_\ell|<2.1$} & \cite{Chatrchyan:2011wt} \\
  \hline
  CMS 8 TeV & 4 & $(1/\sigma)d\sigma/d |\qT|$ & 8000 & 60 - 120  & $|y|<2.1$ & \makecell{$p_{T\ell} > 15$ GeV\\$|\eta_\ell|<2.1$} & \cite{Khachatryan:2016nbe} \\
  \hline
  CMS 13 TeV & 70 & $d\sigma/d |\qT|$ & 13000 & 76 - 106 & \makecell{$|y|<0.4$ \\ $0.4<|y|<0.8$ \\ $0.8<|y|<1.2$\\$1.2<|y|<1.6$\\$1.6<|y|<2.4$} & \makecell{$p_{T\ell} > 25$~GeV\\$|\eta_\ell|<2.4$} & \cite{CMS:2019raw} \\
  \hline
  ATLAS 7 TeV & \makecell{6\\6\\6}& $(1/\sigma)d\sigma/d |\qT|$ & 7000 & 66 - 116 & \makecell{$|y|<1$ \\ $1<|y|<2$ \\ $2<|y|<2.4$}  & \makecell{$p_{T\ell} > 20$~GeV\\$|\eta_\ell|<2.4$} & \cite{Aad:2014xaa} \\
  \hline
  \makecell{ATLAS 8 TeV \\ on-peak} & \makecell{6\\6\\6\\6\\6\\6} & $(1/\sigma)d\sigma/d |\qT|$ & 8000 & 66 - 116  & \makecell{$|y|<0.4$ \\ $0.4<|y|<0.8$ \\ $0.8<|y|<1.2$\\$1.2<|y|<1.6$\\$1.6<|y|<2$\\$2<|y|<2.4$} & \makecell{$p_{T\ell} > 20$~GeV\\$|\eta_\ell|<2.4$} & \cite{Aad:2015auj} \\
  \hline
  \makecell{ATLAS 8 TeV \\ off-peak} & \makecell{4 \\ 8} & $(1/\sigma)d\sigma/d |\qT|$ & 8000 & \makecell{46 - 66 \\ 116 - 150} & $|y|<2.4$ & \makecell{$p_{T\ell} > 20$ GeV\\$|\eta_\ell|<2.4$} & \cite{Aad:2015auj} \\
  \hline
  ATLAS 13 TeV & 6 & $(1/\sigma)d\sigma/d |\qT|$ & 13000 & 66 - 113 & $|y|<2.5$ & \makecell{$p_{T\ell} > 27$ GeV\\$|\eta_\ell|<2.5$} & \cite{ATLAS:2019zci} \\
  \hline
  \hline
  Total & 484 & & & & & & \\
  \hline
\end{tabular}
\caption{DY experimental data sets included in this global fit. Each row contains the number of data points ($N_{\rm dat}$) after kinematic cuts, the measured observable, the center-of-mass energy $\sqrt{s}$, the invariant mass range, the angular variable ($y$ or $x_F$), possible cuts on the final-state leptons, and the published reference.}
\label{t:dataDY}
\end{center}
\end{table}

\begin{table}[t]
\footnotesize
\begin{center}
\renewcommand{\tabcolsep}{0.4pc}
\renewcommand{\arraystretch}{1.2}
\begin{tabular}{|c|c|c|c|c|c|c|c|c|}
  \hline
  Experiment & $N_{\rm dat}$ & Observable & Channels & $Q$ [GeV] & $x$ & $z$ & Phase space cuts & Ref. \\
  \hline
  \hline
  HERMES & 344 & $M(x,z,|\PhT|,Q)$ & \makecell{$p \rightarrow \pi^+$ \\ $p \rightarrow \pi^-$ \\$p \rightarrow K^+$ \\ $p \rightarrow K^-$ \\ $d \rightarrow \pi^+$ \\ $d \rightarrow \pi^-$ \\$d \rightarrow K^+$ \\ $d \rightarrow K^-$ \\} & 1 - $\sqrt{15}$ & \makecell{ \\ $0.023<x<0.6$ \\ (6 bins) \\ \\} & \makecell{$0.1<z<1.1$ \\ (8 bins)} & \makecell{$W^2 > 10$ GeV$^2$\\$ 0.1<y<0.85$} & \cite{HERMES:2012uyd} \\
  \hline
  COMPASS & 1203 & $M(x,z,\PhT^2,Q)$ & \makecell{$d \rightarrow h^+$ \\ $d \rightarrow h^-$ \\} & \makecell{ 1 - 9 \\ (5 bins) \\} & \makecell{ \\ $0.003<x<0.4$ \\ (8 bins) \\ \\} & \makecell{ $0.2<z<0.8$ \\ (4 bins) \\ } & \makecell{$W^2 > 25$ GeV$^2$\\$ 0.1<y<0.9$} & \cite{COMPASS:2017mvk} \\
  \hline
  \hline
  Total & 1547 & & & & & & & \\
  \hline
\end{tabular}
\caption{SIDIS experimental data sets included in this global fit. Each row contains the number of data points ($N_{\rm dat}$) after kinematic cuts, the measured observable, the SIDIS channel, the invariant mass range of the virtual photon, the covered ranges for the invariants $x$ and $z$, possible cuts on the final-state lepton, and the published reference.}
\label{t:dataSIDIS}
\end{center}
\end{table}

\subsection{Fit procedure}
\label{s:fit_procedure}

The agreement between our theoretical predictions and the experimental data is assessed by the usual $\chi^2$ test,
\begin{equation}
\chi^2 = \sum_{i,j}^N (m_i - t_i) V_{ij}^{-1} (m_j - t_j) \, ,
\label{e:chi2}
\end{equation}
where $m_i$ represents the experimental value for data point $i$, $t_i$ denotes the corresponding theoretical prediction, and $V_{ij}$ is the covariance matrix.
When bin-by-bin correlated uncertainties are present, the total $\chi^2$ can be decomposed into two components~\cite{Bacchetta:2019sam,Bacchetta:2022awv}:
\begin{equation}
\label{e:chi2terms}
\chi^2 = \sum_i^N \left( \frac{m_i - \overline{t}_i}{\sigma_i} \right)^2 + \chi^2_{\lambda} = \chi^2_D + \chi^2_{\lambda} \, ,
\end{equation}
where $\chi^2_D$
is given by the standard formula for $N$ experimental data points with statistical and uncorrelated systematic uncertainties summed in quadrature, $\sigma_i^2 = \sigma_{i, {\text{stat}}}^2 + \sigma_{i, {\text{uncor}}}^2$,
but involving theoretical predictions $\overline{t}_i$ for data point $i$
shifted by the correlation uncertainties according to
\begin{equation}
\label{e:th_shifts}
\overline{t}_i = t_i + \sum_{\alpha=1}^k \lambda_\alpha \, \sigma_{i, \text{corr}}^{(\alpha)} \, ,
\end{equation}
where the sum runs upon the sources of correlated uncertainties, $\sigma_{i, \text{corr}}^{(\alpha)}$ represents the $\alpha$-th (fully) correlated uncertainty affecting the $i$-th experimental data point, and $\lambda_\alpha$ denotes the nuisance parameter. The term $\chi^2_{\lambda}$ in Eq.~\eqref{e:chi2terms} is a penalty contribution due to correlated uncertainties and it is entirely determined by the nuisance parameters:
\begin{equation}
\label{e:chilambda}
\chi^2_\lambda = \sum_{\alpha=1}^k \lambda_\alpha^2 \, .
\end{equation}

The optimal values of the nuisance parameters are obtained by minimizing the
total $\chi^2$ in Eq.~\eqref{e:chi2terms} with respect to them. Since the
shifted predictions in Eq.~\eqref{e:th_shifts} offer a better visual
evaluation of the fit quality, we consistently present them for all
observables employed in this global fit.

We performed the analysis by employing the so-called bootstrap method, which
entails fitting a set of several Monte Carlo replicas of the data (100 in our case).
Moreover, we use Monte Carlo sets for collinear PDFs and FFs and
we change the member of the collinear sets for each replica.
The most complete statistical information about the extracted
TMDs is given by the full ensemble of replicas
but, consistently with our 
previous work~\cite{Bacchetta:2022awv}, we use as the most appropriate
estimator of the fit quality the $\chi^2$ value of the best fit for the \textit{central} replica
($\chi^2_0$), defined as the replica obtained by fitting experimental
data without fluctuations.

\section{Results}
\label{s:results}

\subsection{Flavor-independent nonperturbative parametrization}
\label{s:baseline}

In this Section, we describe our new simultaneous extraction of TMD PDFs and TMD FFs
similar to the MAPTMD22 one~\cite{Bacchetta:2022awv}, where the models for these two
nonperturbative objects are considered the same for each quark flavor. This provides us with a reference to which the core results
of this paper will be compared.
The main innovation of this new extraction is the choice
of the collinear PDF sets to build the TMDs: we use LHAPDF sets delivered as Monte Carlo
ensembles~\cite{Buckley:2014ana}. This choice allows us to assign a specific member of the collinear sets to each
TMD replica, which leads to a robust estimate of the uncertainty of the
extracted TMD distributions, as already suggested in Ref.~\cite{Bury:2022czx}. We use the NNPDF3.1 set (NNPDF31\_nnlo\_pch\_as\_0118)~\cite{NNPDF:2017mvq}
for PDFs, and a variation of the baseline MAPFF1.0 NNLO set~\cite{AbdulKhalek:2022laj}
for FFs. The variation consists in the choice of the parametrization
scale (1 GeV in our new set, 5 GeV in the baseline). In this way, we avoid
complications related to backward evolution to the scale $\mu_b$ that appears in
the expression of experimental observables in TMD factorization, because $\mu_b$
can be as low as 1 GeV.

Then, we repeat the analysis with the same settings but with a different
approach for the model of TMD FFs. Specifically, we consider a more flexible
model that separates the parametrization of the fragmentation of a quark into a
pion from the one into a kaon.
Such a separation was explored so far only in Ref~\cite{Signori:2013mda}.
In the following, we denote these two reference
extractions as MAPTMD24 Flavor Independent (MAPTMD24 FI) and MAPTMD24 Hadron Dependent
(MAPTMD24 HD).

For both these analyses, the model of the nonperturbative part
of the TMDs is the same as in the
MAPTMD22 extraction~\cite{Bacchetta:2022awv}. Thus, the parametrization of
TMD PDFs is
\begin{equation}
\label{e:f1NP}
f_{1\, NP}(x, \bT^2; \zeta, Q_0) =
\frac{
g_1(x)\, e^{ - g_1(x) \frac{\bT^2}{4}} +
\lambda^2\, g_{2}^2(x)\, \bigg[ 1 - g_{2}(x) \frac{\bT^2}{4} \bigg]\, e^{ - g_{2}(x) \frac{\bT^2}{4}} +
\lambda_2^2\, g_{3}(x)\, e^{ - g_{3}(x) \frac{\bT^2}{4}}
}{
g_1(x) +  \lambda^2\, g_{2}^2(x) + \lambda_2^2\, g_{3}(x)
} \,
\bigg[ \frac{\zeta}{Q_0^2} \bigg]^{g_K(\bT^2)/2}\, ,
\end{equation}
corresponding to the Fourier transform of the sum of two Gaussians and a
Gaussian weighted by $\kperp^2$.

The expression of the model for the TMD FFs is
\begin{align}
\label{e:D1NP}
D_{1\, NP}(z, \bT^2; \zeta, Q_0) =
\frac{
g_4(z)\, e^{ - g_4(z) \frac{\bT^2}{4z^2}} +
\frac{\lambda_F}{z^2}\, g_{5}^2(z)\, \bigg[ 1 -
g_{5}(z) \frac{\bT^2}{4z^2} \bigg]\, e^{ - g_{5}(z) \frac{\bT^2}{4z^2}}
}{
g_4(z) +  \frac{\lambda_F}{z^2}\, g_{5}^2(z)
} \,
\bigg[ \frac{\zeta}{Q_0^2} \bigg]^{g_K(\bT^2)/2}\, ,
\end{align}
corresponding to the Fourier transform of the sum of a Gaussians and a
Gaussian weighted by $\Pperp^2$.

The $g_i$ functions describe the widths of the distributions
 and include a dependence on $x$ and $z$:
\begin{align}
\label{e:gi_func_PDF}
& g_{\{1,2,3\}}(x) = N_{\{1,2,3\}} \frac{x^{\sigma_{\{1,2,3\}}}(1-x)^{\alpha^2_{\{1,2,3\}}}}{\hat{x}^{\sigma_{\{1,2,3\}}}(1-\hat{x})^{\alpha^2_{\{1,2,3\}}}} \, ,
\\
\label{e:gi_func_FF}
& g_{\{4,5\}}(z) = N_{\{4,5\}} \frac{(z^{\beta_{\{1,2\}}}+\delta^2_{\{1,2\}})(1-z)^{\gamma^2_{\{1,2\}}}}{(\hat{z}^{\beta_{\{1,2\}}}+\delta^2_{\{1,2\}})(1-\hat{z})^{\gamma^2_{\{1,2\}}}} \, ,
\end{align}
where $\hat{x} = 0.1$, $\hat{z} = 0.5$, and $N_i \, (i=1-5),\, \sigma_j, \, \alpha_j \, (j=1-3),\, \beta_i, \, \delta_i, \, \gamma_i \, (i=1,2),$ are free parameters.

Finally, the nonperturbative part of the Collins-Soper kernel is parametrized
as
\begin{equation}
  g_K(\bT^2) = - g_2^2\, \frac{\bT^2}{2} \, .
\label{e:CSkernelNP}
\end{equation}
This function governs the nonperturbative contribution $(\zeta_f/Q_0^2)^{g_{K}/2}$
to the TMD evolution, where $Q_0$ is the scale at
which this contribution is parametrized; we set $Q_0=1$~GeV.

The functional forms in Eqs.~\eqref{e:f1NP}-\eqref{e:gi_func_FF} are largely
arbitrary. We choose to parametrize the nonperturbative parts of TMDs in terms of
Gaussians and weighted Gaussians in transverse-momentum space because they are
guaranteed to be positive at the initial scale $Q_0 = 1$ GeV.
The widths of the Gaussians, expressed by
Eqs.~\eqref{e:gi_func_PDF}-\eqref{e:gi_func_FF}, depend on $x$ or $z$
and vanish as $x$ or $z$ approach one. Our choice of the functional form is
also inspired by model calculations of TMD PDFs (see,
\textit{e.g.}, Refs.~\cite{Bacchetta:2008af,Pasquini:2008ax,Avakian:2010br,Burkardt:2015qoa,Gutsche:2016gcd,Maji:2017bcz,Alessandro:2021cbg,Signal:2021aum})
and TMD FFs (see, \textit{e.g}, Refs.~\cite{Bacchetta:2007wc,Matevosyan:2011vj}). Many of
these models predict the existence of terms that behave similarly to Gaussians
and weighted Gaussians. The details of their functional dependence
are related to the correlation between the spin of the quarks and
their transverse momentum. In the case of fragmentation functions, a different
role can be played by different fragmentation channels.
For example, a pion in the final state can be produced by the direct
fragmentation of the active quark in the hard process, or by the
decay of hadronic resonances, such as the $\rho$ meson. The interplay of these
two channels can generate different nontrivial features in the shape of
the extracted TMD FFs.

After trying several parameter configurations,
we noticed that it is possible to set $\sigma_2 = \sigma_3$ in Eq.~\eqref{e:gi_func_PDF}
without deteriorating the quality of the fit.
With this last assumption, the fit involves 20 free parameters:
10 for the nonperturbative part of the TMD PDFs,
9 for the nonperturbative part of the TMD FFs, and 1 for the nonperturbative
part of the Collis--Soper kernel.

We fitted
100 Monte Carlo replicas of the experimental data.
We obtain for the central replica a $\chi^2$ per data point  $\chi^2_0 / N_{\text{dat}} = 1.40$.
This result is not compatible with the one of the MAPTMD22 extraction
($\chi^2_0 / N_{\text{dat}} = 1.06$).
In order to understand the origin of this deterioration, we investigated the impact of  different combinations of
collinear PDFs (MMHT2014~\cite{Harland-Lang:2014zoa} and NNPDF3.1~\cite{NNPDF:2017mvq})
and FFs (DSS14-17~\cite{deFlorian:2014xna,deFlorian:2017lwf} and MAPFF1.0~\cite{AbdulKhalek:2022laj}). In Tab.~\ref{t:chisquare_diff_sets},
we report the values of $\chi^2_0 / N_{\text{dat}}$ for each scenario.
\begin{table}[h]
\begin{center}
    \begin{tabular}{|l|c|c|c|}
  \hline
  \multicolumn{1}{|c|}{ } & \multicolumn{3}{c|}{ Data set $\chi^2_0/ N_{\text{dat}}$} \\
  \hline
  Collinear sets & DY total & SIDIS total  &  \bf{Total} \\
  \hline
  \hline
    MMHT + DSS (MAPTMD22) & 1.66 & 0.87 & \bf{1.06} \\
    \hline
    NNPDF + DSS & 1.62 & 0.90 & \bf{1.07} \\
    \hline
    MMHT + MAPFF & 1.58 & 1.33 & \bf{1.39} \\
    \hline
        NNPDF + MAPFF (MAPTMD24 FI) & 1.58  & 1.34 & \bf{1.40} \\
    \hline
        NNPDF + MAPFF (MAPTMD24 HD) & 1.57  & 1.08 & \bf{1.19} \\
    \hline
    \end{tabular}
    \caption{Breakdown of the values of $\chi_0^2 / N_{\text{dat}}$ for different choices of collinear PDF and FF sets.}
    \label{t:chisquare_diff_sets}
    \end{center}
\end{table}

The results in Tab.~\ref{t:chisquare_diff_sets} clearly show that a change in the collinear PDF set from MMHT to NNPDF produces
a negligible effect on the quality of the fit.
 This is reasonable because in the kinematic region covered by
the global dataset included in this analysis the two considered PDF sets are well constrained and
compatible with each other.\footnote{We remark that in
Ref.~\cite{Bury:2022czx}, where also other sets of PDFs were taken into
account, the authors concluded that the choice of collinear PDF sets led to a
significant difference in the description of experimental data and required a
change in the functional form of the nonperturbative components.}

In contrast,
our results are significantly affected
by the choice of collinear FFs. In fact, the $\chi_0^2 / N_{\text{dat}}$ becomes larger
when moving from DSS to MAPFF. Unsurprisingly, this deterioration affects the description
of SIDIS data, without significant impact on the description of Drell-Yan data.
The increase of the $\chi_0^2 / N_{\text{dat}}$ value
is mostly due to the MAPFF collinear set being affected by lower
uncertainties as compared to the DSS one.

\begin{figure}[h]
\centering
\includegraphics[width=1.0\textwidth]{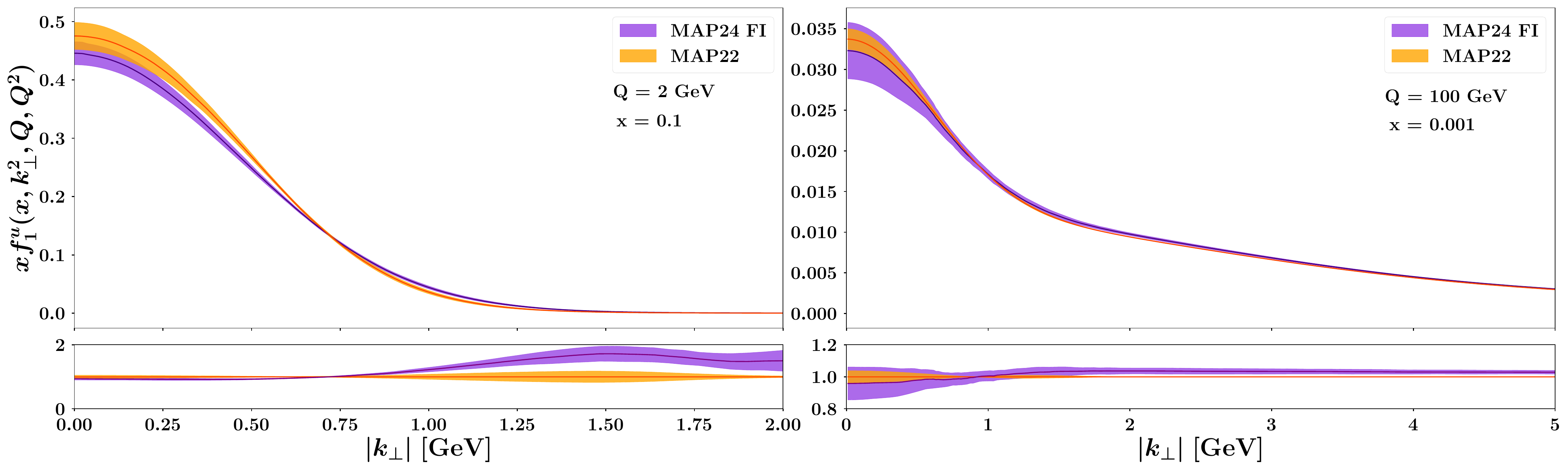}
\caption{Comparison between the unpolarized TMD PDFs of the up quark in a proton extracted in the MAPTMD22 fit
(orange) and the MAPTMD24 Flavor Independent fit (purple), as functions of the partonic transverse momentum
$|\kperp|$ at  $\mu = \sqrt{\zeta} = Q = 2$ GeV, $x =0.1$ (left panel) and $\mu = \sqrt{\zeta} = Q = 100$ GeV,
$x =0.001$ (right panel). Lower panels show the ratio of MAPTMD24 Flavor Independent to MAPTMD22. The uncertainty bands represent the 68\% C.L.}
\label{f:tmdpdf_baseline_vs_map22}
\end{figure}

In Fig.~\ref{f:tmdpdf_baseline_vs_map22}, we show the unpolarized TMD PDFs of the up quark
in a proton extracted in MAPTMD22 (orange) and MAPTMD24 FI (purple) as functions
of the partonic transverse momentum $|\kperp|$ at  $\mu = \sqrt{\zeta} = Q = 2$ GeV and $x =0.1$
(left panel), and $\mu = \sqrt{\zeta} = Q = 100$ GeV and $x =0.001$ (right panel).
The plots evidently show that the TMD PDFs extracted with two different choices
of collinear PDF sets are compatible with each other in the kinematic region covered by experimental
data. We note that the MAPTMD24 uncertainty bands, corresponding to the 68\%
confidence level (C.L.), are equal to or larger
than the MAPTMD22 ones, as a consequence of the fact that each replica of the MAPTMD24 fit
is associated to a different member of the collinear PDF set, while in the MAPTMD22 fit
all TMD replicas were associated to the same member.

\begin{figure}[h]
\centering
\includegraphics[width=1.0\textwidth]{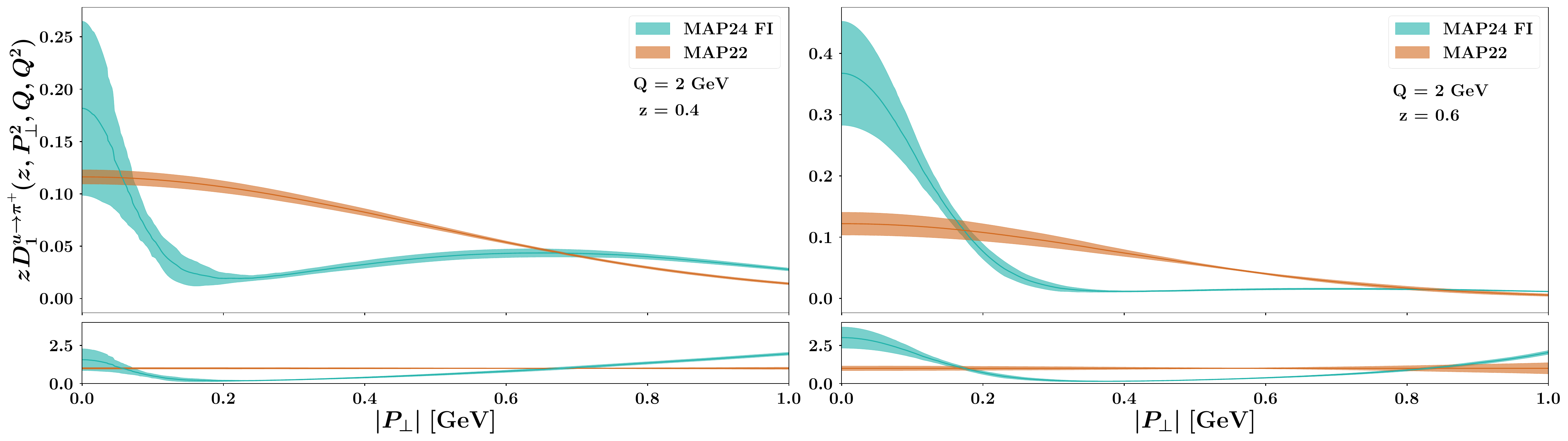}
\caption{Comparison between the TMD FFs for an up quark fragmenting into a $\pi^+$ extracted in the MAPTMD22 fit (brown)
and the MAPTMD24 Flavor Independent fit (light blue), as functions of the partonic transverse momentum $|\Pperp|$ at
$\mu = \sqrt{\zeta} = Q = 2$ GeV $z =0.4$ (left panel), and $z =0.6$ (right panel). Lower panels show the ratio of MAPTMD24 Flavor Independent to MAPTMD22. The uncertainty bands represent the 68\% C.L.}
\label{f:tmdff_baseline_vs_map22}
\end{figure}

In Fig.~\ref{f:tmdff_baseline_vs_map22}, we display the unpolarized TMD FFs for an
up quark fragmenting into a $\pi^+$ extracted in the MAPTMD22 (brown)
and MAPTMD24 FI (light blue) fits, as functions of the pion transverse momentum $|\Pperp|$ at
$\mu = \sqrt{\zeta} = Q = 2$ GeV and $z =0.4$ (left panel), and $z =0.6$ (right panel).
We note significant differences both in shape and normalization, which can be traced
back to the different choice of the collinear FF set (see Tab.~\ref{t:chisquare_diff_sets}).
However, there was no need to change the
functional form of the nonperturbative parametrization, since it turned out to be sufficiently flexible to
accommodate the differences caused by changing the collinear FF set.
The MAPTMD24 FI fragmentation function has a second smaller
peak at intermediate $|\Pperp|$, especially in the low-$z$ region. This feature is
present also in the MAPTMD22 fit, but the size of the peak is smaller and
its position shifted to higher $|\Pperp|$ values.
As anticipated in model descriptions of fragmentation functions, this behavior could be induced by the interference of different channels in the fragmentation process where the detected hadron could be produced directly or through the decay of heavier resonances.
The TMD FFs could be better constrained by
data from double-inclusive hadron production in electron-positron annihilation~\cite{Bacchetta:2015ora}.
Important constraints could be obtained also from
different processes, such as single-inclusive hadron production in electron-positron annihilation
with the reconstruction of the thrust or jet axis~\cite{Boglione:2017jlh,Belle:2019ywy,Soleymaninia:2019jqo,Boglione:2022nzq}.

Since the flavor-independent ansatz for the nonperturbative part of TMDs does not provide
a sufficiently good description of the data, as an intermediate step toward a flavor-dependent
extraction we consider a flavor-independent but hadron-dependent ansatz. Namely,
we allow the non-perturbative
parts of the TMD FF for pions to differ from those for kaons. We employ the same functional
form of Eq.~\eqref{e:D1NP} but with different parameters for pions and
kaons. In this version of the extraction, denoted as MAPTMD24 HD,
we have a total
of 29 free parameters: 1 for the Collins--Soper kernel, 10 for the TMD PDF, 9 for the TMD FF in pions, and 9 for the TMD FF
in kaons.

Because of the increased flexibility, we achieve a significantly better
description of the data, obtaining
$\chi_0^2 / N_{\text{dat}} = 1.19$ (see Tab.~\ref{t:chisquare_diff_sets}).
The SIDIS data are now described much better than in the MAPTMD24 FI case,
while the description of the DY data is
almost unaffected.

\begin{figure}[h]
\centering
\includegraphics[width=1.0\textwidth]{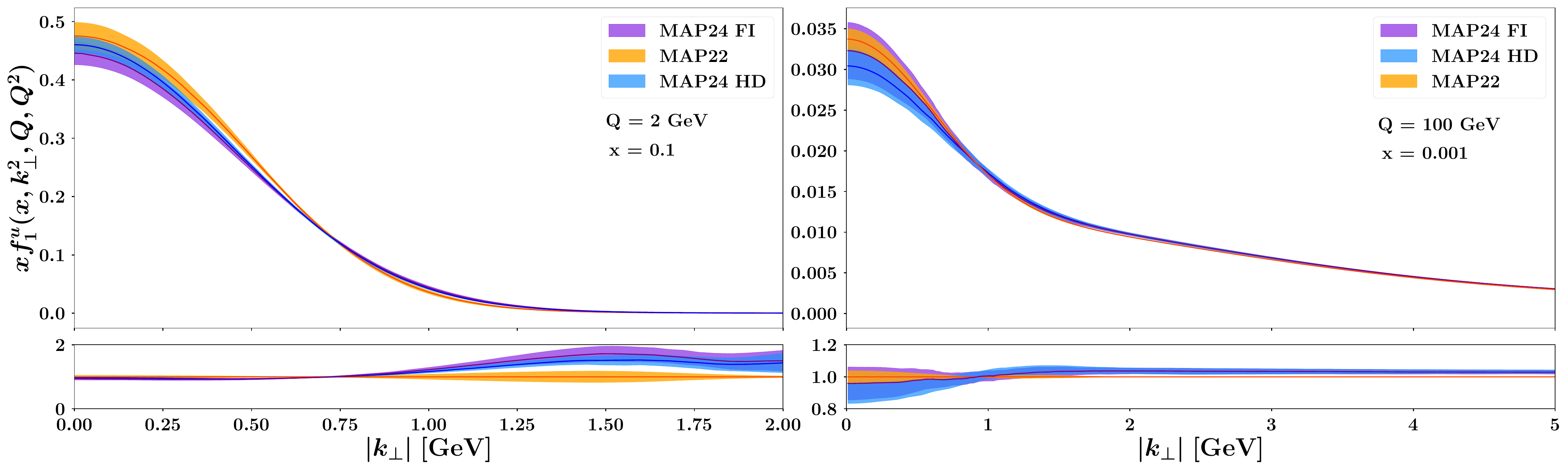}
\caption{Comparison between the TMD PDFs of the up quark in a proton extracted
  in the MAPTMD22 fit (orange), the
MAPTMD24 FI fit (purple) and the MAPTMD24 HD fit (blue), as functions of the partonic
transverse momentum $|\kperp|$ at  $\mu = \sqrt{\zeta} = Q = 2$ GeV and $x =0.1$ (left panel), and
$\mu = \sqrt{\zeta} = Q = 100$ GeV and $x =0.001$ (right panel). Lower panels show the ratio of MAPTMD24 FI and MAPTMD24 HD to MAPTMD22. The uncertainty bands represent the 68\% C.L.}
\label{f:tmdpdf_baseline_vs_map22_vs_hadrondep}
\end{figure}

In Fig.~\ref{f:tmdpdf_baseline_vs_map22_vs_hadrondep}, we show the unpolarized TMD PDFs of
the up quark in a proton extracted in the MAPTMD22 fit (orange), the
MAPTMD24 FI fit (purple) and the MAPTMD24 HD fit (blue), as functions of the partonic
transverse momentum $|\kperp|$ at  $\mu = \sqrt{\zeta} = Q = 2$ GeV and $x =0.1$ (left panel), and
$\mu = \sqrt{\zeta} = Q = 100$ GeV and $x =0.001$ (right panel). All three extractions
are compatible with each other.

\begin{figure}[h]
\centering
\includegraphics[width=1.0\textwidth]{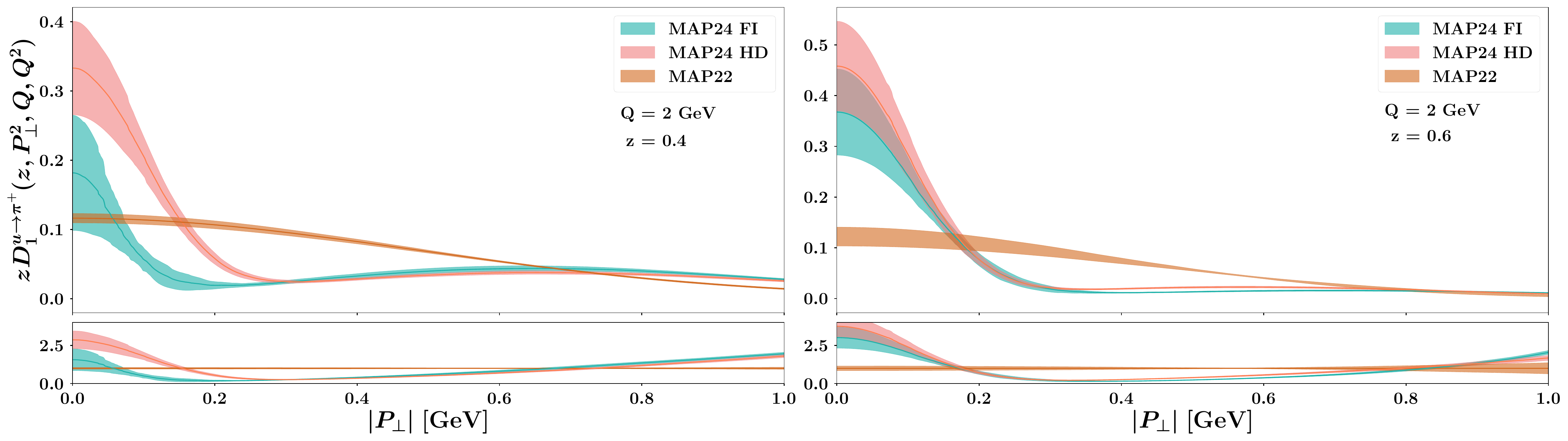}
\caption{Comparison between the TMD FFs for an up quark fragmenting into a $\pi^+$ in the MAPTMD22 fit (brown), the
MAPTMD24 FI fit (light blue) and the MAPTMD24 HD fit (pink), as functions of the hadronic transverse
momentum $|\Pperp|$ at  $\mu = \sqrt{\zeta} = Q = 2$ GeV and $z =0.4$ (left panel), and $z =0.6$ (right panel).
Lower panels show the ratio of MAPTMD24 FI and MAPTMD24 HD to MAPTMD22. The uncertainty bands represent the 68\% C.L.}
\label{f:tmdff_baseline_vs_map22_vs_hadrondep}
\end{figure}

In Fig.~\ref{f:tmdff_baseline_vs_map22_vs_hadrondep}, we show the unpolarized
TMD FFs for an up quark fragmenting into a $\pi^+$ in the MAPTMD22 fit (brown), the
MAPTMD24 FI fit (light blue) and the MAPTMD24 HD fit (pink), as functions of the hadronic transverse
momentum $|\Pperp|$ at  $\mu = \sqrt{\zeta} = Q = 2$ GeV, and $z =0.4$ (left panel),
and $z =0.6$ (right panel).
The MAPTMD24 distributions are more strongly peaked
at $|\Pperp| = 0$ and also have a noticeable bump at higher $|\Pperp|$
but there is a sharp difference between them, particularly at smaller
values of $z$.

\begin{figure}[h]
\centering
\includegraphics[width=1.0\textwidth]{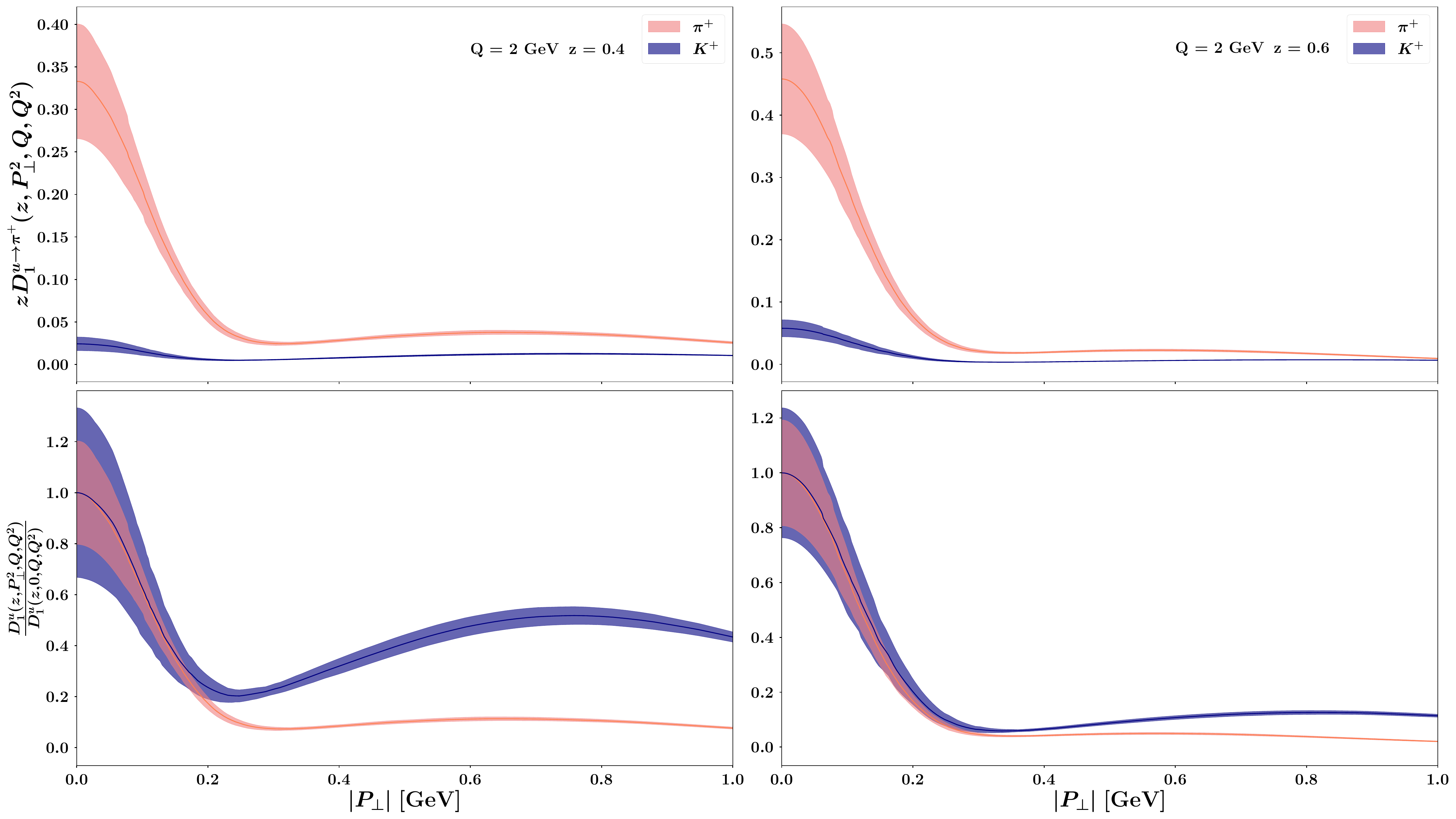}
\caption{Comparison between the TMD FFs obtained in the MAPTMD24 HD fit for an up quark
fragmenting into a $\pi^+$ (pink) and a $K^+$ (blue), as functions of the hadronic transverse momentum
$|\Pperp|$ at  $\mu = \sqrt{\zeta} = Q = 2$ GeV and $z =0.4$ (left panel), and $z =0.6$ (right panel).
In the lower panels, the TMD FFs normalized to the central replica at $|\Pperp| = 0$.
The uncertainty bands represent the 68\% C.L.}
\label{f:tmdff_hadrondep_comparison}
\end{figure}

In the upper panels of Fig.~\ref{f:tmdff_hadrondep_comparison}, we display the
unpolarized TMD FFs of the MAPTMD24 HD fit
for an up quark fragmenting into a $\pi^+$ (pink) and a $K^+$ (blue), as functions of the hadronic transverse momentum
$|\Pperp|$ at  $\mu = \sqrt{\zeta} = Q = 2$ GeV and $z =0.4$ (left panel), and $z =0.6$ (right panel). In the lower panels, we show
the TMD FFs normalized to the values of the corresponding central replica at $|\Pperp| = 0$. The lower panels
clearly indicate that the fragmentations into pions
and kaons exhibit distinctly different behaviors. In particular, the kaon FF
displays at intermediate $|\Pperp|$ a large second peak, emphasized at low $z$.

\subsection{Flavor-dependent nonperturbative parametrization}
\label{s:flavor}

In this section, we present the main result of this work, namely the extraction with a flavor-dependent approach of TMD PDFs for unpolarized quarks in the proton and TMD FFs for final pions and kaons. We will refer to this extraction
as MAPTMD24 FD or simply MAPTMD24. This work represents a significant upgrade
compared to the MAPTMD22 fit and similar studies, since it is the first time that a
global analysis of SIDIS and DY data with flavor dependence is performed.
We follow the same strategy as in the hadron-dependent extraction
discussed in the previous section, \textit{i.e.}, we use the same
functional form as in the flavor-blind case, Eqs.~\eqref{e:f1NP}-\eqref{e:D1NP}, but with different parameters for different flavors. In particular, for TMD PDFs we independently parametrize the following flavors: $u$, $\bar{u}$, $d$, $\bar{d}$, and $sea$, where $sea$ includes $s$, $\bar{s}$, $c$, $\bar{c}$, $b$, and $\bar{b}$. For simplicity, in the following the $sea$ channel of TMD PDFs will be denoted as $s$.

For TMD FFs, we independently parametrize five different cases, as proposed
in the exploratory study of Ref.~\cite{Signori:2013mda} where charge conjugation and isospin
symmetries had been assumed.
First, we separate the fragmentation processes where the final hadron is a pion or a kaon.
Then, the fragmentation functions used to describe each process are classified as
\textit{favored} if the fragmenting quark belongs to the valence content of
the final state hadron, and \textit{unfavored} otherwise.
Additionally, for the fragmentation into a $K^+$ we independently parametrize the favored fragmentation functions for the $u$ and anti-strange $\bar{s}$ quarks (similarly, for $K^-$ the favored channels involve the $\bar{u}$ and strange $s$ quarks). In total, we have 5 sets of parameters for the following channels:

\begin{itemize}
\item{favored pion TMD FFs: $u \to \pi^+$, $d \to \pi^-$, $\bar{d} \to
  \pi^+$, $\bar{u} \to \pi^-$}
\item{unfavored pion TMD FFs: $\bar{u}, d, s, \bar{s} \to \pi^+$, $u, \bar{d}, s, \bar{s} \ \to \pi^-$}
\item{favored strange kaon TMD FFs:  $\bar{s} \to K^+$, $s \to K^-$}
\item{favored kaon TMD FFs:   $u \to K^+$, $\bar{u} \to K^-$}
\item{unfavored kaon TMD FFs: $\bar{u}, d, \bar{d}, s \to K^+$, $u, d,
  \bar{d}, \bar{s} \to K^-$}  \, .
\end{itemize}

In total, the MAPTMD24 fit involves 96 free parameters: 1 for the
nonperturbative part of the Collins-Soper kernel, 50 ($5$ flavors $ \times 10$ parameters) for the nonperturbative part of the TMD PDFs, and 45 ($5$ channels $\times 9$ parameters) for the nonperturbative part of the
TMD FFs.

We fitted 100 Monte
Carlo replicas of the experimental data
and we obtained the global $\chi^2_0 / N_{\text{dat}} = 1.08$ (see Tab.~\ref{t:chitable}), indicating that we are able to
simultaneously describe the experimental data coming from both SIDIS and DY processes in
an excellent way. It
is noteworthy that by allowing for the possibility that flavors behave
differently in transverse momentum space, we achieve a better description
compared to both MAPTMD24 FI ($\chi^2_0 = 1.40$) and
MAPTMD24 HD ($\chi^2_0 = 1.19$) scenarios.
The description improves for both SIDIS and DY data.

We report in App.~\ref{appendixA} the plots of the comparison between
experimental data and theoretical predictions for most of the included
data sets, with the blue bands representing the 68\% C.L. The plots show a very good
agreement for all experiments.

\begin{table}[h]
\begin{center}
\begin{tabular}{|l|c|c|c|c|}
 \hline
 \multicolumn{1}{|c|}{ } & \multicolumn{4}{|c|}{N$^3$LL} \\
 \hline
 Data set & $N_{\rm dat}$ & $\chi^2_D$ & $\chi^2_{\lambda}$ & $\chi^2_0$ \\
 \hline
 \hline
 Tevatron total & 71 & 1.10 & 0.07 & 1.17 \\
 \hline
 LHCb total & 21 & 3.56 & 0.96 & 4.52 \\
 \hline
 ATLAS total & 72 & 3.54 & 0.82 & 4.36 \\
 \hline
 CMS total & 78 & 0.38 & 0.05 & 0.43 \\
 \hline
 PHENIX 200 & 2 & 2.76 & 1.04 & 3.80 \\
 \hline
 STAR 510 & 7 & 1.12 & 0.26 & 1.38 \\
 \hline
 \hline
 {\it DY collider total} & 251 & 1.37 & 0.28 & 1.65 \\
 \hline
 \hline
 E288 200 GeV & 30 & 0.13 & 0.40 & 0.53 \\
 \hline
 E288 300 GeV & 39 & 0.16 & 0.26 & 0.42 \\
 \hline
 E288 400 GeV & 61 & 0.11 & 0.08 & 0.19 \\
 \hline
 E772 & 53 & 0.88 & 0.20 & 1.08 \\
 \hline
 E605 & 50 & 0.70 & 0.22 & 0.92 \\
 \hline
 \hline
 {\it DY fixed-target total} & 233 & 0.63 & 0.31 & 0.94 \\
 \hline
 \hline
 {\it DY total} & 484 & 1.02 & 0.29 & 1.31 \\
 \hline
 \hline
 HERMES total & 344 & 0.81 & 0.24 & 1.05 \\
 \hline
 COMPASS total & 1203 & 0.67 & 0.27 & 0.94 \\
 \hline
 \hline
 {\it SIDIS total} & 1547 & 0.70 & 0.26 & 0.96 \\
 \hline
 \hline
 {\bf Total} & {\bf 2031} & {\bf 0.81} & {\bf 0.27} & {\bf 1.08} \\
 \hline
\end{tabular}
\caption{Breakdown of the values of $\chi^2$ normalized to the number of data
 points $N_{\text{dat}}$ that survive the kinematic cuts for all
 datasets considered in the MAPTMD24 fit.
 The $\chi^2_D$ refers to uncorrelated
 uncertainties, $\chi^2_\lambda$ is the penalty term due to correlated
 uncertainties, $\chi^2_0$ is the sum of $\chi^2_D$ and $\chi^2_\lambda$ (see text).}
\label{t:chitable}
\end{center}
\end{table}

The values of the nonperturbative parameters and their uncertainties are
reported in Tab.~\ref{t:parameters} of
App.~\ref{appendixB}. All parameters are well constrained and not compatible
with zero. We observe no strong correlations among them
(see Fig.~\ref{f:correlation_matrix} in App.~\ref{appendixB}).

\begin{figure}[h]
\centering
\includegraphics[width=0.9\textwidth]{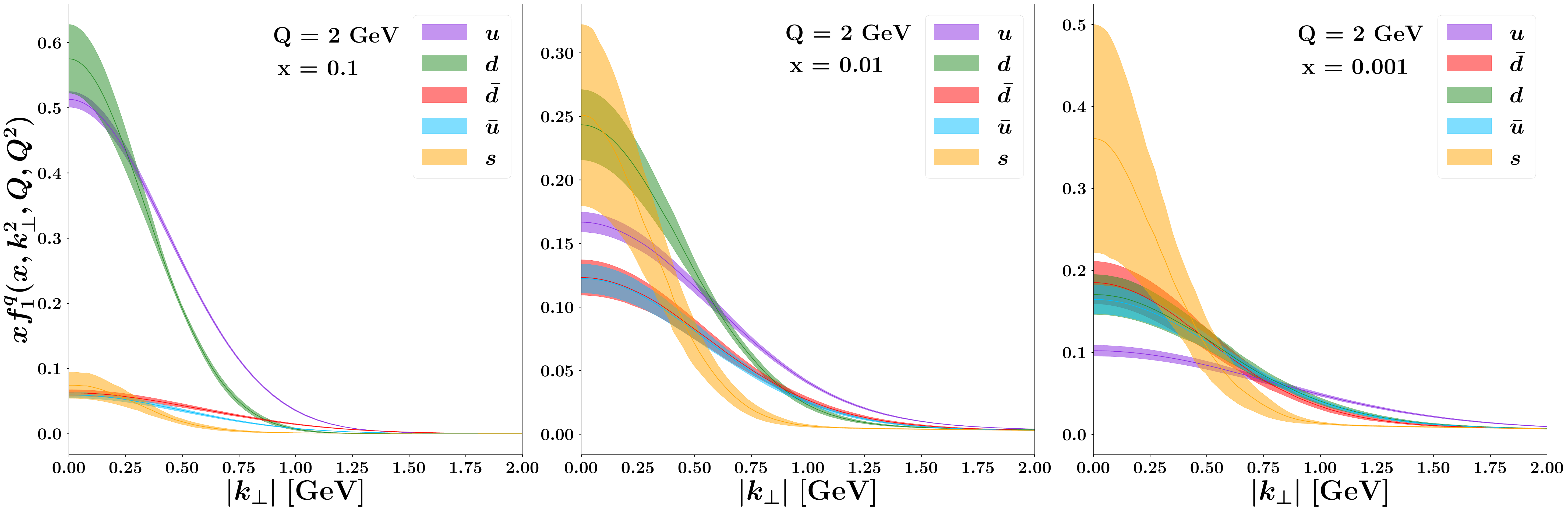}
\caption{Comparison between the unpolarized TMD PDFs extracted in the MAPTMD24 fit with a flavor dependent approach,
for a up (purple), anti-up (light blue), down (green), anti-down (red), and $sea$ (orange) quark, as
functions of the partonic transverse momentum $|\kperp|$ at $\mu = \sqrt{\zeta} = Q = 2$ GeV and
$x =0.1$ (left panel), $x =0.01$ (central panel), and $x =0.001$ (right panel).
The uncertainty bands represent the 68\% C.L.}
\label{f:tmdpdf_fl_dep}
\end{figure}

\begin{figure}[h]
\centering
\includegraphics[width=0.9\textwidth]{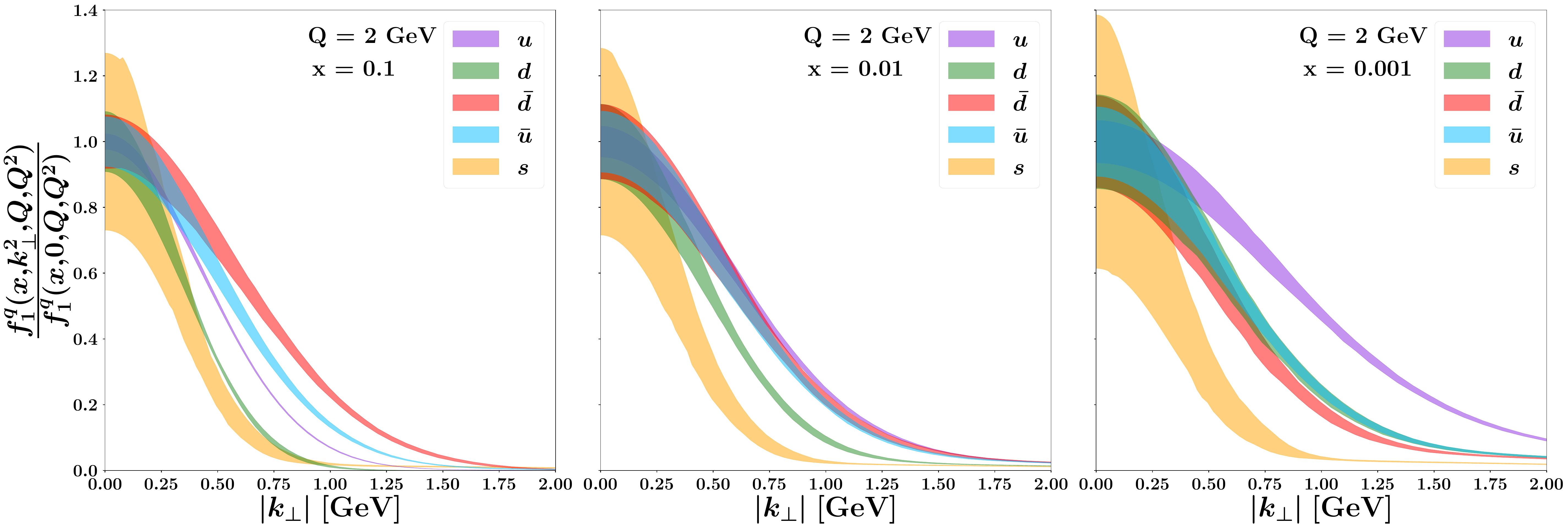}
\caption{Comparison between the normalized unpolarized TMD PDFs extracted in the MAPTMD24 fit with a flavor-dependent approach,
for a up (purple), anti-up (light blue), down (green), anti-down (red), and $sea$ (orange) quark, as
functions of the partonic transverse momentum $|\kperp|$ at $\mu = \sqrt{\zeta} = Q = 2$ GeV and
$x =0.1$ (left panel), $x =0.01$ (central panel), and $x =0.001$ (right panel).
The uncertainty bands represent the 68\% C.L.}
\label{f:tmdpdf_fl_dep_norm}
\end{figure}

\subsubsection{TMDs}

We now discuss the TMD PDFs and FFs extracted from the MAPTMD24 FD fit at
N${}^3$LL accuracy.

Figure~\ref{f:tmdpdf_fl_dep} displays
the unpolarized TMD PDFs for the various independent flavors, as functions of the partonic transverse momentum
$|\kperp|$ at $\mu = \sqrt{\zeta} = Q = 2$ GeV and $x =0.1$ (left panel), $x
=0.01$ (central panel), and $x =0.001$ (right panel). The uncertainty bands
represent the 68\% C.L.

We note that at $x=0.1$ the contributions of the up and down
quarks dominate. The $d$-quark TMD PDF is larger at low values of $|\kperp|$ and
decreases more rapidly than the $u$-quark one.
At small $x$, the contributions from the sea quarks increase and become
dominant at low $|\kperp|$ values.
Furthermore, at medium to low $x$ the $\bar{u}$-quark and $\bar{d}$-quark TMD PDFs
behave in a similar way, while the $u$-quark and $d$-quark ones are very different.

In Fig.~\ref{f:tmdpdf_fl_dep_norm}, using the same notation as above, we show
the normalized TMD PDFs, \textit{i.e.}, divided by the value of the corresponding
central replica at $|\kperp| = 0$.
This representation allows one to better visualize the difference in
shape among various flavors.

At $x=0.1$ (left panel), the TMD PDFs of the $sea$ ($s$) and $d$ quarks show the sharpest decrease in $|\kperp|$,
while the $\bar{d}$ quark is the widest. At $x=0.001$, the $s$ quark is still narrow, while the $u$ quark is the widest.
As $x$ becomes smaller, the TMD PDFs of $u$ and $d$ become much wider while there are no significant differences in the
other TMD PDFs.

\begin{figure}[h]
\centering
\includegraphics[width=0.9\textwidth]{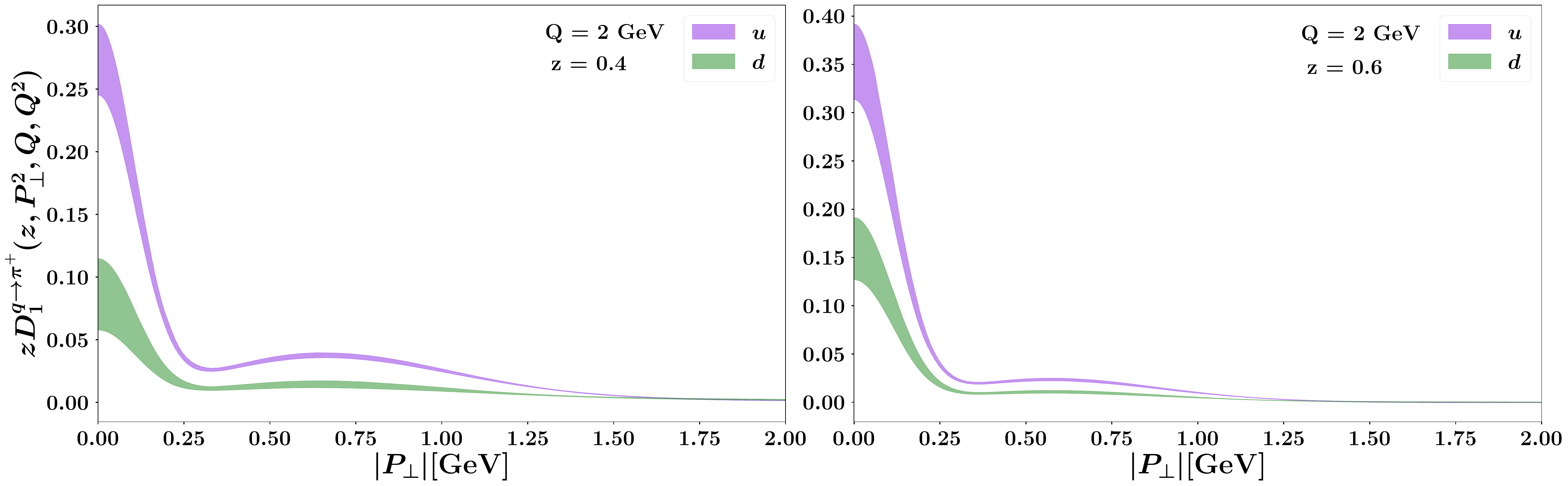}
\caption{Comparison between the unpolarized TMD FFs for the fragmentation into a $\pi^+$ of up (purple) and down (green) quarks, extracted in the MAPTMD24 fit with a flavor dependent approach, as
functions of the hadronic transverse momentum $|\Pperp|$ at $\mu = \sqrt{\zeta} = Q = 2$ GeV and
$z =0.4$ (left panel), and $z =0.6$ (right panel).
The uncertainty bands represent the 68\% C.L.}
\label{f:tmdff_pi_fl_dep}
\end{figure}

Moreover, the distribution of quarks not belonging to the valence content of the proton
appears to be the least constrained with large uncertainty bands for all $x$ values,
as expected from the lack of experimental data directly sensitive to sea
quarks. On the contrary, at larger $x$ (left panel) the uncertainty bands of the TMD PDFs for up and down quarks are very narrow, due to
the large amount of SIDIS data in combination with high-precision DY data. It is useful to remark that  the uncertainties for all flavors
increase as $x$ decreases, confirming the need for experimental data in this kinematic region.

In Fig.~\ref{f:tmdff_pi_fl_dep}, we display the unpolarized TMD FFs for the fragmentation into a
$\pi^+$ of up (purple) and down (green) quarks, as functions of the hadronic
transverse momentum $|\Pperp|$ at
$\mu = \sqrt{\zeta} = Q = 2$ GeV and $z =0.4$ (left panel), and $z =0.6$ (right panel).
We note that the favored fragmentation channel (in this example,
$ u \rightarrow \pi^+$) dominates over the unfavored one. Also, both TMD FFs show a second bump at
intermediate $|\Pperp|$
which decreases in size at larger $z$, as already observed in Sec.~\ref{s:baseline}.

\begin{figure}[h]
\centering
\includegraphics[width=0.9\textwidth]{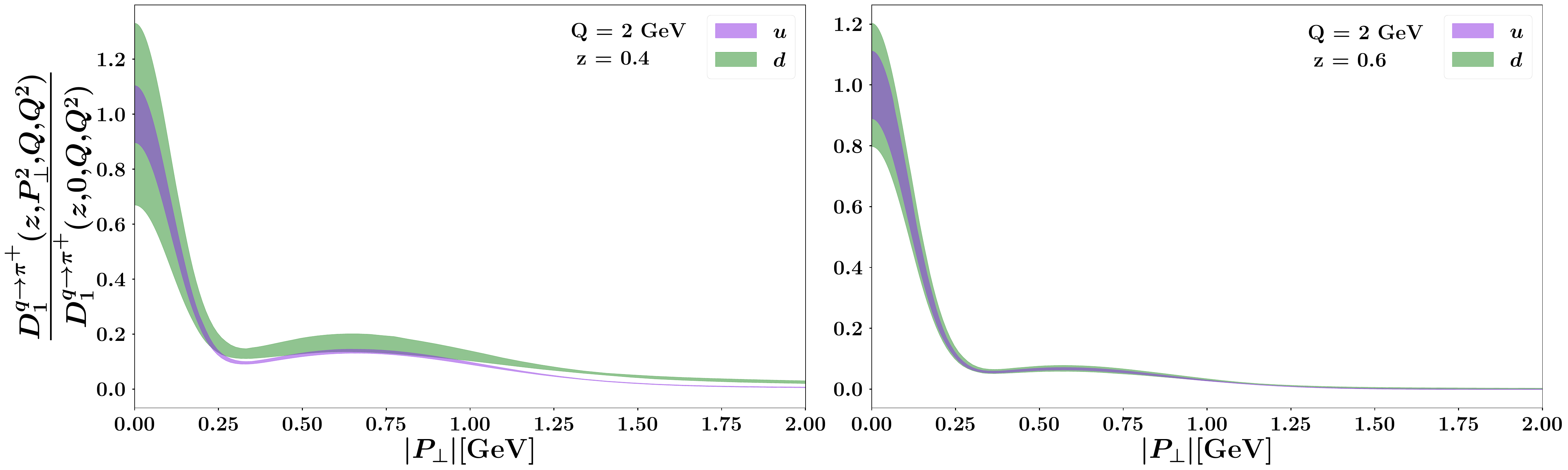}
\caption{Comparison between the normalized unpolarized TMD FFs for the fragmentation into a $\pi^+$ of up (purple) and down (green) quarks, extracted in the MAPTMD24 fit with a flavor dependent approach, in the same conditions and with same notation as in the previous figure.}
\label{f:tmdff_pi_fl_dep_norm}
\end{figure}

In Fig.~\ref{f:tmdff_pi_fl_dep_norm}, we display the same TMD FFs of the previous figure but normalized to each corresponding central replica at $|\Pperp| = 0$. The unfavored channel (here, $d \rightarrow \pi^+$) is affected by larger
error bands. This is mainly due to the larger uncertainties in the
corresponding collinear FFs.
There is generally no significant difference between favored and unfavored
channels at high $z$, probably due to the limited sensitivity of SIDIS
data in that kinematic region.

\begin{figure}[h]
\centering
\includegraphics[width=0.9\textwidth]{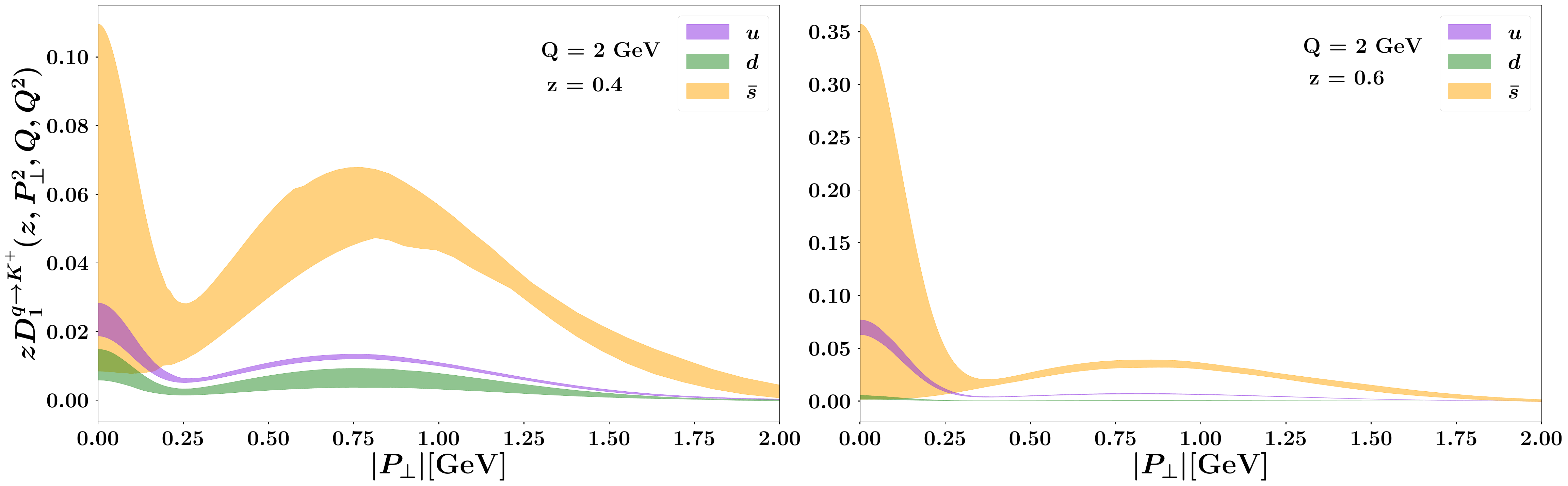}
\caption{Comparison between the unpolarized TMD FFs for the fragmentation of up (purple), down (green), and anti-strange (orange) quarks into a $K^+$, extracted in the MAPTMD24 fit with a flavor dependent approach, as
functions of the hadronic transverse momentum $|\Pperp|$ at $\mu = \sqrt{\zeta} = Q = 2$ GeV and
$z =0.4$ (left panel),  and $z =0.6$ (right panel).
The uncertainty bands represent the 68\% C.L.}
\label{f:tmdff_ka_fl_dep}
\end{figure}

\begin{figure}
\centering
\includegraphics[width=0.9\textwidth]{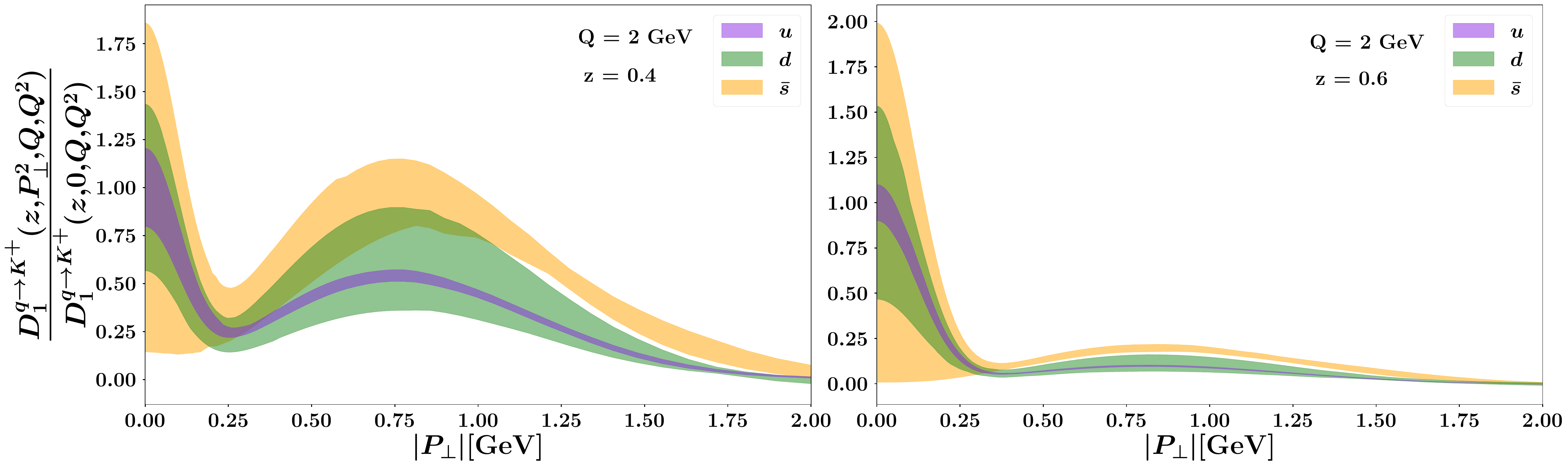}
\caption{Comparison between the normalized unpolarized TMD FFs for the fragmentation of up (purple), down (green), and anti-strange (orange) quarks into a $K^+$, extracted in the MAPTMD24 fit with a flavor dependent approach, in the same kinematic conditions and with same notation as in the previous figure.}
\label{f:tmdff_ka_fl_dep_norm}
\end{figure}

In Fig.~\ref{f:tmdff_ka_fl_dep}, we show the unpolarized TMD FFs for the fragmentation of quarks $u$, $d$, and $\bar{s}$ into a $K^+$ in the same
kinematic regions and with same conventions as in  Fig.~\ref{f:tmdff_pi_fl_dep}. Similarly, in Fig.~\ref{f:tmdff_ka_fl_dep_norm} we show the normalized versions, as we did in Fig.~\ref{f:tmdff_pi_fl_dep_norm} for the fragmentation into a $\pi^+$. We note that in general the extracted TMD FFs for kaons are affected by larger uncertainties than for pions.
Also, the bump at intermediate $|\Pperp|$ is more pronounced than in the case of pions, as was also observed with the hadron-dependent MAPTMD24 HD fit (see Fig.~\ref{f:tmdff_hadrondep_comparison}).
Due to the size of the corresponding collinear FFs, the fragmentation channel
$\bar{s} \rightarrow K^+$ is dominant, also in the normalized case.
An interesting feature of our extraction is that the two
favored channels ($u \rightarrow K^+$ and $\bar{s} \rightarrow K^+$) are quite
different from each other.
The large uncertainties in the
$\bar{s} \rightarrow K^+$ fragmentation channel may be related to the fact that this TMD FF appears in the SIDIS cross section through the convolution with a TMD PDF of a $sea$ quark, which is small and has large uncertainties in our extraction.

\subsubsection{Impact of PDF uncertainties}

In Figs.~\ref{f:E288_plot}-\ref{f:COMP_plot_P} of App. A, we note that the uncertainty bands of the MAPTMD24 FD predictions are larger than
those from the MAPTMD22 fit, as it can be realized by inspecting the corresponding Figs.~4-11 of Ref.~\cite{Bacchetta:2022awv}. This is due to a more flexible parametrization but also to the fact that for MAPTMD24 we consider different members of collinear PDF and FF Monte Carlo sets for each TMD replica. In fact, also the error bands of the MAPTMD24 FI fit are larger than in MAPTMD22, even though the fitting function is the same.
Hence, in MAPTMD24 we have a more accurate assessment of the uncertainty in the normalization of our predictions. For a better visualization of this effect, in the following we show the comparison with data of the results from the MAPTMD22 (blue), MAPTMD24 FI (green) and MAPTMD24 FD (red) fits for selected bins of the SIDIS multiplicities and DY cross sections.

\begin{figure}[h]
\centering
\includegraphics[width=0.48\textwidth]{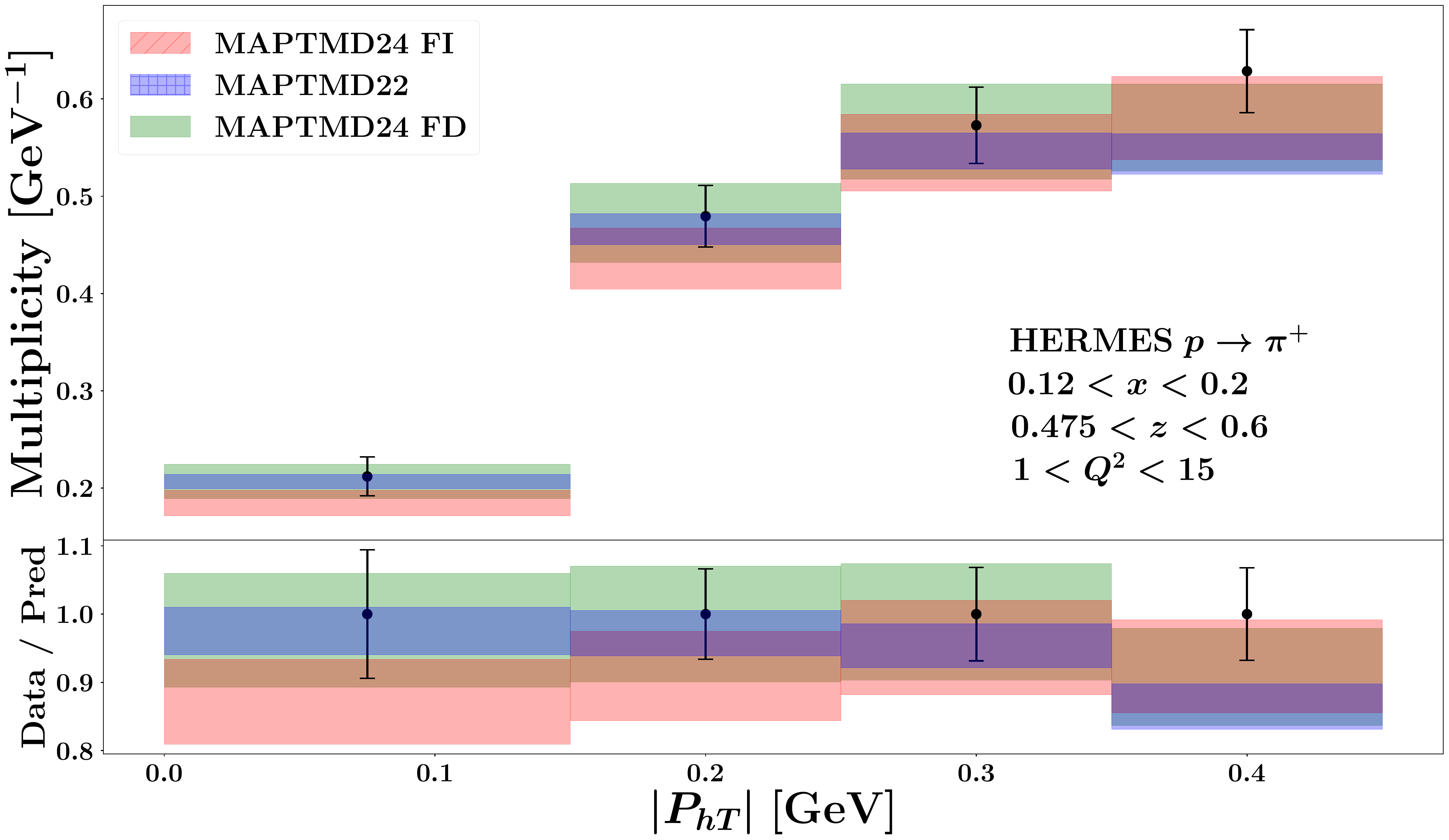} \hspace{0.3cm}
\includegraphics[width=0.48\textwidth]{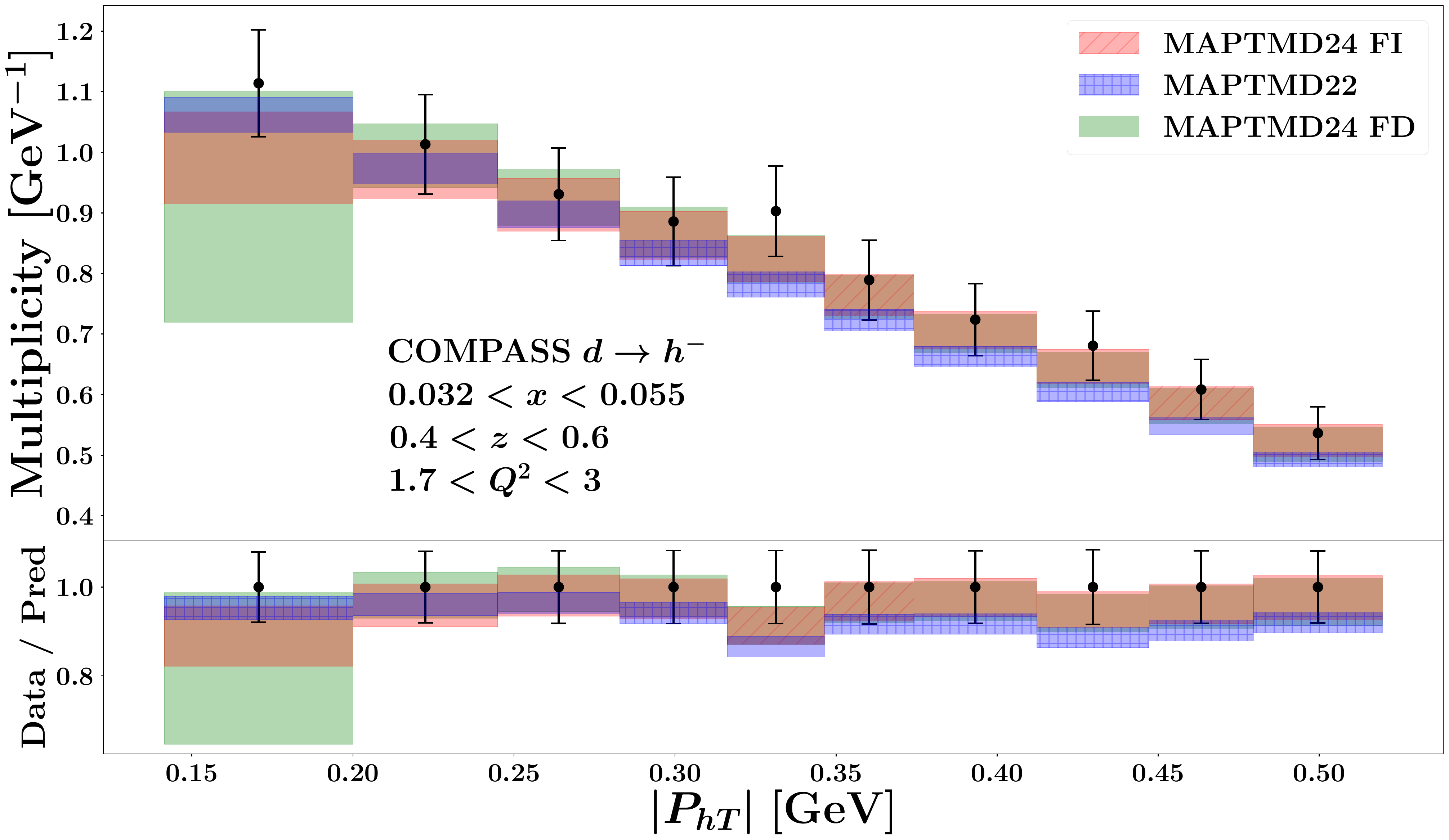}
\caption{Comparison of SIDIS multiplicities  as function of $|\PhT|$, obtained by MAPTMD22 (blue), MAPTMD24 FI (green) and MAPTMD24 FD (red) fits and measured by the $\hermes$ collaboration for $\pi^+$ off proton in the $0.12 < x < 0.2, \, 0.475 < z < 0.6, \, 1< Q^2 < 15$ GeV$^2$ bin (left plot) and by the $\compass$ collaboration for negative charged hadrons off deuteron in the $0.032 < x < 0.055, \, 0.4 < z < 0.6, \, 1.7 < Q^2 < 3$ GeV$^2$ bin (right plot). Error bands at 68\% C.L.}
\label{fig:SIDIS_fitcomp}
\end{figure}

In Fig.~\ref{fig:SIDIS_fitcomp}, we show the comparison between both the MAPTMD24 FI and MAPTMD24 FD fits and our previous MAPTMD22 fit for the  SIDIS multiplicity as function of $|\PhT|$, measured by the $\hermes$ collaboration for $\pi^+$ production off proton target in the $0.12 < x < 0.2, \, 0.475 < z < 0.6, \, 1 < Q^2 < 15$ GeV$^2$ bin (left plot), and by the $\compass$ collaboration for negative charged hadrons off deuteron in the $0.032 < x < 0.055, \, 0.4 < z < 0.6, \, 1.7 < Q^2 < 3$ GeV$^2$ bin (right plot). The error bands (all at 68\% C.L.) of the MAPTMD24 fits are evidently larger than the MAPTMD22 ones, and give a more accurate estimate of the uncertainty on this observable.

\begin{figure}[h]
\centering
\includegraphics[width=0.48\textwidth]{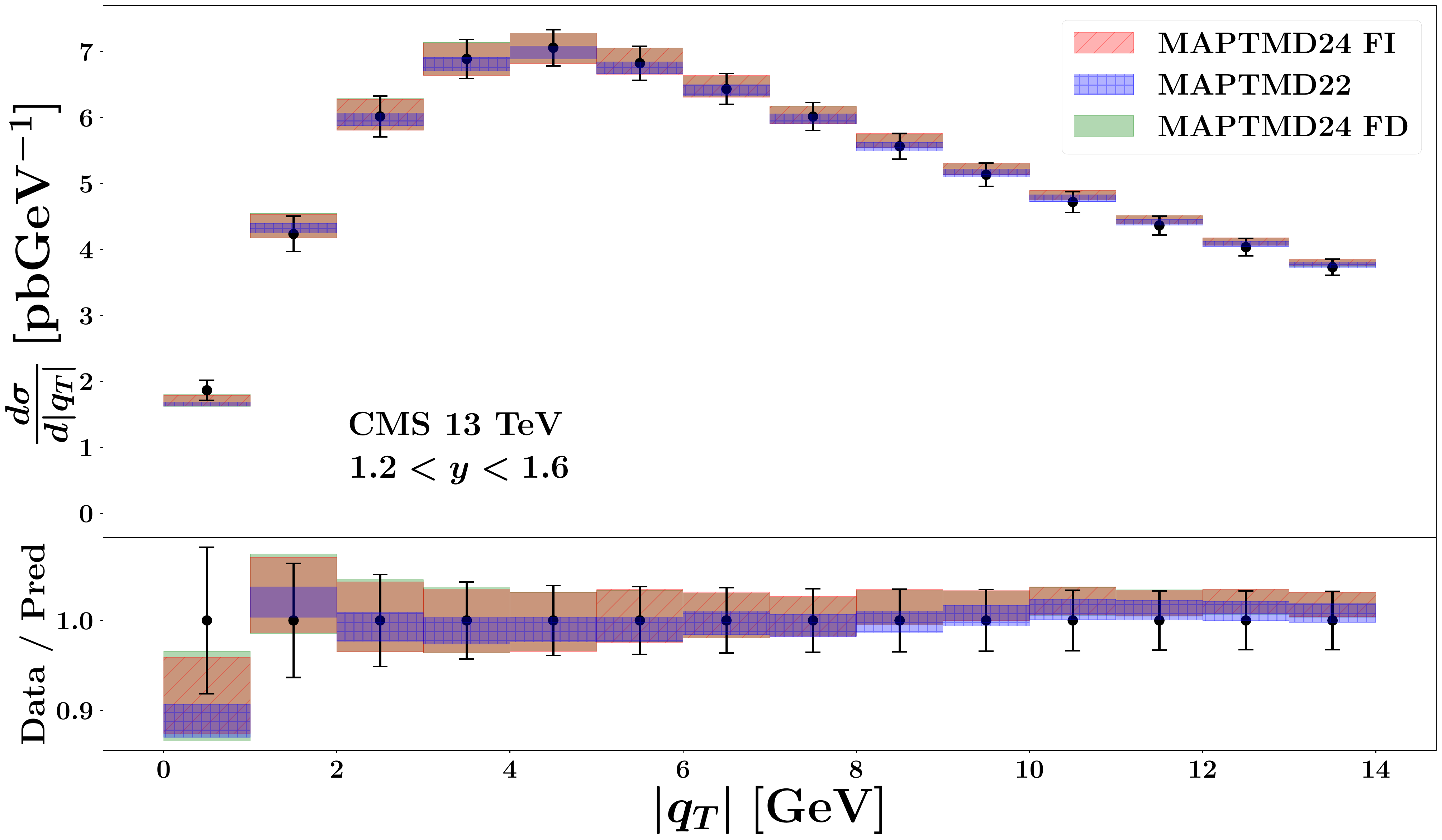} \hspace{0.3cm}
\includegraphics[width=0.48\textwidth]{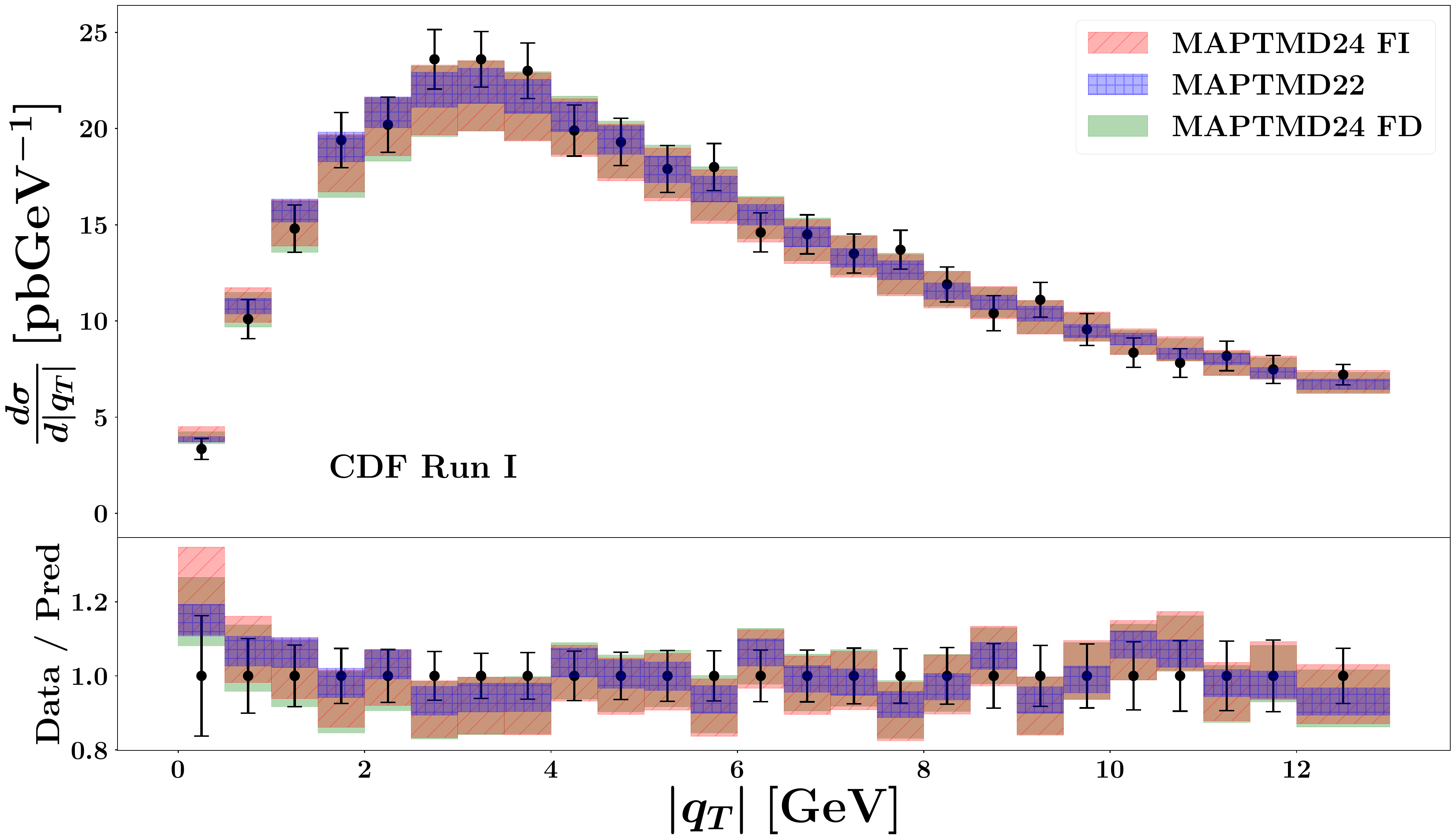}
\caption{Same as in the previous figure but for the DY unpolarized cross section as function of $|\qT|$, measured by the \textsc{CMS} collaboration at 13 TeV in the $1.2 < y < 1.6$ bin (left plot) and by the \textsc{CDF} collaboration in Run I (right plot).}
\label{fig:DY_fitcomp}
\end{figure}

In Fig.~\ref{fig:DY_fitcomp}, we show the same comparison as in previous figure but for the DY unpolarized cross section as function of $|\qT|$, measured by the \textsc{CMS} collaboration at 13 TeV in the $1.2 < y < 1.6$ bin (left plot) and by the \textsc{CDF} collaboration in Run I (right plot). As for the width of the error bands, the same previous comment applies. It is also worth noting that for the DY process the MAPTMD24 FD uncertainties are very similar to the MAPTMD24 FI ones: the DY observables, being related to the sum upon all flavors of quark-antiquark contributions, are not significantly affected by the flavor dependence.

\begin{figure}[h]
\centering
\includegraphics[width=0.8\textwidth]{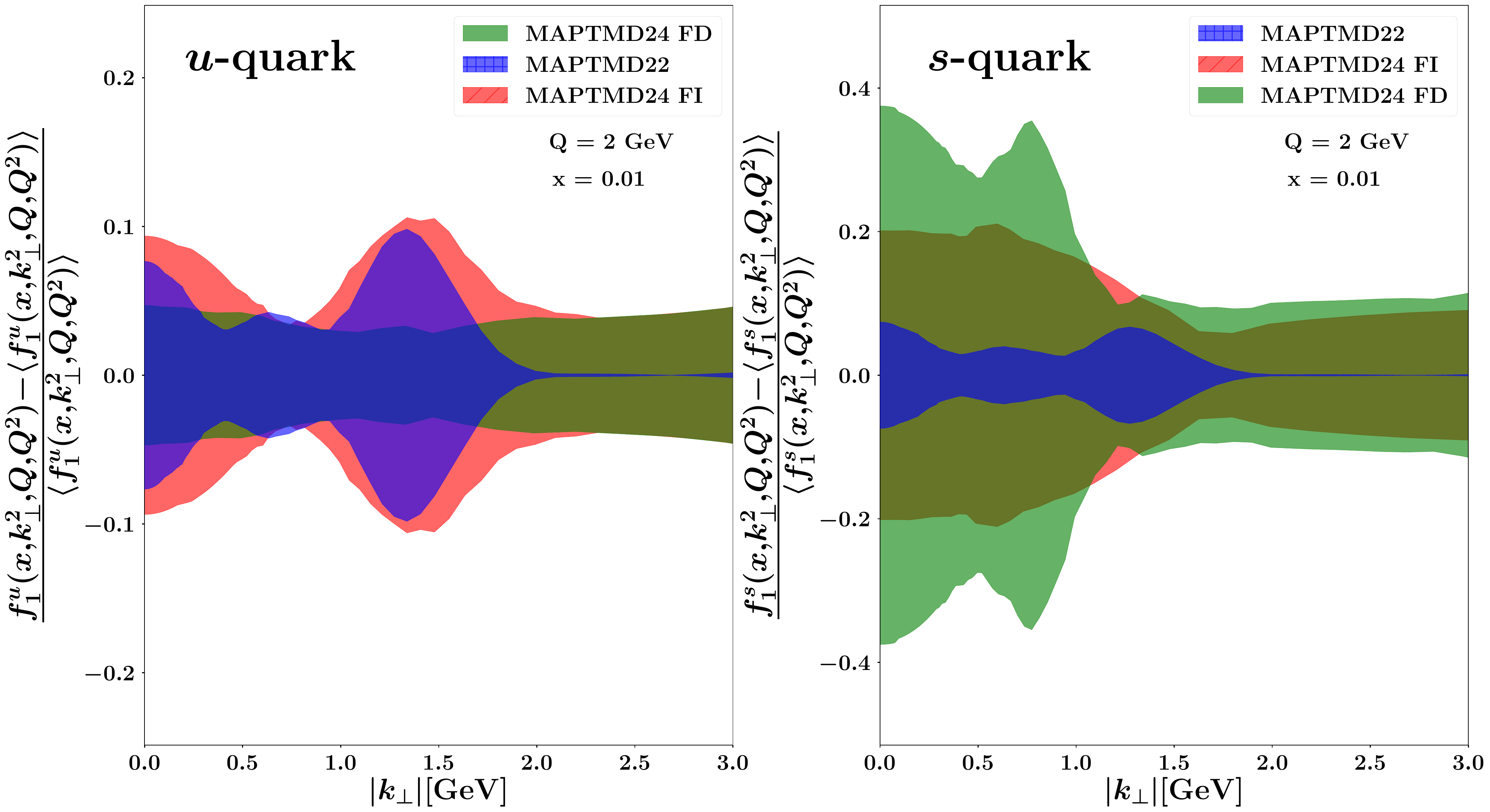}
\caption{Comparison among the relative uncertainties of the MAPTMD24 FI (green), MAPTMD24 FD (red), MAPTMD22 (blue) fits for the up quark (left panel) and $sea$ (right panel), as functions of the partonic transverse momentum $|\kperp|$ at $\mu = \sqrt{\zeta} = Q = 2$ GeV and $x =0.01$. The uncertainty bands represent the 68\% C.L.}
\label{f:tmdpdf_error_band}
\end{figure}

It is useful to perform the same comparison at the level of the extracted TMDs. In Fig.~\ref{f:tmdpdf_error_band}, we compare the error bands at 68\% C.L. of the TMD PDFs for the $u$ quark (left plot) and $sea$ ($s$) quark (right plot), extracted from the MAPTMD24 FI (green), MAPTMD24 FD (red), and MAPTMD22 (blue) fits. The uncertainties are relative to the corresponding average value of all fit replicas and are plotted as functions of $|\kperp|$ at $Q = 2$ GeV and $x =0.01$. For the $u$ quark, the error bands are similar in the low-$|\kperp|$ region, but at high $|\kperp|$ the MAPTMD24 uncertainties are larger because each TMD replica is matched onto a different replica of the collinear PDFs. For the $s$ quark, the MAPTMD24 error bands are larger than MAPTMD22 over the whole $|\kperp|$ range, due to the large uncertainties in the collinear PDFs which affect both the integral of the TMD and its large $|\kperp|$ tail.

\subsubsection{Collins-Soper kernel}

\begin{figure}
\centering
\includegraphics[width=0.7\textwidth]{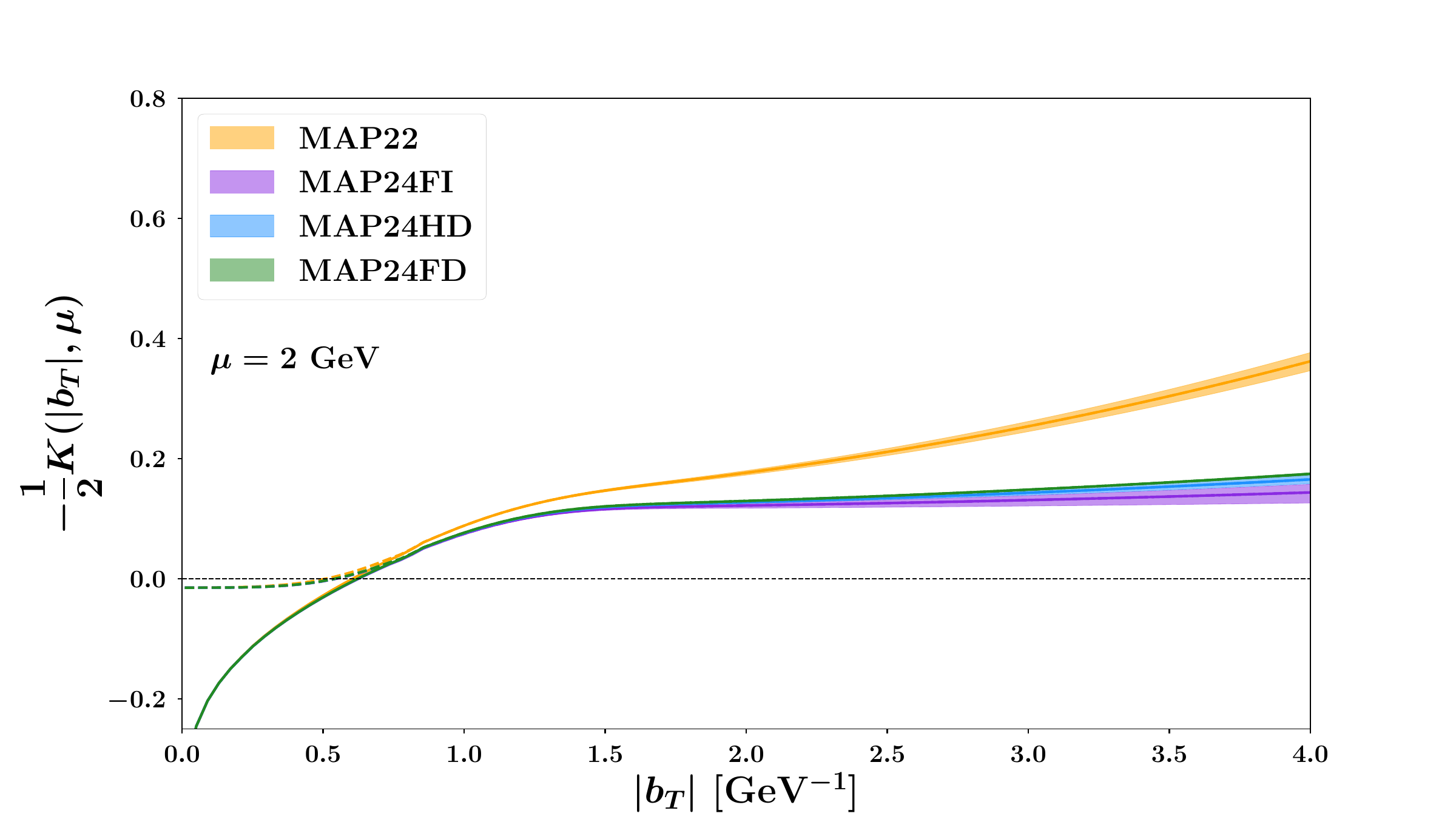}
\caption{The Collins--Soper kernel as a function of $|\bT|$ at the
scale $\mu=2$ GeV from the three versions of the present analysis (MAPTMD24
FI, MAPTMD24 HD, and MAPTMD24 FD), compared with
the MAPTMD22 result~\cite{Bacchetta:2022awv}. The uncertainty bands represent the 68$\%$ C.L. Dashed lines show
the effect of including the $b_{\text{min}}$-prescription of Eq.~\eqref{e:bminmax}.}
\label{f:CSKernel}
\end{figure}

In Fig.~\ref{f:CSKernel}, we show the result for the Collins--Soper (CS)
 kernel obtained in our MAPTMD24 extraction at N$^3$LL with a flavor-dependent approach, compared to our previous
 MAPTMD22 results. The form of the CS kernel at
 low values of $|\bT|$ is unchanged, as it depends on perturbative
 ingredients.
The behavior at high $|\bT|$ is determined by the combination of the
$b_{*}$ prescription and the parametrization of the nonperturbative component of TMD evolution in Eq.~\eqref{e:CSkernelNP}.

In our new MAPTMD24 extraction, the value of the parameter $g_2$ is smaller than
in MAPTMD22: it is approximately 0.12, about half as big as the MAPTMD22 result
($\approx 0.25$). Because of this difference, the new MAPTMD24 CS kernel is
flatter than the MAPTMD22 one.
This feature is not related to the flavor
dependence of the new extraction, because it is present also in the MAPTMD24 FI
and MAPTMD24 HD scenarios. Instead, it is due to the differences in the perturbative ingredients between
the present work and the MAPTMD22 analysis, already discussed in Sec.~\ref{ss:TMDs}. A milder dependence on $|\bT|$ is obtained also by assuming in Eq.~\eqref{e:CSkernelNP} a constant or a linear dependence of the CS kernel on $|\bT|$~\cite{Collins:2014jpa,Hautmann:2020cyp,Boglione:2023duo}.

The size of the error band on the CS kernel is small and similar to the MAPTMD22 one. It is
possible that our fit procedure leads to an underestimation of
the errors, especially for the CS kernel, since its functional form is
particularly rigid and determined by a single
parameter (see Eq.~\eqref{e:CSkernelNP}).

Our result can be compared with other recent extractions in the
literature. The ART23 extraction~\cite{Moos:2023yfa} included DY data only and obtained a CS kernel similar to the
MAPTMD22, which is therefore steeper than our MAPTMD24
result. Ref.~\cite{Aslan:2024nqg} obtained a result, based on a smaller set of
DY data
and a simplified analysis, with larger error bands that are compatible with
MAPTMD22, ART23 and also MAPTMD24.
The result obtained in
Ref.~\cite{Isaacson:2023iui}, obtained with DY data only, is also compatible with MAPTMD22 and ART23, and
about 1.5 sigma away from our present results.

Apart from data-driven extractions, there have been several computations of
the CS kernel in lattice
QCD~\cite{Shanahan:2020zxr,LatticeParton:2020uhz,Schlemmer:2021aij,Li:2021wvl,Shanahan:2021tst,LatticePartonLPC:2022eev,Zhang:2022xuw,Shu:2023cot,LatticePartonLPC:2023pdv,Avkhadiev:2023poz,Spanoudes:2024kpb,Avkhadiev:2024mgd,Bollweg:2024zet}.
The error bars are still relatively
large and there are sizeable differences between different computations.
Our MAPTMD24 extraction is compatible with the recent
work of Ref.~\cite{Bollweg:2024zet}.

\subsubsection{Average squared transverse momenta}

In order to measure the effective width of the TMDs, in this section we
study their average squared transverse momentum
at specific values of $x$ and $\mu = \sqrt{\zeta} = Q$, defined as~\cite{Boer:2011xd,Boer:2014bya}:

\begin{align}
\label{e:avkp2}
\langle \kperp^2 \rangle^{q}(x,Q) & =
\frac{\int d^2 \kperp\, \kperp^2\, f_1^{q}(x,\kperp^2,Q,Q^2)}{\int d^2 \kperp\, f_1^{q}(x,\kperp^2,Q,Q^2)} =
\frac{2M^2\, \hat{f}_1^{q\, (1)}(x,|\bT|,Q,Q^2)}{\hat{f}_1^{q}(x,|\bT|,Q,Q^2)}\bigg|_{\modbT=0} \, ,
\end{align}
where the Fourier transform $\hat{f}_1^q$ of the TMD PDF has been defined in Eq.~\eqref{eq:FTdef}, and the first Bessel moment of the TMD PDF $\hat{f}_1^{q\, (1)}$ is defined as~\cite{Boer:2011xd}:
\begin{equation}
\label{e:BesMom_f1}
\hat{f}_1^{q\, (1)}(x,|\bT|,Q,Q^2) = \frac{2\pi}{M^2}\,
\int_0^{+\infty} d|\kperp|\, \frac{\kperp^2}{|\bT|}\, J_1\big( |\kperp| |\bT|
\big)\, f_1^q(x,\kperp^2,Q,Q^2) =
-\frac{2}{M^2} \frac{\partial}{\partial \bT^2}
 \hat{f}_1^{q\,}(x,|\bT|,Q,Q^2) \, .
\end{equation}

As discussed in Ref.~\cite{Bacchetta:2022awv}, we shift the value of $|\bT|$ in Eq.~\eqref{e:avkp2} from 0 to $\modbT=2.0\, b_{\text{max}}$, a value well inside the nonperturbative region~\cite{Boer:2014bya}, that ensures meaningful values for the average squared transverse momenta that must be finite, positive across all the $x$ and $Q$ values considered in this fit, and dominated by the small-$|\kperp|$ region of the TMDs:
\begin{equation}
\label{e:avkp2_reg}
\langle \kperp^2 \rangle^{q}_r (x,Q) =
\frac{2M^2\, \hat{f}_1^{q\, (1)}(x,|\bT|,Q,Q^2)}{\hat{f}_1^{q}(x,|\bT|,Q,Q^2)}\bigg|_{\modbT=2.0\, b_{\text{max}}} \, ,
\end{equation}
where we denote with the subscript $r$ the \emph{regularized} definition of the average squared momenta.

The same arguments can be applied to the \emph{regularized} average squared transverse momentum produced in the fragmentation of a given quark $q$ into the final state hadron $h$~\cite{Boer:2011xd,Boer:2014bya,Bacchetta:2019qkv,Bacchetta:2022awv}:
\begin{align}
\label{e:avPp2_reg}
\langle \Pperp^2 \rangle^{q \to h}_r (z,Q) & =
\frac{2\, z^2\, M_h^2\, \hat{D}_1^{q \to h\, (1)}(z,|\bT|,Q,Q^2)}{\hat{D}_1^{q \to h}(z,|\bT|,Q,Q^2)}\bigg|_{\modbT=2.0\, b_{\text{max}}} \, ,
\end{align}
where the Fourier transform $\hat{D}_1^{q \to h}$ of the TMD FF is defined in Eq.~\eqref{eq:FTdefFF}, and the first Bessel moment of the TMD FF $\hat{D}_1^{q \to h\, (1)}$ is defined as~\cite{Bacchetta:2019qkv}:
\begin{equation}
 \begin{split}
\label{e:BesMom_D1}
\hat{D}_1^{q \to h\, (1)}(z,|\bT|,Q,Q^2) &= \frac{2\pi}{M_h^2}\,
\int_0^{+\infty} \frac{d|\Pperp|}{z}\, \frac{|\Pperp|}{z}\,
\frac{|\Pperp|}{z|\bT|}\, J_1\big( |\bT| |\Pperp|/z \big)\, D_1^{q \to
 h}(z,\Pperp^2,Q,Q^2)
\\
&=
-\frac{2}{M_h^2} \frac{\partial}{\partial \bT^2}
\hat{D}_1^{q\to h}(z,|\bT|,Q,Q^2) \, .
\end{split}
\end{equation}

\begin{figure}[h]
\centering
\includegraphics[width=1.0\textwidth]{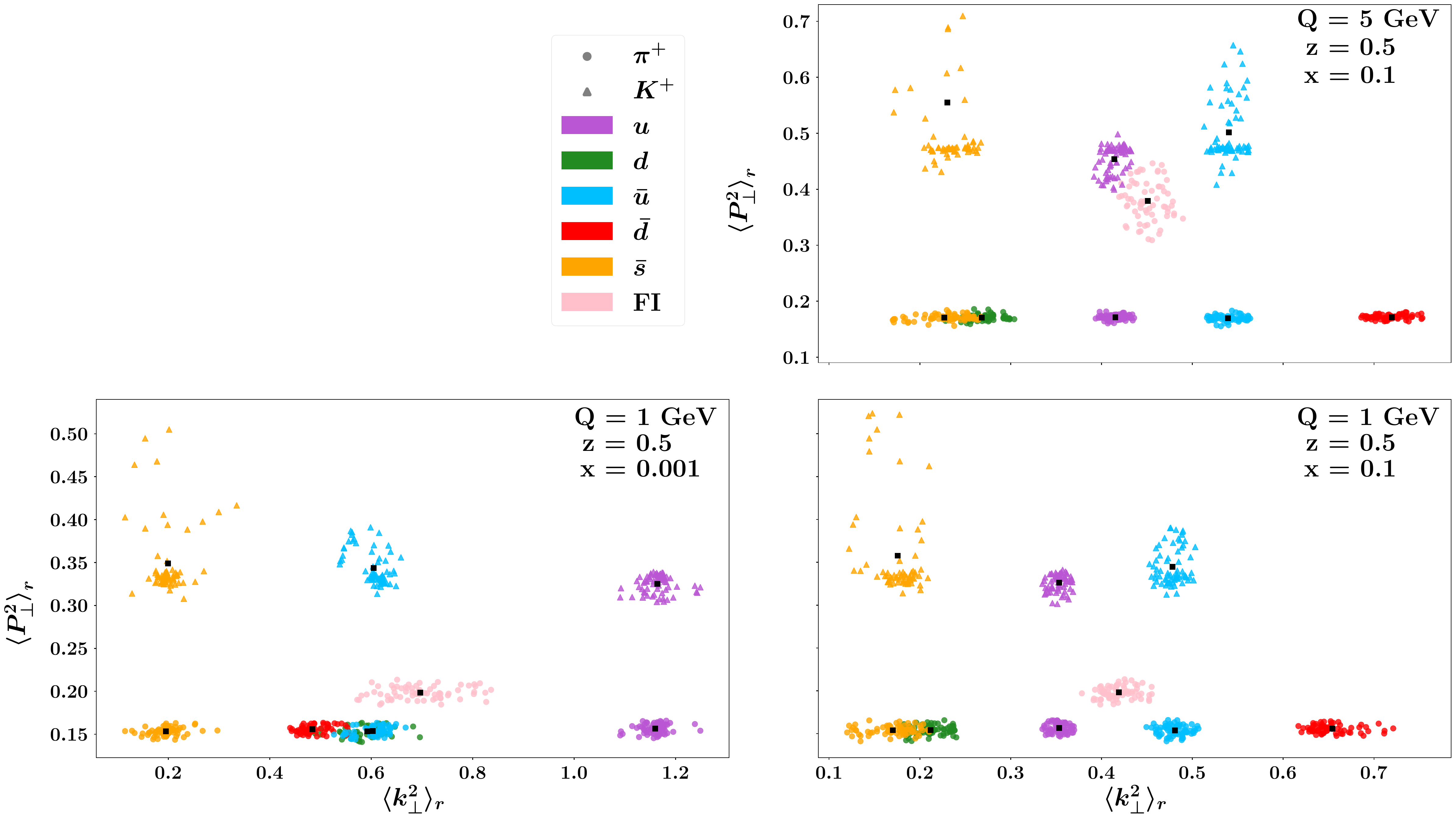}
\caption{Scatter plot of average squared transverse momenta for the unpolarized TMD PDF at $x=0.1$ (right panels),
$x=0.001$ (left panel) and for the unpolarized TMD FF for fragmentation into $\pi^+$ (circle) or into $K^+$ (triangle) at
$z = 0.5$. In the upper panel, TMDs are evaluated at $Q = 5$ GeV, in the lower panels at $Q = 1$ GeV. Different colors for different flavors as indicated in the legend. Black squares represent the mean value for the different clusters. The 68\% C.L. of the different ensembles of replicas is reported.}
\label{f:scatter_plots}
\end{figure}

In Fig.~\ref{f:scatter_plots}, we display the scatter plot of
$\langle\Pperp^2 \rangle^{q \to h}_r$ at $z=0.5$ versus $\langle
\kperp^2\rangle^{q}_r$ for different flavors $q$. Lower panels show the
results at $Q = 1$ GeV, the upper-right panel at $Q = 5$ GeV. The
$\langle \kperp^2 \rangle^{q}_r$ in the right panels are evaluated at
$x = 0.1$, while in the left panel at $x = 0.001$. In the upper-left
corner we display the legend of the various scatter plots with different color codes for the different flavors: the circles refer to $\langle \Pperp^2 \rangle^{q \to \pi^+}_r$ for the fragmentation into $\pi^+$ pions, while the triangles are for $\langle \Pperp^2 \rangle^{q\to K^+}_r$ into $K^+$ kaons. The black squares refer to the mean value of each cluster of colored points.
We display only the 68\% C.L. of the different ensembles of replicas.

The pink cluster, representing the replicas of the MAPTMD24 FI fit, appears along the $x$ axis in an
intermediate position with respect to other clusters, indicating that
the nonperturbative component of the TMD PDFs in the flavor-independent
approach is approximately an average across different flavors.
Similarly, its position along the $y$ axis is an average between the positions
of the clusters of pions and kaons.
The clusters for the fragmentation into kaons appear at higher average squared transverse momenta than for pions, and are more spread.
For different values of $x$, the ordering of the various flavors
changes. All these features reflect the results of the MAPTMD24 FD fit that we already commented, in particular the outcome in Fig.~\ref{f:tmdpdf_fl_dep_norm}. Finally, both the values of $\langle \kperp^2 \rangle^{q}_r$ and $\langle \Pperp^2 \rangle^{q \to h}_r$
increase as $Q$ increases, since the evolution equations generate a broadening of the transverse momentum distributions. A similar trend is observed also in parton-shower-based Monte Carlo generators of collider events, either after underlying-event tuning~\cite{CMS:2019csb} or after including TMD effects into the shower according to the parton branching model~\cite{Bubanja:2023nrd}.

\section{Conclusions}
\label{s:conclusions}

In this paper, we performed an extraction of transverse-momentum-dependent parton distribution
and fragmentation functions from a comprehensive set of 2031 experimental
data points from the 
Drell-Yan (DY) process and semi-inclusive deep-inelastic scattering
(SIDIS), with the main goal of unraveling the distinctions among 
different quark flavors. It is the first time that
the flavor-dependent nature of Transverse Momentum
Distributions (TMDs) is fully taken into consideration in a global fit.

Our study builds upon previous work by incorporating state-of-the-art theory
results reaching N$^3$LL accuracy, and adopting the fitting framework used in
our past works, available through the {\tt NangaParbat} public code.\footnote{The code and a collection of final results will be made publicly available by the MAP collaboration at \href{https://github.com/MapCollaboration}{https://github.com/MapCollaboration}.} As done in Ref.~\cite{Bury:2022czx} for DY, we used Monte Carlo replicas of collinear PDFs and FFs. This
enabled an accurate portrayal of the flavor-specific characteristics of TMDs
and their uncertainties, at least within the choices for prescriptions and
functional forms that we adopted.

After reviewing the formalism in Sec.~\ref{s:formalism} and the analysis framework in
Sec.~\ref{s:analysis}, we presented three extractions with three different approaches. In
Sec.~\ref{s:baseline}, we discussed a Flavor
Independent extraction (MAPTMD24 FI) and a Hadron Dependent one (MAPTMD24 HD), characterized by different
fragmentation functions for different final-state hadrons. They constitute a baseline
to assess the relevance of a flavor-depedent fit. We adopted the same choices as
in our previous extraction (MAPTMD22), but we used two Monte Carlo
sets of collinear PDFs and FFs in order to fully account for their uncertainties.
We obtained $\chi_0^2 / N_{dat} = 1.40$ and 1.19 for the two extractions, respectively.

Section~\ref{s:flavor} presents the core of our analysis,
where we separately parametrized five TMD PDFs ($u$, $\bar{u}$, $d$,
$\bar{d}$, and $sea$) and five TMD FFs (favored and unfavored pion
fragmentation, favored, unfavored and $s$-quark kaon fragmentation). We extracted a
total of 96 free parameters. This flavor-dependent extraction (MAPTMD24 FD)
reached $\chi_0^2 / N_{dat} = 1.08$.
Therefore, the MAPTMD24 FD fit demonstrates superior capability in
simultaneously describing data from both SIDIS and DY processes, and is able
to capture the nontrivial interplay between quark flavors and their transverse
momentum distributions.

The extracted TMD PDFs and FFs offer valuable insights into the
three-dimensional structure of hadrons, revealing distinctive flavor-dependent
behaviors across different kinematic regimes.
In particular, the $u$-quark TMD PDF results to be the most constrained
among all flavors, and it is the widest at small and intermediate $x$.
On the other hand, an
examination of TMD FFs demonstrates the importance of distinguishing between
favored and unfavored channels, particularly evident for kaon fragmentations.

We also obtained a new determination of the Collins–Soper kernel, which provides crucial insights
into TMD evolution. Our MAPTMD24 result shows a lower slope at large $b_T$
compared to other recent
results~\cite{Bacchetta:2022awv,Moos:2023yfa,Aslan:2024nqg,Isaacson:2023iui}. Further
precise,
multidimensional data sets
spanning a wide $Q^2$ range will be invaluable to
further investigate these differences.

Overall, our study represents a significant step forward in the quest for a
comprehensive understanding of the flavor-dependent structure of hadrons in
momentum space.
Our findings pave the way for more refined theoretical predictions and
improved interpretations of experimental phenomena in
high-energy physics.

\begin{acknowledgments}
We thank Emanuele Nocera for stimulating discussions.
This work is supported by the European Union's Horizon 2020 programme
under grant agreement No.~824093 (STRONG2020) and by the European Union ``Next
Generation EU'' program through the Italian PRIN 2022 grant
n.~20225ZHA7W. This material is also based upon work supported by the U.S. Department of Energy, Office of Science, Office of Nuclear Physics under contract DE-AC05-06OR23177. C.B. is supported by the DOE contract DE-AC02-06CH11357.
\end{acknowledgments}

\appendix
\section{Quality of global fit}
\label{appendixA}

\begin{figure}[h]
\centering
\includegraphics[width=1.0\textwidth]{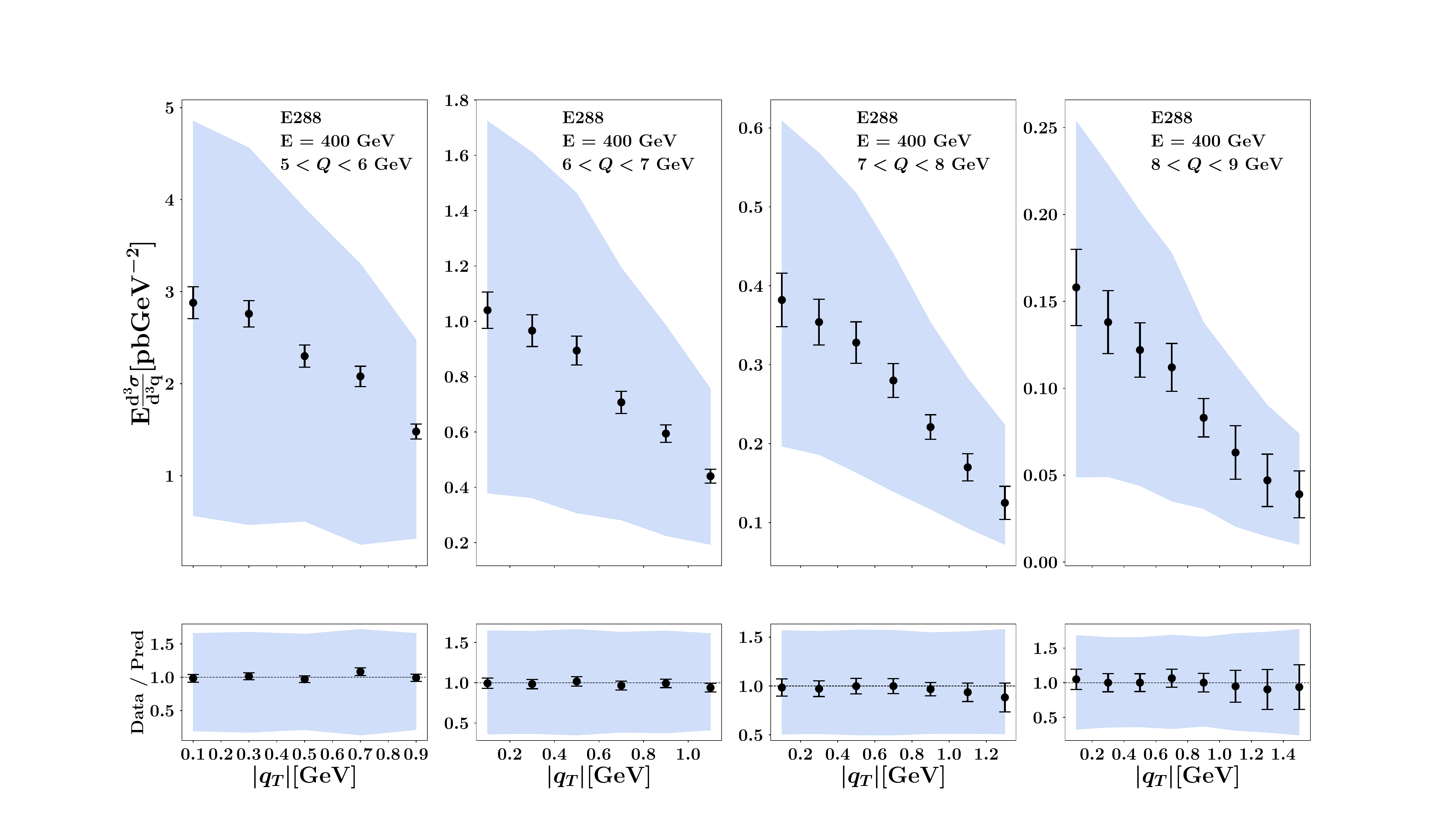}
\caption{Upper panel: comparison between data and theoretical predictions for the DY cross section differential in $|\qT|$ for the \textsc{E288} dataset at $E_{beam} = 400$ GeV for different $Q$ bins; uncertainty bands correspond to the 68\% C.L. Lower panel: ratio between experimental data and theoretical cross section.} 
\label{f:E288_plot}
\end{figure}

\begin{figure}[h]
\centering
\includegraphics[width=1.0\textwidth]{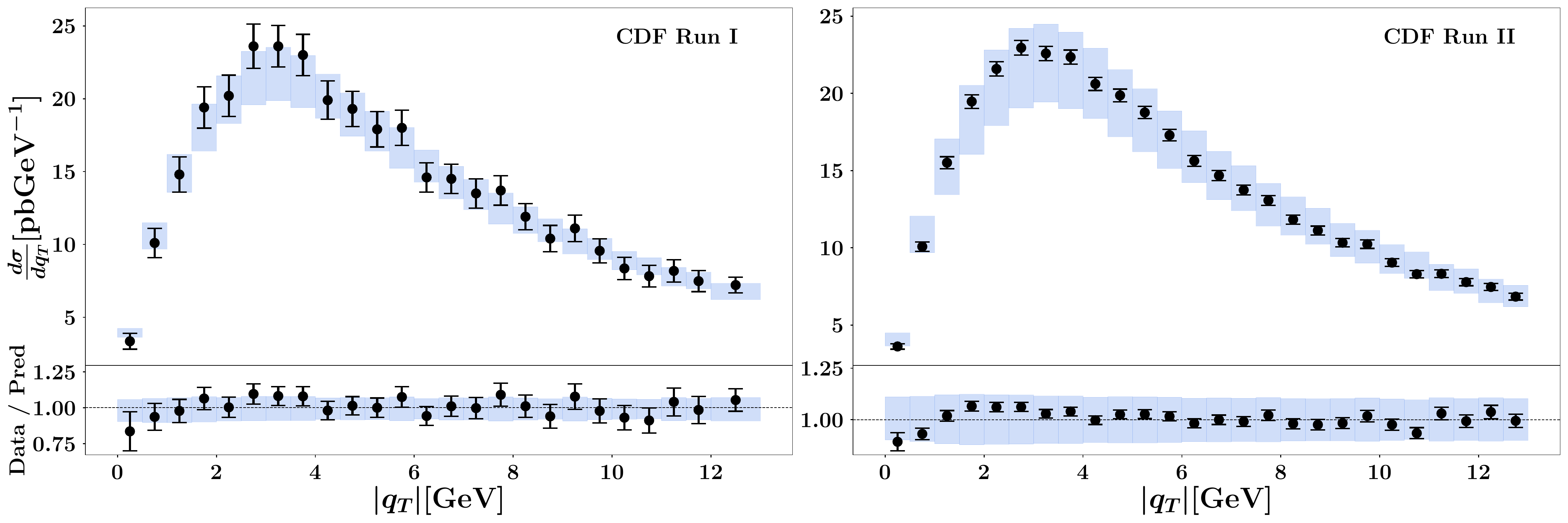}
\caption{Upper panels: comparison between experimental data and theoretical
  predictions for the cross section differential in $|\qT|$ for $Z$ bosons produced in
  $p\bar{p}$ collisions at the Tevatron from \textsc{CDF} Run I (left panel) and run II (right panel); uncertainty bands correspond to the 68\% CL. Lower panel: ratio between
  experimental data and theoretical results.}
\label{f:CDF_plot}
\end{figure}

\begin{figure}[h]
\centering
\includegraphics[width=1.0\textwidth]{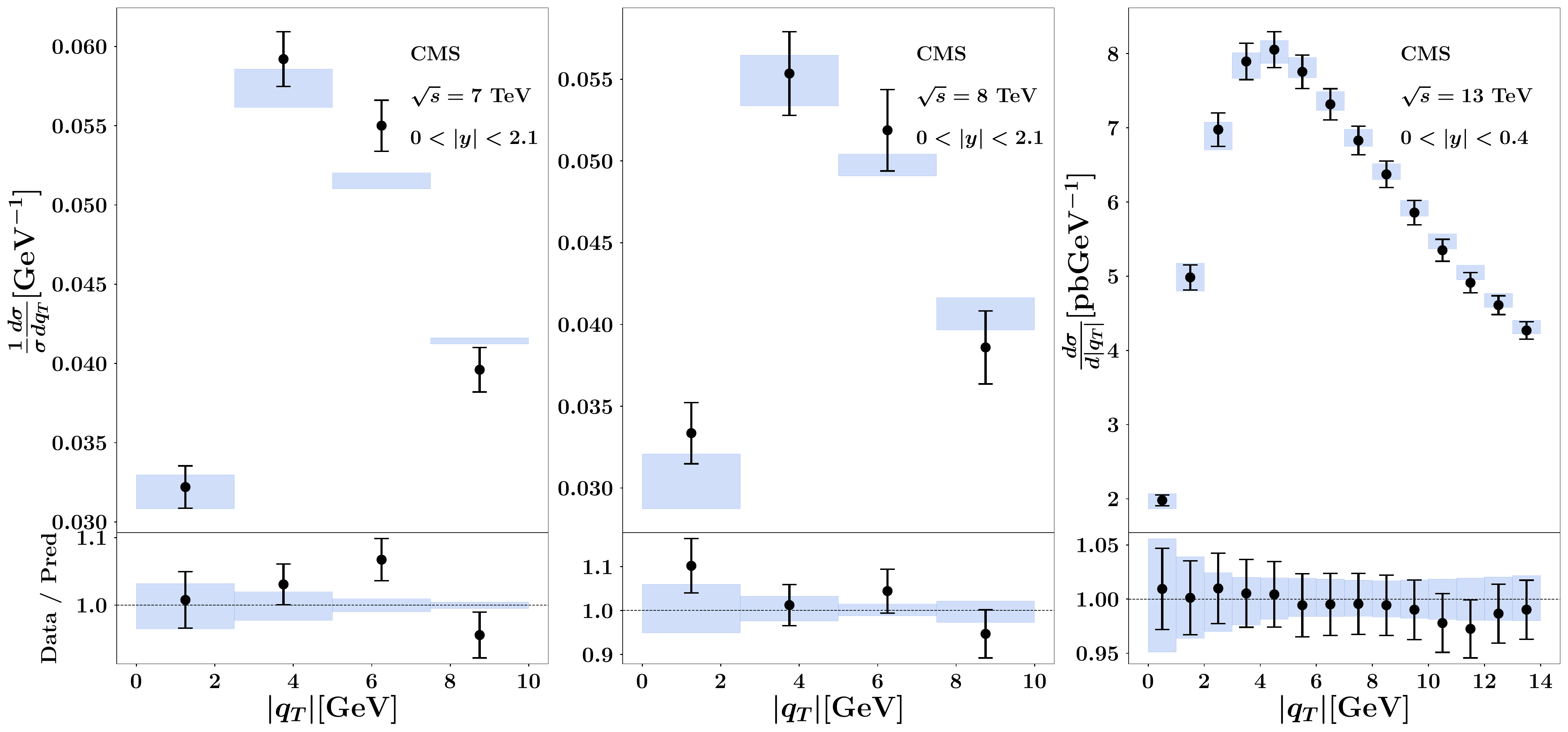}
\caption{Same as in previous figure but for $Z$ boson production in $pp$ collisions measured by the CMS Collaboration. From left to right: increasing $\sqrt{s} =$ 7,  8, 13 TeV, respectively. For $\sqrt{s} =$ 7, 8 TeV, the results are normalized to the fiducial cross section.}
\label{f:CMS_plot}
\end{figure}

\begin{figure}[h]
\centering
\includegraphics[width=1.0\textwidth]{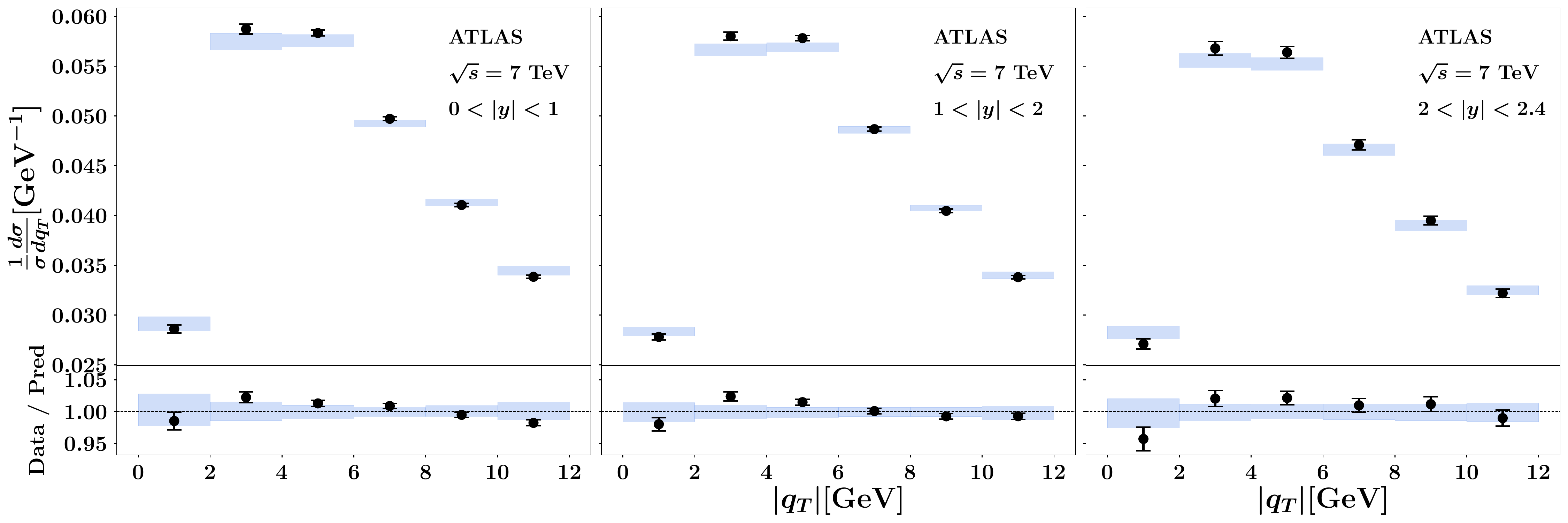}
\caption{Same as in the left and central panels of previous figure, but
  for \textsc{ATLAS} kinematics at $\sqrt{s} = 7$ TeV. From left to right, results
  at increasing rapidity.}
\label{f:ATLAS_plot}
\end{figure}

\begin{figure}[h]
\centering
\includegraphics[width=0.8\textwidth]{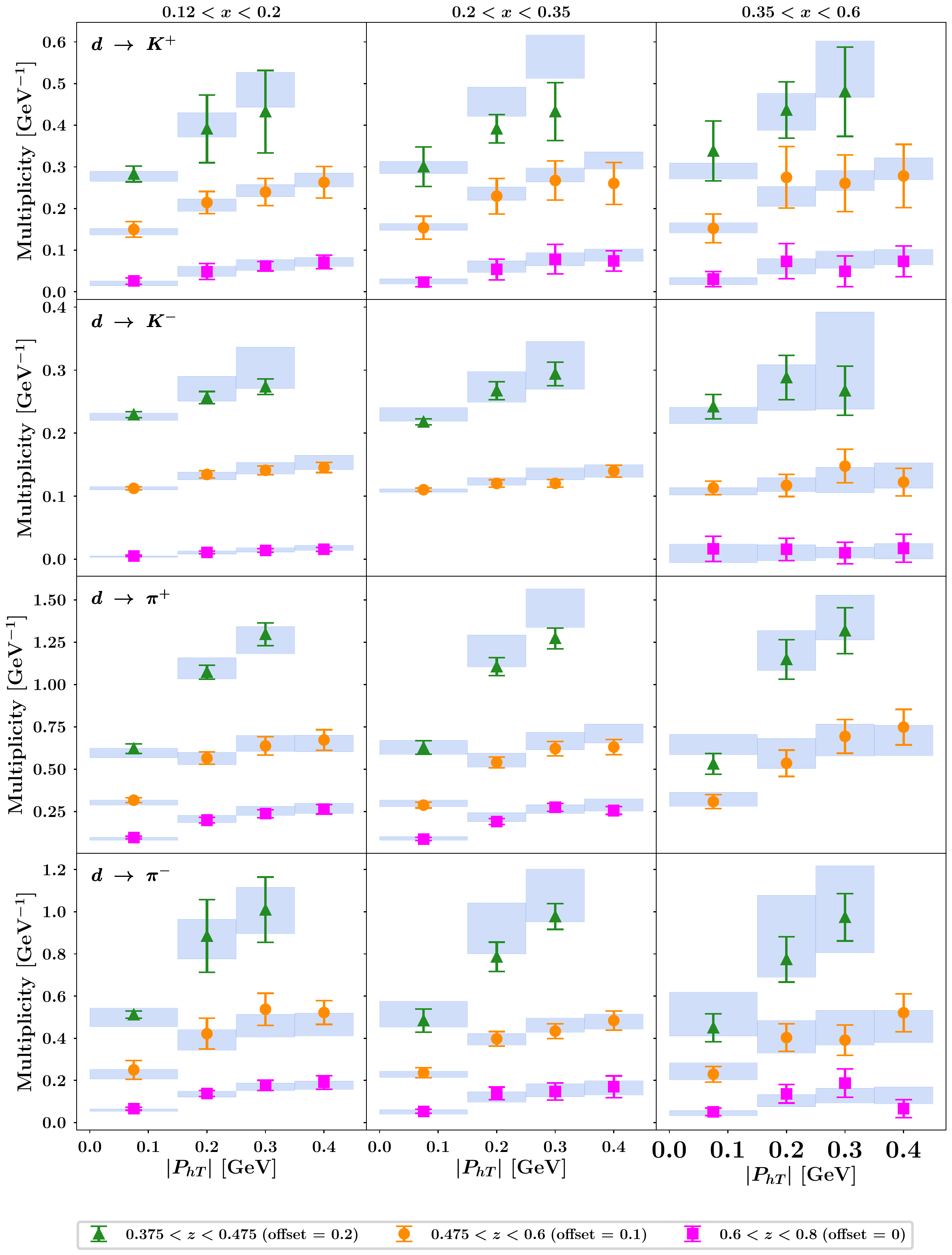}
\caption{Comparison between data and theoretical predictions for the $\hermes$ multiplicities for the production of charged pions and kaons off a deuteron target for different $x$ and $z$ bins as a function of the transverse momentum $|\PhT|$ of the final-state hadron. For better visualization, each $z$ bin is shifted by the indicated offset.}
\label{f:HER_plot_D}
\end{figure}

\begin{figure}[h]
\centering
\includegraphics[width=0.8\textwidth]{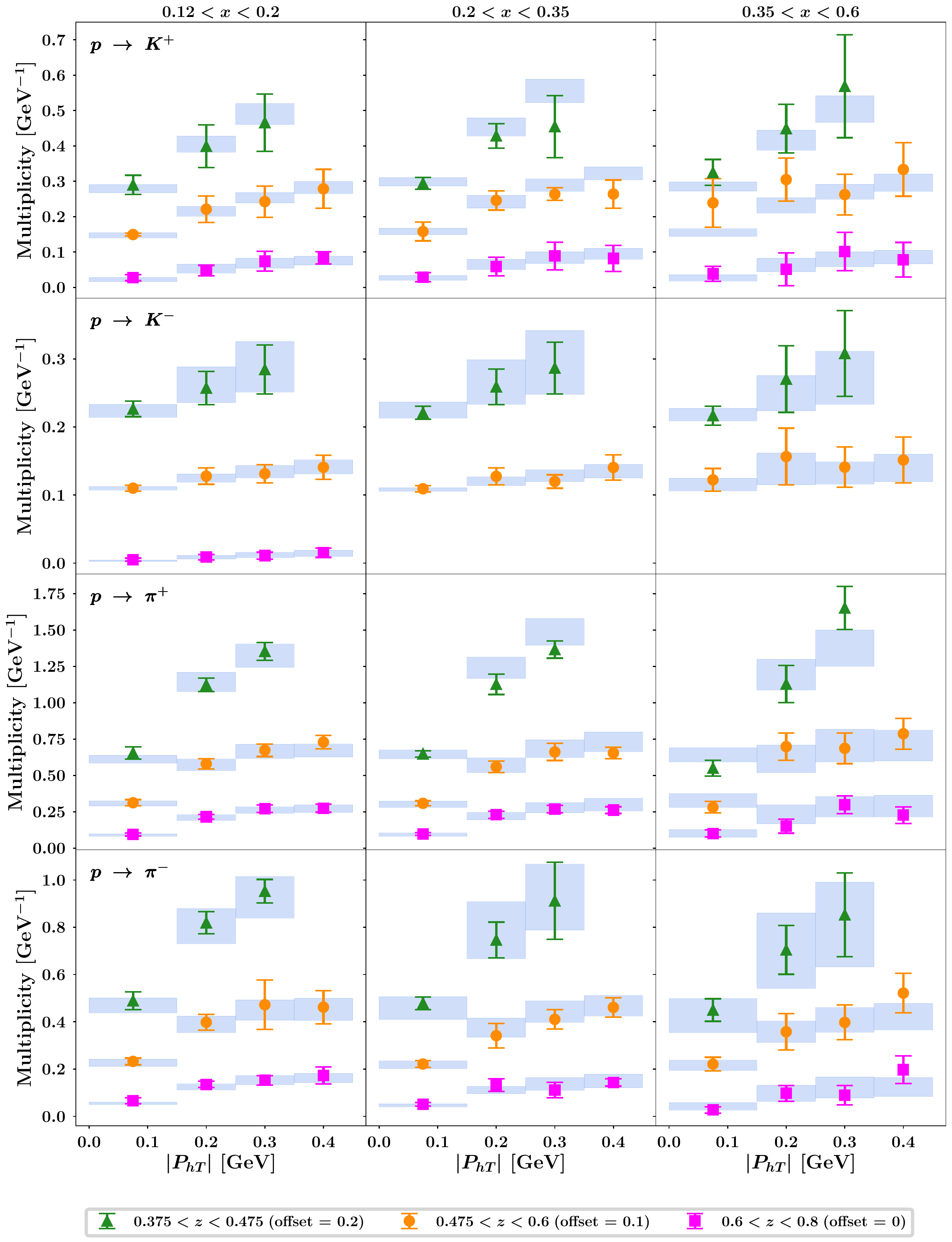}
\caption{Comparison between data and theoretical predictions for the $\hermes$ multiplicities for the production of charged pions and kaons off a proton target for different $x$ and $z$ bins as a function of the transverse momentum $|\PhT|$ of the final-state hadron. For better visualization, each $z$ bin is shifted by the indicated offset.}
\label{f:HER_plot_P}
\end{figure}

\begin{figure}[h]
\centering
\includegraphics[width=1.0\textwidth]{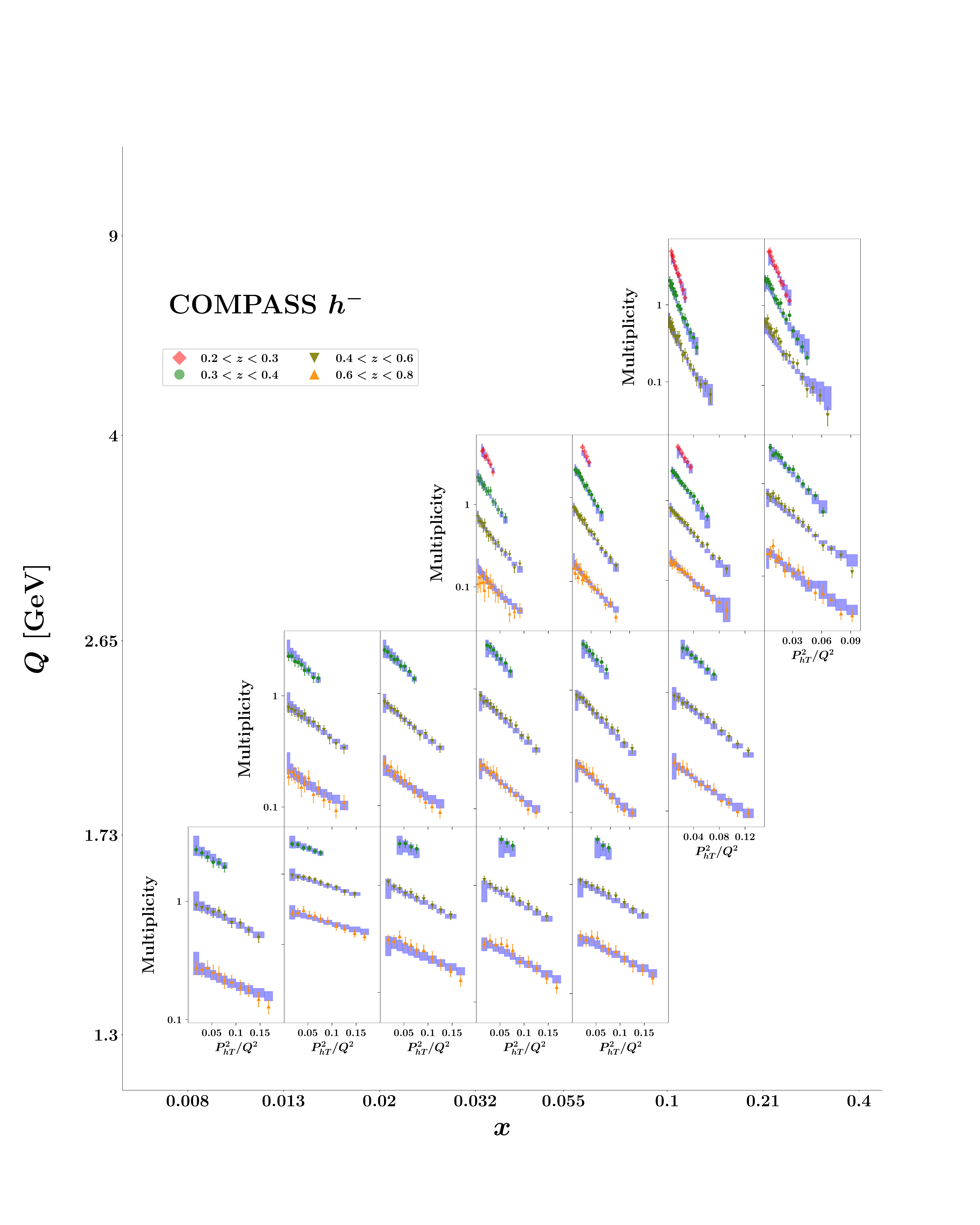}
\caption{Comparison between data and theoretical predictions for the $\compass$ multiplicities for the production of negative charged hadrons off a deuteron target. For each $Q, x$ bin,  the multiplicities are displayed as functions of $\PhT^2/Q^2$ for different $z$ bins surviving kinematic cuts, as indicated in the legend.}
\label{f:COMP_plot_M}
\end{figure}

\begin{figure}[h]
\centering
\includegraphics[width=1.0\textwidth]{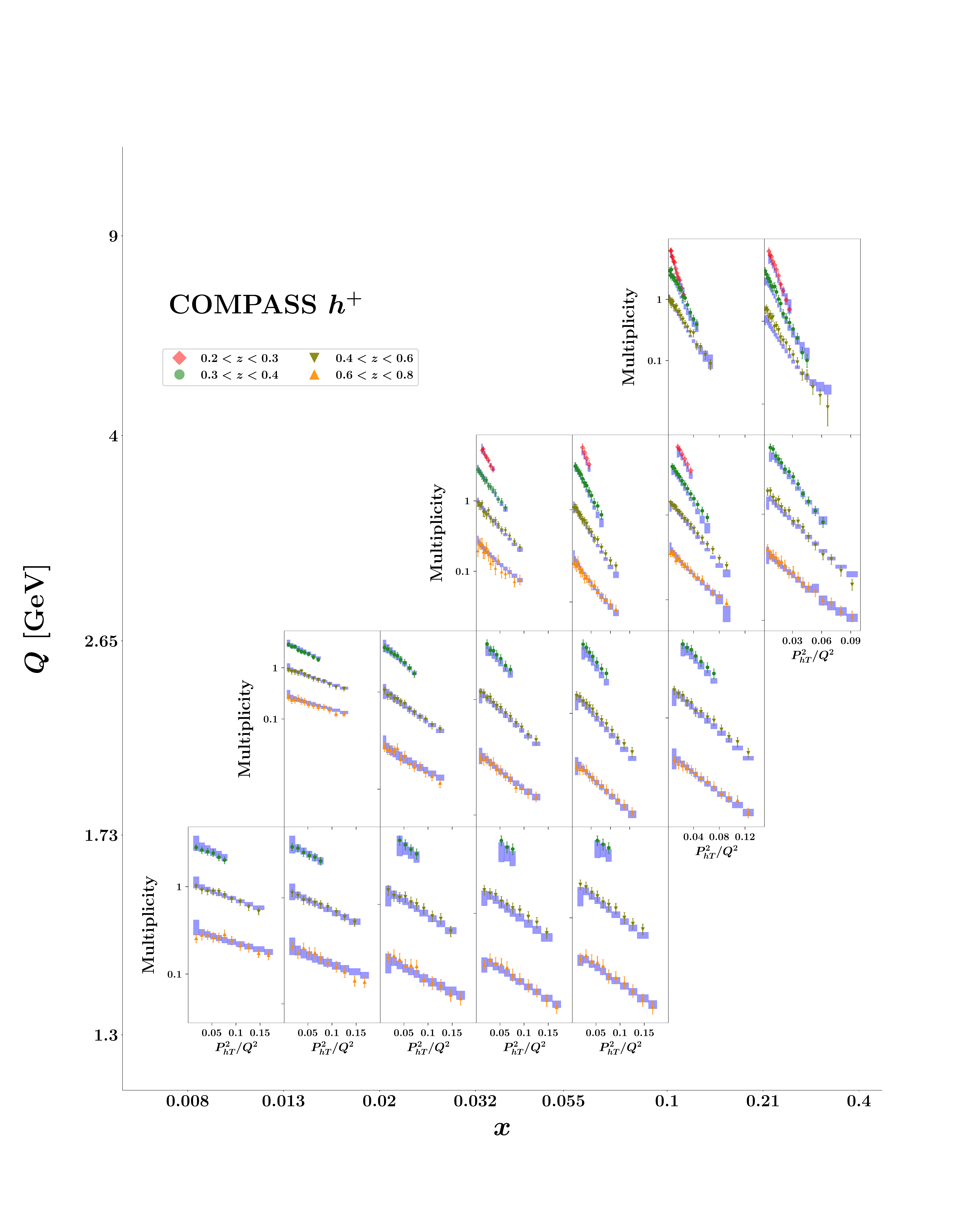}
\caption{Same as in the previous figure but for the production of positive charged hadrons off a deuteron target.}
\label{f:COMP_plot_P}
\end{figure}

In this Appendix, we present in Figs.~\ref{f:E288_plot}-\ref{f:COMP_plot_P} the quality of our fit (MAPTMD24 FD) for most of the used data. The blue error bands represent the 68\% C.L. of the theoretical predictions. 

\section{Nonperturbative parameters}
\label{appendixB}
\begin{figure}
\centering
\includegraphics[width=1.0\textwidth]{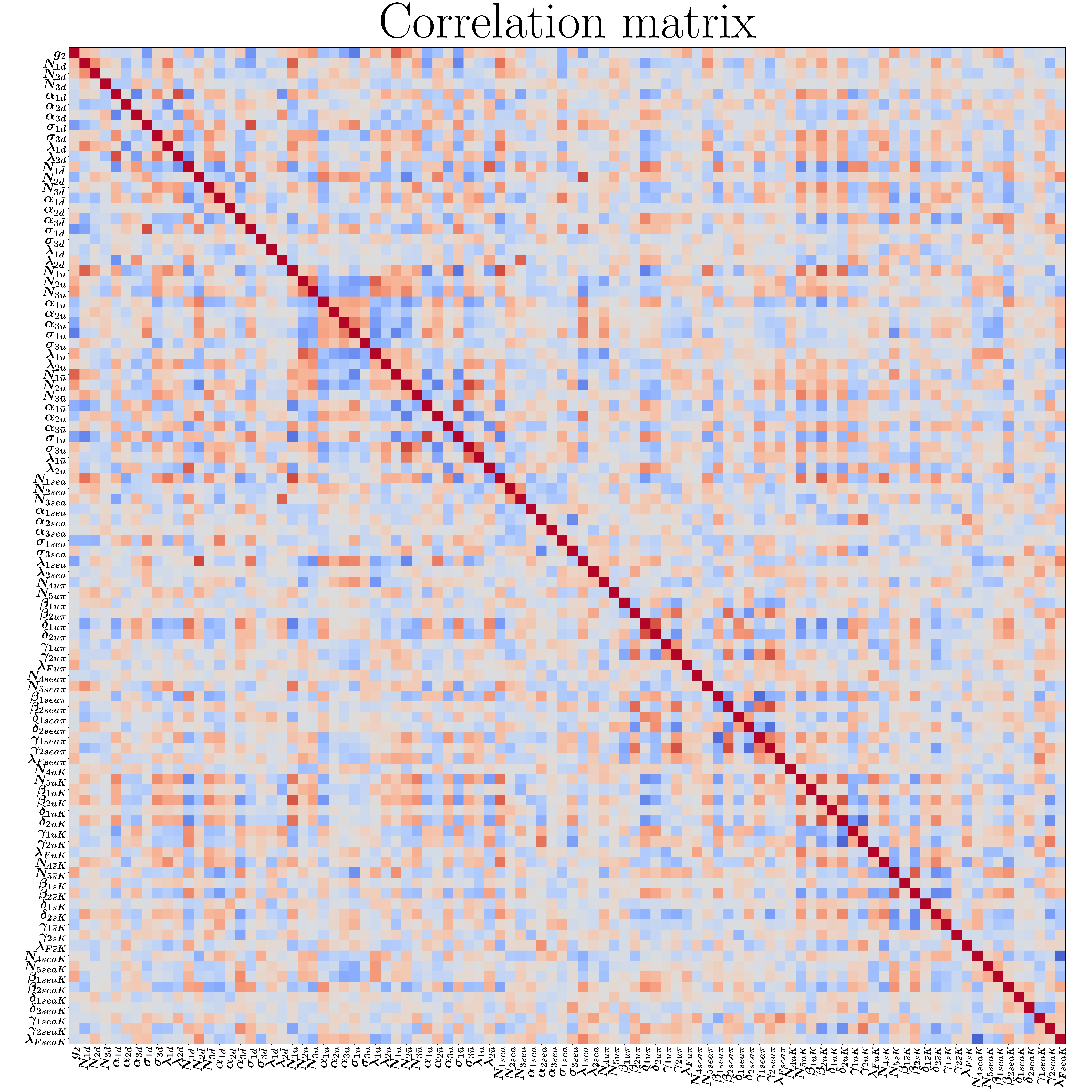}
\caption{ Graphical representation of the correlation matrix for the free parameters of the MAPTMD24 FD fit; color code ranges from blue (-1) to red (+1).}
\label{f:correlation_matrix}
\end{figure}

In Tabs.~\ref{t:FIparam}, \ref{t:HDparam}, and \ref{t:parameters}
we report the tables with the central values of the
fitted
parameters for the MAPTMD24 FI, MAPTMD24 HD, and MAPTMD24 FD extractions. For the
latter one, in Fig.~\ref{f:correlation_matrix} we also show a graphical representation of the correlation matrix.

\begin{table}
\parbox{.45\linewidth}{
\centering
\begin{tabular}{|c|c|}
  \hline
  \textbf{Parameter} & \textbf{Average over replicas} \\
\hline
$g_2 \ [\text{GeV}]$ & 0.080 $\pm$ 0.030 \\
\hline
$N_1 \ [\text{GeV}^2]$ & 0.42 $\pm$ 0.022 \\
\hline
$N_{2} \ [\text{GeV}^2]$ & 0.022 $\pm$ 0.003 \\
\hline
$N_{3} \ [\text{GeV}^2]$ & $(49 \pm 7.8) \times 10^{-5}$ \\
\hline
$\alpha_1$ & 0.21 $\pm$ 0.20 \\
\hline
$\alpha_2$ & 5.42 $\pm$ 0.074 \\
\hline
$\alpha_3$ & 2.27 $\pm$ 0.34 \\
\hline
$\sigma_1$ & -0.11 $\pm$ 0.03 \\
\hline
$\sigma_3$ & 10.16 $\pm$ 0.34 \\
\hline
$\lambda_1 \ [\text{GeV}^{-1}]$ & 0.48 $\pm$ 0.060 \\
\hline
$\lambda_2 \ [\text{GeV}^{-1}]$ & 0.095 $\pm$ 0.016 \\
\hline
$N_4 \ [\text{GeV}^2]$ & $(107 \pm 6.0) \times 10^{-5}$ \\
\hline
$N_{5} \ [\text{GeV}^2]$ & 0.11 $\pm$ 0.0036 \\
\hline
$\beta_1$ & 11.62 $\pm$ 0.22 \\
\hline
$\beta_2$ & 4.34 $\pm$ 0.17 \\
\hline
$\delta_1$ & 0.0023 $\pm$ 0.0021 \\
\hline
$\delta_2$ & 0.19 $\pm$ 0.012 \\
\hline
$\gamma_1$ & 1.27 $\pm$ 0.055 \\
\hline
$\gamma_2$ & 0.16 $\pm$ 0.15 \\
\hline
$\lambda_F \ [\text{GeV}^{-2}]$ & 0.16 $\pm$ 0.010 \\
\hline
\end{tabular}
\caption{Mean value and error related to the 68\% C.L. over the Monte Carlo replicas of the free parameters in the flavor-blind MAPTMD24 FI fit. \label{t:FIparam}}
}
\hfill
\parbox{.45\linewidth}{
\centering
\begin{tabular}{|c|c|}
  \hline
  \textbf{Parameter} & \textbf{Average over replicas} \\
\hline
$g_2  [\text{GeV}]$ & $0.11 \pm 0.016$ \\
\hline
$N_1  [\text{GeV}^2]$ & $0.40 \pm 0.014$ \\
\hline
$N_2  [\text{GeV}^2]$& $0.020 \pm 0.0022$ \\
\hline
$N_3  [\text{GeV}^2]$& $(3.8 \pm 1.5) \times 10^{-4}$ \\
\hline
$\alpha_1$ & $0.40 \pm 0.24$ \\
\hline
$\alpha_2$ & $5.4 \pm 0.026$ \\
\hline
$\alpha_3$ & $2.2 \pm 0.076$ \\
\hline
$\sigma_1$ & $-0.12 \pm 0.018$ \\
\hline
$\sigma_3$ & $10 \pm 0.030$ \\
\hline
$\lambda_1  [\text{GeV}^{-1}]$ & $0.48 \pm 0.089$ \\
\hline
$\lambda_2  [\text{GeV}^{-1}]$ & $0.084 \pm 0.0054$ \\
\hline
$N_{4\pi}  [\text{GeV}^2]$ & $(85 \pm 6.0) \times 10^{-5}$ \\
\hline
$N_{5\pi}  [\text{GeV}^2]$ & $0.096 \pm 0.0015$ \\
\hline
$\beta_{1\pi}$ & $5.1 \pm 0.28$ \\
\hline
$\beta_{2\pi}$ & $2.0 \pm 0.070$ \\
\hline
$\delta_{1\pi}$ & $0.0027 \pm 0.0027$ \\
\hline
$\delta_{2\pi}$ & $0.19 \pm 0.00075$ \\
\hline
$\gamma_{1\pi}$ & $1.4 \pm 0.059$ \\
\hline
$\gamma_{2\pi}$ & $0.88 \pm 0.038$ \\
\hline
$\lambda_{F\pi} [\text{GeV}^{-2}]$ & $0.082 \pm 0.0049$ \\
\hline
$N_{4K}  [\text{GeV}^2]$ & $(72 \pm 8.8) \times 10^{-5}$ \\
\hline
$N_{5K}  [\text{GeV}^2]$ & $0.15 \pm 0.0053$ \\
\hline
$\beta_{1K}$ & $8.5 \pm 0.52$ \\
\hline
$\beta_{2K}$ & $3.9 \pm 0.21$ \\
\hline
$\delta_{1K}$ & $0.0072 \pm 0.0065$ \\
\hline
$\delta_{2K}$ & $0.19 \pm 0.0095$ \\
\hline
$\gamma_{1K}$ & $1.3 \pm 0.14$ \\
\hline
$\gamma_{2K}$ & $0.18 \pm 0.15$ \\
\hline
$\lambda_{FK} [\text{GeV}^{-2}]$ & $0.16 \pm 0.021$ \\
\hline
\end{tabular}
\caption{Mean value and error related to the 68\% C.L. over the Monte Carlo replicas of the free parameters in the hadron-dependent MAPTMD24 HD fit. \label{t:HDparam}}
}

\end{table}

\begin{table}
\centering
\resizebox{1\textwidth}{!}{%
\begin{tabular}{|c|c|c|c|c|c|}
\hline
Parameter & Value & Parameter & Value & Parameter & Value \\ \hline
$g_2$ [GeV] & $0.12 \pm 0.0033$ & \multicolumn{4}{|c|}{} \\ \hline \hline
$N_{1d}$ [GeV$^2$] & $0.21 \pm 0.017$ & $N_{2d}$ [GeV$^2$] & $0.015 \pm 0.0013$ & $N_{3d}$ [GeV$^2$] & $(40 \pm 2.2) \times 10^{-4}$ \\ \hline
$\alpha_{1d}$ & $0.86 \pm 0.11$ & $\alpha_{2d}$ & $5.5 \pm 0.041$ & $\alpha_{3d}$ & $2.38 \pm 0.032$  \\ \hline
$\sigma_{1d}$ & $-0.21 \pm 0.013$ & $\sigma_{2d} = \sigma_{3d}$ & $9.91 \pm 0.061$ & \multicolumn{2}{|c|}{} \\ \hline
$\lambda_{1d}$ [GeV$^{-1}$] & $0.32 \pm 0.038$ & $\lambda_{2d}$ [GeV$^{-1}$] & $0.052 \pm 0.0022$ & \multicolumn{2}{|c|}{} \\ \hline
$N_{1\bar{d}}$ [GeV$^2$] & $0.68 \pm 0.038$ & $N_{2\bar{d}}$ [GeV$^2$] & $0.0037 \pm 0.0037$ & $N_{3\bar{d}}$ [GeV$^2$] & $(5.9 \pm 5.8)\times 10^{-5}$ \\ \hline
$\alpha_{1\bar{d}}$ & $0.64 \pm 0.18$ & $\alpha_{2\bar{d}}$ & $5.69 \pm 0.64$ & $\alpha_{3\bar{d}}$ & $1.57 \pm 0.53$ \\ \hline
$\sigma_{1\bar{d}}$ & $0.075 \pm 0.012$ & $\sigma_{2\bar{d}} = \sigma_{3\bar{d}}$ & $10.19 \pm 0.09$ & \multicolumn{2}{|c|}{} \\ \hline
$\lambda_{1\bar{d}}$ [GeV$^{-1}$] & $0.7 \pm 0.67$ & $\lambda_{2\bar{d}}$ [GeV$^{-1}$] & $0.051 \pm 0.0071$ & \multicolumn{2}{|c|}{} \\ \hline
$N_{1u}$ [GeV$^2$] & $0.35 \pm 0.0063$ & $N_{2u}$ [GeV$^2$] & $0.019 \pm 0.00015$ & $N_{3u}$ [GeV$^2$] & $(355 \pm 4.5) \times 10^{-6}$ \\ \hline
$\alpha_{1u}$ & $0.18 \pm 0.1$ & $\alpha_{2u}$ & $5.42 \pm 0.0037$ & $\alpha_{3u}$ & $2.14 \pm 0.0068$ \\ \hline
$\sigma_{1u}$ & $-0.26 \pm 0.0079$ & $\sigma_{2u} = \sigma_{3u}$ & $10.17 \pm 0.011$ & \multicolumn{2}{|c|}{} \\ \hline
 $\lambda_{1u}$ [GeV$^{-1}$] & $0.49 \pm 0.0037$ & $\lambda_{2u}$ [GeV$^{-1}$] & $0.081 \pm 0.0009$ & \multicolumn{2}{|c|}{} \\ \hline
$N_{1\bar{u}}$  [GeV$^2$] & $0.48 \pm 0.0074$ & $N_{2\bar{u}}$ [GeV$^2$] & $0.022 \pm 0.00037$ & $N_{3\bar{u}}$ [GeV$^2$] & $(21 \pm 1.5) \times 10^{-5}$ \\ \hline
 $\alpha_{1\bar{u}}$ & $0.95 \pm 0.077$ & $\alpha_{2\bar{u}}$ & $5.38 \pm 0.0099$ & $\alpha_{3\bar{u}}$ & $1.77 \pm 0.052$ \\ \hline
$\sigma_{1\bar{u}}$ & $-0.026 \pm 0.01$ & $\sigma_{2\bar{u}} = \sigma_{3\bar{u}}$ & $10.21 \pm 0.02$ & \multicolumn{2}{|c|}{} \\ \hline
$\lambda_{1\bar{u}}$ [GeV$^{-1}$] & $0.53 \pm 0.0067$ & $\lambda_{2\bar{u}}$ [GeV$^{-1}$] & $0.11 \pm 0.0055$ & \multicolumn{2}{|c|}{} \\ \hline
$N_{1sea}$ [GeV$^2$] & $0.16 \pm 0.035$ & $N_{2sea}$ [GeV$^2$] & $0.029 \pm 0.0027$ & $N_{3sea}$ [GeV$^2$] & $0.0039 \pm 0.002$ \\ \hline
$\alpha_{1sea}$ & $0.65 \pm 0.48$ & $\alpha_{2sea}$ & $5.24 \pm 0.032$ & $\alpha_{3sea}$ & $1.48 \pm 0.74$ \\ \hline
$\sigma_{1sea}$ & $-0.018 \pm 0.022$ & $\sigma_{2sea} = \sigma_{3sea}$ & $10.72 \pm 0.037$ & \multicolumn{2}{|c|}{} \\ \hline
$\lambda_{1sea}$ [GeV$^{-1}$] & $2.43 \pm 0.97$ & $\lambda_{2sea}$ [GeV$^{-1}$] & $0.015 \pm 0.0083$ & \multicolumn{2}{|c|}{} \\ \hline \hline
$N_{4u\pi}$ [GeV$^2$] & $(82 \pm 1.8) \times 10^{-5}$ & $N_{5u\pi}$ [GeV$^2$] & $0.095 \pm 0.0008$ & $\beta_{1u\pi}$ & $5.19 \pm 0.066$ \\ \hline
$\beta_{2u\pi}$ & $2.3 \pm 0.041$ & $\delta_{1u\pi}$ & $0.017 \pm 0.0084$ & $\delta_{2u\pi}$ & $0.19 \pm 0.0049$ \\ \hline
$\gamma_{1u\pi}$ & $1.46 \pm 0.015$ & $\gamma_{2u\pi}$ & $0.8 \pm 0.0095$ & $\lambda_{Fu\pi}$ [GeV$^{-2}$] & $0.089 \pm 0.003$ \\ \hline
$N_{4sea\pi}$ [GeV$^2$] & $(83 \pm 2.4) \times 10^{-5}$ & $N_{5sea\pi}$ [GeV$^2$] & $0.094 \pm 0.0012$ & $\beta_{1sea\pi}$ & $5.38 \pm 0.21$ \\ \hline
$\beta_{2sea\pi}$ & $2.31 \pm 0.072$ & $\delta_{1sea\pi}$ & $0.022 \pm 0.0064$ & $\delta_{2sea\pi}$ & $0.19 \pm 0.0044$ \\ \hline
$\gamma_{1sea\pi}$ & $1.44 \pm 0.026$ & $\gamma_{2sea\pi}$ & $0.8 \pm 0.012$ & $\lambda_{Fsea\pi}$ [GeV$^{-2}$] & $0.086 \pm 0.004$ \\ \hline \hline
$N_{4uK}$ [GeV$^2$] & $(87 \pm 5.7) \times 10^{-5}$ & $N_{5uK}$ [GeV$^2$] & $0.14 \pm 0.0026$ & $\beta_{1uK}$ & $8.52 \pm 0.081$ \\ \hline
$\beta_{2uK}$ & $3.86 \pm 0.19$ & $\delta_{1uK}$ & $0.0061 \pm 0.0035$ & $\delta_{2uK}$ & $0.19 \pm 0.0059$ \\ \hline
$\gamma_{1uK}$ & $1 \pm 0.041$ & $\gamma_{2uK}$ & $0.19 \pm 0.054$ & $\lambda_{FuK}$ [GeV$^{-2}$] & $0.14 \pm 0.0048$ \\ \hline
$N_{4\bar{s}K}$ [GeV$^2$] & $(4.5 \pm 3.7) \times 10^{-4}$ & $N_{5\bar{s}K}$ [GeV$^2$] & $0.16 \pm 0.016$ & $\beta_{1\bar{s}K}$ & $7.17 \pm 1.4$ \\ \hline
$\beta_{2\bar{s}K}$ & $5.1 \pm 1.04$ & $\delta_{1\bar{s}K}$ & $1.51 \pm 1.51$ & $\delta_{2\bar{s}K}$ & $0.16 \pm 0.033$ \\ \hline
$\gamma_{1\bar{s}K}$ & $0.71 \pm 0.42$ & $\gamma_{2\bar{s}K}$ & $0.36 \pm 0.19$ & $\lambda_{F\bar{s}K}$ [GeV$^{-2}$] & $0.34 \pm 0.2$ \\ \hline
$N_{4seaK}$ [GeV$^2$] & $(78 \pm 2.8) \times 10^{-5}$ & $N_{5seaK}$ [GeV$^2$] & $0.15 \pm 0.0059$ & $\beta_{1seaK}$ & $8.63 \pm 0.24$ \\ \hline
$\beta_{2seaK}$ & $4.19 \pm 0.14$ & $\delta_{1seaK}$ & $0.0075 \pm 0.0051$ & $\delta_{2seaK}$ & $0.2 \pm 0.0029$ \\ \hline
$\gamma_{1seaK}$ & $0.96 \pm 0.036$ & $\gamma_{2seaK}$ & $0.17 \pm 0.092$ & $\lambda_{FseaK}$ [GeV$^{-2}$] & $0.15 \pm 0.0055$ \\ \hline
\end{tabular}%
}
\caption{Table of the 96 free parameters in the flavor-dependent MAPTMD24 FD fit. For each parameter, the mean value and the error related to the 68\% C.L. are reported.}
\label{t:parameters}
\end{table}


\clearpage

\bibliographystyle{JHEP}
\bibliography{MAPTMD24.bib}
\end{document}